\documentclass[12pt]{article}
\pdfoutput=1
\usepackage{epsfig,amsfonts,amsthm}
\usepackage{amsmath,amssymb}
\usepackage{array}
\newcommand{\be}{\begin{equation}}
\newcommand{\ee}{\end{equation}}
\newcommand{\bea}{\begin{eqnarray}}
\newcommand{\eea}{\end{eqnarray}}

\renewcommand{\Re}{\mathrm{Re }}
\renewcommand{\Im}{\mathrm{Im }}
\newcommand{\doublet}[2]{ \left( \begin{array}{c}#1 \\ #2 \end{array}\right) }

\newcommand{\Z}{\mathbb{Z}}

\providecommand{\RR}{{\mathbb{R}}}

\def\lsim{\mathrel{\rlap{\lower4pt\hbox{\hskip1pt$\sim$}}
    \raise1pt\hbox{$<$}}}         
\def\gsim{\mathrel{\rlap{\lower4pt\hbox{\hskip1pt$\sim$}}
    \raise1pt\hbox{$>$}}}         

\def\beq{\begin{equation}}
\def\eeq{\end{equation}}
\def\bea{\begin{eqnarray}}
\def\eea{\end{eqnarray}}

\def\<{\left\langle}
\def\>{\right\rangle}

\newcommand{\bt}{\begin{tabular}}
\newcommand{\et}{\end{tabular}}

\allowdisplaybreaks[2]
\begin{document}
\bibliographystyle{OurBibTeX}

\title{\hfill ~\\[-30mm]
                  \textbf{Three-Higgs-doublet models: symmetries, potentials and Higgs boson masses
                }        }
\date{}
\author{\\[-5mm]
Venus~Keus\footnote{E-mail: {\tt V.Keus@soton.ac.uk}} $^{1,2,3}$,\ 
Stephen F. King\footnote{E-mail: {\tt king@soton.ac.uk}} $^{1}$,\ 
Stefano~Moretti\footnote{E-mail: {\tt S.Moretti@soton.ac.uk}} $^{1,3}$ 
\\ \\
  \emph{\small $^1$ School of Physics and Astronomy, University of Southampton,}\\
  \emph{\small Southampton, SO17 1BJ, United Kingdom}\\
  \emph{\small  $^2$ Department of Physics, Royal Holloway, University of London,}\\
  \emph{\small Egham Hill, Egham TW20 0EX, United Kingdom}\\
  \emph{\small $^3$ Particle Physics Department, Rutherford Appleton Laboratory,}\\
  \emph{\small Chilton, Didcot, Oxon OX11 0QX, United Kingdom}\\[4mm]}

\maketitle

\begin{abstract}
\noindent
{We catalogue and study three-Higgs-doublet models in terms of all possible allowed symmetries
(continuous and discrete, Abelian and non-Abelian), where such symmetries
may be identified as flavour symmetries of quarks and leptons.
We analyse the potential in each case, and derive the conditions under which 
the vacuum alignments $(0,0,v)$, $(0,v,v)$ and $(v,v,v)$ are minima of the potential.
For the alignment $(0,0,v)$, relevant for dark matter models, we calculate the corresponding 
physical Higgs boson mass spectrum. Motivated by supersymmetry, we extend the analysis
to the case of three up-type Higgs doublets and three down-type Higgs doublets (six doublets in total).
Many of the results are also applicable to flavon
models where the three Higgs doublets are replaced by three electroweak singlets.} 
 \end{abstract}
\thispagestyle{empty}
\vfill
\newpage
\setcounter{page}{1}

\section{Introduction}

The discovery of a Higgs boson by the Large Hadron Collider (LHC) \cite{Aad:2012tfa}, whose properties are
consistent with those predicted by the Standard Model (SM), 
indicates that at least one Higgs doublet must be responsible for electroweak symmetry breaking (EWSB).
However, there is no special reason why there should be only one Higgs doublet in Nature,
and it is entirely possible that there could be additional Higgs doublets, accompanied by further Higgs bosons
which could be discovered in the next run of the LHC.

The simplest example of $N$-Higgs-doublet models (NHDMs) is the class of two-Higgs-doublet models 
(2HDMs) \cite{Gunion:1989we},
a special example being that predicted by the minimal supersymmetric standard model 
(MSSM) \cite{Chung:2003fi}.  Of course, the 
general class of 2HDMs is much richer than the MSSM example, and indeed all possible types 
of 2HDMs have been well studied in the literature \cite{Chang:2013ona}. However, 2HDMs generally face severe phenomenological problems with flavour changing neutral currents (FCNCs) and 
possible charge breaking vacua, and it is common to consider restricted classes of models controlled
by various symmetries.

The introduction of symmetries into 2HDMs provides a welcome restriction to the
rather unwieldy general Higgs potential, as well as solutions to the phenomenological challenges mentioned above. For example, the remaining symmetry of the potential after EWSB can have the effect of stabilising the lightest Higgs boson, which can become a possible Dark Matter (DM) candidate. 
In 2HDMs, the full list of possible 
symmetries of the potential is now known \cite{2HDM}: the symmetry group can be $Z_2$,
$(Z_2)^2$, $(Z_2)^3$, $O(2)$, $O(2) \times Z_2$, or $O(3)$\footnote{The maximum number of distinct symmetries in 2HDMs is 13 if the custodial symmetries are included \cite{Battye,Pilaftsis:2011ed}. In this paper we shall not consider custodial symmetries in 3HDM.}.
In 2HDMs these symmetries can be conserved or spontaneously violated after the EWSB, depending on the coefficients of the potential.

Generalising these results to NHDMs is technically difficult, although there has been
some recent progress in this direction \cite{Ivanov:2011ae,Ivanov:2010ww,Ivanov:2010wz,Mohapatra:2011gp}.
For example, with more than two doublets, there exist symmetries which are always spontaneously violated after EWSB, dubbed ''frustrated symmetries" in analogy with a similar phenomenon in condensed-matter physics
\cite{Ivanov:2010zx}. The idea of stabilising the lightest scalar via a preserved 
$Z_p$ symmetry (where $p$ is an integer) has also been put forward 
for NHDMs \cite{Ivanov:2012hc}.

The case of three-Higgs-doublet models (3HDMs) is particularly promising for several reasons. 
To begin with, it is the next simplest example beyond 2HDMs, which has been exhaustively studied in the
literature. Furthermore, 3HDMs are more tractable than NHDMs, and all possible finite symmetries (but not all continuous ones) have been identified \cite{Ivanov:2012fp}.
Finally, and perhaps most intriguingly, 3HDMs may shed light on the flavour problem, namely the problem
of the origin and nature of the three families of quarks and leptons, including neutrinos, and their 
pattern of masses, mixings and CP violation. 
It is possible that the three families of quarks and leptons could be described by the same symmetries that describe the three Higgs doublets \cite{Felipe:2013ie}. 
In such models this family symmetry could be spontaneously broken along with the electroweak symmetry, although some remnant subgroup could survive, thereby stabilising a possible scalar DM candidate. 
For certain symmetries it is possible to find a vacuum expectation value (VEV) alignment that respects the original symmetry of the potential which will then be responsible for the stabilization of the DM candidate as in \cite{Ivanov:2012hc}.

Despite the motivations above, 3HDMs remain rather poorly understood, at least when compared to 2HDMs,
as can be clearly seen by the list of outstanding problems stated in \cite{Ivanov:2012fp}.
The outstanding problems include \cite{Ivanov:2012fp}:
the completion of the classification of possible symmetries to include continuous symmetry groups; the possible symmetry breaking patterns of the vacuum state for each choice of symmetry group; additional symmetries of the potential which are not symmetries of the kinetic terms. We would add to this list: the possible different vacuum alignments for each choice of symmetry group; the calculation of the Higgs boson mass spectrum for 
each such vacuum alignment; the possible DM candidates in the case where 
there is a preserved symmetry; the application of 3HDMs to 
quark and lepton flavour models \cite{Felipe:2013ie};
the extension of 3HDMs to the case of supersymmetric (SUSY) models
which motivates having three up-type Higgs doublets and three down-type Higgs doublets (six doublets in total).
Finally the possible generalised CP symmetries and their breaking patterns have not yet been thoroughly studied.

The purpose of the present paper is to consider some of the aforementioned aspects of 
3HDMs. We shall complete the classification of 3HDMs in terms of all possible Abelian symmetries (continuous and discrete) and all possible discrete non-Abelian symmetries. 
We analyse the potential in each case, and derive the conditions under which 
the vacuum alignments $(0,0,v)$, $(0,v,v)$ and $(v,v,v)$ are minima of the potential.
For the alignment $(0,0,v)$, which is of particular interest because of its relevance for dark matter models and the absence of FCNCs, we calculate the corresponding 
physical Higgs boson mass spectrum.
This will lead to phenomenological constraints on the parameters
in the potential, and for certain parameter choices it is possible that there could be additional
light Higgs bosons which may have evaded detection at LEP. 
It is possible that the 125 GeV Higgs boson could decay
into these lighter Higgs bosons, providing striking new signatures for Higgs decays at the LHC. 
Motivated by SUSY, we then extend also the analysis
to the case of three up-type Higgs doublets and three down-type Higgs doublets (six doublets in total)
for the case of low ratio of VEVs $\tan \beta = v_u/v_d $. 
The results could be applicable to various SUSY models with three up-type and down-type 
Higgs families including MSSM
where both Higgs types transform as triplets of $A_4$ \cite{Morisi:2011pt} 
and a version of the E$_6$SSM \cite{King:2005jy} where both Higgs types transform 
as triplets of $\Delta (27)$ \cite{Howl:2009ds}.
In both these examples, only the Higgs fields of the third family are (predominantly) assumed to develop VEVs
which motivates the vacuum alignment $(0,0,v)$.
Many of the results are also applicable to flavon
models where the three Higgs doublets are replaced by three electroweak singlets.

The layout of the remainder of the paper is as follows.
In section~\ref{syms-in-3hdm} we discuss the possible symmetries and symmetry breaking 
patterns in 3HDMs, giving the most general invariant potential, outlining the method we use for finding the 
local minima of the potential, and motivating the relevance of the particular alignment $(0,0,v)$ for DM.
In section~\ref{all-abelian} we systematically discuss the 3HDM potentials which respect Abelian symmetries,
deriving the conditions under which 
the vacuum alignments $(0,0,v)$, $(0,v,v)$ and $(v,v,v)$ are minima of the potential and, 
for the alignment $(0,0,v)$, calculating the corresponding 
physical Higgs boson mass spectrum.
In section~\ref{finite-unitary} we perform an analogous analysis for 
the 3HDM potentials which respect non-Abelian finite symmetries.
In section~\ref{6HDM} we discuss our method for extending the results to the case
of six-Higgs-doublet models (6HDMs), relevant for supersymmetric models, where the method is exemplified for $\tan \beta = 1$.
Section~\ref{conclusion} concludes the paper.
We also include two Appendices where we review how 
highly symmetric potentials can be treated using orbit spaces and geometric minimisation,
which are powerful techniques enabling statements to be made about the global minimum
of the potential.

\section{Symmetries and Symmetry Breaking in 3HDMs}\label{syms-in-3hdm}

When studying the symmetries of the potential, one focuses on the reparametrisation transformations, 
which mix the three different Higgs doublets $\phi_a$, where $a=1,\cdots 3$.
These transformations keep the kinetic term invariant, and are either unitary (Higgs-family transformation) or anti-unitary (generalised CP-transformation), however we shall only consider the former here.

The kinetic terms of 3HDMs are invariant under the unitary transformations
\be 
U: \quad \phi_a \mapsto U_{ab}\phi_b\qquad 
\ee
with a $3\times 3$ unitary matrix $U_{ab}$, where $a,b=1,\cdots 3$.
These transformations form the group $U(3)$. However,
the overall phase factor multiplication is already taken into account by the $U(1)_Y$ from the gauge group.
The kinetic terms of 3HDMs are therefore invariant under the $SU(3)$ group of reparametrisation transformations.
The group $SU(3)$ has a non-trivial center $Z(SU(3))= \Z_3$ generated by the diagonal matrix $\exp(2\pi i/3)\cdot 1_3$, where $1_3$ is the identity matrix. Therefore, the group of physically distinct unitary reparametrisation transformations respected by the kinetic terms $G_0$ is
\be
G_0= PSU(3) \simeq SU(3)/\Z_3.
\ee

In general, the Higgs potential $V$ in 3HDMs will respect a symmetry $G$ which is some subgroup
of $G_0$.
If a symmetry $G$ is imposed on the potential $V$, it sometimes happens that it accidentally respects
a larger symmetry. 
When a potential is symmetric under a group $G$ and not under any larger group containing $G$, then the group 
$G$ is called a realisable group \cite{Ivanov:2011ae}.
A non-realisable group imposed on a potential, on the other hand, leads to a larger symmetry group of the potential. Clearly, the true symmetry properties of the potentials are reflected in realisable groups.

The full list of possible realisable symmetries $G$ of the scalar potential $V$ of 3HDMs have been found \cite{Ivanov:2011ae, Ivanov:2012fp}. For the reader's convenience, we briefly review the methodology applied to obtain these symmetries in Appendix \ref{methodology}. Here we only list all such symmetry group which consist of the continuous Abelian symmetry groups,
\be 
U(1), \quad U(1) \times U(1) , \quad U(1) \times Z_2,
\label{continuous-abelian}
\ee
the finite Abelian symmetry groups,
\be 
Z_2, \quad Z_3, \quad Z_4,\quad Z_2 \times Z_2, 
\label{finite-abelian}
\ee
and the finite non-Abelian symmetry groups,
\bea
\label{finine-non-abelian}
&&D_6,\quad D_8, \quad T\simeq A_4,\quad O\simeq S_4, \\
&& (Z_3\times Z_3)\rtimes Z_2 \simeq \Delta(54)/ Z_3,\quad (Z_3\times Z_3)\rtimes Z_4\simeq \Sigma(36). \nonumber
\eea

A scalar 3HDM potential symmetric under a group $G$ can be written as 
\be 
V = V_0 + V_G
\ee
where $V_0$ is invariant under any phase rotation and $V_G$ is a collection of extra terms ensuring the symmetry group $G$. The most general invariant part of the potential has the following form
\be
V_0 = \sum^3_i \left[- |\mu^2_i| (\phi_i^\dagger \phi_i) + \lambda_{ii} (\phi_i^\dagger \phi_i)^2\right] 
+ \sum^3_{ij}\left[\lambda_{ij}(\phi_i^\dagger \phi_i) (\phi_j^\dagger \phi_j) + 
\lambda'_{ij}(\phi_i^\dagger \phi_j) (\phi_j^\dagger \phi_i)\right] .
\label{general-pot}
\ee
For this potential to have a stable vacuum (bounded from below) the following conditions are required:
\be
\lambda_{ii}>0, \qquad \lambda_{ij}+\lambda'_{ij}>-2\sqrt{\lambda_{ii}\lambda_{jj}}, \qquad i\neq j =1,2,3. 
\ee
The potential $V_0$ and associated stability conditions above 
are common to all the cases, Abelian and non-Abelian,
which only differ by $V_G$.

After constructing a potential symmetric under a realisable group, we shall find the minima of the potential by explicit calculation in each case by: 
\begin{itemize}
\item
Parametrising the VEVs by $v_i$
\item
Expanding the potential around the minimum point and calculating $V(v_i)$
\item
Setting all $\partial V / \partial v_i$ to zero and solving these equations for $v_i$
\item
Constructing the Hessian from $\partial^2 V / \partial v_i \partial v_j$ and requiring it to be positive definite 
\end{itemize}
We shall follow this standard method for minimising potentials symmetric under all groups listed in (\ref{continuous-abelian}), (\ref{finite-abelian}) and (\ref{finine-non-abelian}).

In this paper we shall present the 3HDM potentials symmetric under each of these groups, 
examine the conditions for stability of the three possible minima in each case,
\be 
(0,0,v), \quad (0,v,v), \quad (v,v,v).
\ee
We shall focus on one of these minima, $(0,0,v)$ and study the mass spectrum around this point. 
The vacuum $(0,0,v)$ may either respect or breaks the original symmetry of the potential. 

The motivation for considering the vacuum alignment $(0,0,v)$ is that such models can be interpreted as 
DM models in which the weakly interacting massive particle (WIMP) is identified as the lightest neutral scalar boson.
If we define the three Higgs doublets as
\be 
\phi_1= \doublet{$\begin{scriptsize}$ H^+_1 $\end{scriptsize}$}{\frac{H^0_1+iA^0_1}{\sqrt{2}}} ,\quad 
\phi_2= \doublet{$\begin{scriptsize}$ H^+_2 $\end{scriptsize}$}{\frac{H^0_2+iA^0_2}{\sqrt{2}}} , \quad 
\phi_3= \doublet{$\begin{scriptsize}$ H^+_3 $\end{scriptsize}$}{\frac{v+H^0_3+iA^0_3}{\sqrt{2}}}  
\label{explicit-fields}
\ee
then the vacuum $(0,0,v)$ corresponds to having 
two inert doublets ($\phi_1$ and $\phi_2$) and one active doublet ($\phi_3$).
``Inert'' means not only zero VEV but also no couplings to fermions.
To be precise, if the symmetry of the potential after EWSB is $G$, we assign a quantum number to each doublet according to the generator of $G$. To make sure that the entire Lagrangian and not only the scalar potential is $G$ symmetric, we set the $G$ quantum number of all SM particles equal to the $G$ quantum number of the only doublet that couples to them i.e. the active doublet $\phi_3$. With this charge assignment FCNCs are avoided as the extra doublets are forbidden to couple to fermions by $G$ conservation.
In each case, with the vacuum alignment $(0,0,v)$, the CP-even/odd neutral fields resulting from the inert doublets ($H^0_1, H^0_2, A^0_1, A^0_2$) could in principle be dark matter (DM) candidates since only the active doublet acquires a VEV and couples to the fermions. To stabilize the DM candidate from decaying into SM particles, we make use of the remnant symmetry of the potential after EWSB.

For the special cases of the non-Abelian symmetries listed in Eq.~(\ref{finine-non-abelian}), when the symmetry of the potential is sufficiently large, a powerful geometric method for minimising the potential has been introduced \cite{Degee:2012sk}.
From this list of finite non-Abelian symmetries in (\ref{finine-non-abelian}), the following symmetries are ``frustrated''
\footnote{In the case of 2HDMs, it is always possible to find a vacuum alignment which respects the symmetry of the potential. However, in NHDMs with $N>2$ when the potential is highly symmetric (having $M_i=0$ when written in terms of the bilinears as in Eq.~(\ref{potential})) any vacuum alignment breaks the symmetry of the potential.
Therefore, these symmetry groups can never be conserved after EWSB. The origin of this phenomena and some 3HDM examples were discussed in \cite{Ivanov:2010zx}. These symmetries are not specific to doublets. They can arise when the representation of the electroweak group has lower dimensionality than the horizontal (Higgs family) space, i.e. more than one singlet, more than two doublets, more than three triplets, etc.}
since they are inevitably broken after EWSB:
\be 
A_4,\quad S_4 , \quad  \Delta(54)/ Z_3,\quad  \Sigma(36) .
\label{frustrated}
\ee
For each of these four cases we rewrite the potential in terms of the bilinears and using the geometric method introduced in \cite{Degee:2012sk} find all minima of the potential. The results of this method are presented in Appendix \ref{Appendix-geometric} using the orbit space formalism briefly reviewed in Appendix \ref{orbit}.


\section{Analysis of Abelian 3HDMs}
\label{all-abelian}

In this section we study all Abelian symmetries in 3HDM potentials. Table \ref{Abelian-table} lists all these symmetry groups, their generators and their corresponding potentials.

\begin{table} [ht]
\begin{footnotesize}
\begin{center}
\begin{tabular}{|m{2cm} || m{5.5cm}| m{6.5cm}|} \hline
\mbox{Symmetry}          & \mbox{Diagonal Generators}   &  \mbox{Potential}   \\[3mm] \hline  \hline
$U(1) \times U(1)$ & $(e^{-i\alpha}, e^{i\alpha}, 1), \quad (e^{-2i\beta /3}, e^{i\beta / 3}, e^{i\beta / 3})$   & $V_0$        \\[3mm] \hline
$U(1)$             & $(e^{-i\alpha}, e^{i\alpha}, 1)$  & $V_0 + \lambda_1(\phi^\dagger_1 \phi_3)(\phi^\dagger_2 \phi_3) +h.c.$     \\[3mm] \hline  
$U(1) \times Z_2$  & $ (e^{-2i\beta /3}, e^{i\beta / 3}, e^{i\beta / 3}), \quad(-1,-1,1)$   & $V_0 + \lambda_1(\phi^\dagger_2 \phi_3)^2 +h.c.$  \\[3mm] \hline  
$Z_2$              & $(-1,-1,1)$  & $V_0 -\mu'^2_{12}(\phi_1^\dagger\phi_2)+\lambda_{1}(\phi_1^\dagger\phi_2)^2 +
                   \lambda_2(\phi_2^\dagger\phi_3)^2+\lambda_3(\phi_3^\dagger\phi_1)^2+h.c. $ \\[3mm] \hline  
$Z_3$              & $(\omega,\omega^2,1)$   & $ V_0 + \lambda_{1}(\phi_1^\dagger\phi_2)(\phi_3^\dagger\phi_2) + 
                   \lambda_{2}(\phi_2^\dagger\phi_3)(\phi_1^\dagger\phi_3) + \lambda_{3}(\phi_3^\dagger\phi_1)(\phi_2^\dagger\phi_1)+h.c. $  \\[3mm] \hline  
$Z_4$              & $(i,-i,1)$   & $V_0 +\lambda_1 (\phi_3^\dagger \phi_1)(\phi_3^\dagger \phi_2) + \lambda_2 (\phi_1^\dagger \phi_2)^2+h.c. $ \\[3mm] \hline  
$Z_2 \times Z_2$   & $(-1,1,1), \quad(1,-1,1)$   & $V_0+ \lambda_1 (\phi_1^\dagger \phi_2)^2 + \lambda_2(\phi_2^\dagger \phi_3)^2 + \lambda_3 (\phi_3^\dagger \phi_1)^2+h.c. $ \\[3mm] \hline  
\end{tabular}
\end{center}
\end{footnotesize}
\caption{\footnotesize All Abelian symmetries realisable in the scalar sector of 3HDMs. $V_0$ is the phase invariant part of the potential presented in Eq.~(\ref{U(1)XU(1)-3HDM}).}
\label{Abelian-table}
\end{table}

\subsection{U(1)$\times$U(1) symmetric 3HDM potential}\label{Common-V0}

The $U(1) \times U(1) \subset PSU(3)$ group is generated by \be
U_1(1) = \left(\begin{array}{ccc}
e^{-i\alpha} & 0 & 0 \\
0 & e^{i\alpha} & 0   \\
0 & 0 & 1   \\
\end{array}
\right),\quad  
U_2(1) =\left(\begin{array}{ccc}
e^{-2i\beta /3} & 0 & 0 \\
0 & e^{i\beta/3} & 0   \\
0 & 0 &  e^{i\beta/3}  \\
\end{array}
\right)
\label{U(1)s}
\ee
where $\alpha,\beta \in [0,2\pi)$.

The most general $U(1) \times U(1)$-symmetric 3HDM potential has the following form:
\bea
\label{U(1)XU(1)-3HDM}
V_0 &=& - \mu^2_{1} (\phi_1^\dagger \phi_1) -\mu^2_2 (\phi_2^\dagger \phi_2) - \mu^2_3(\phi_3^\dagger \phi_3) \\
&&+ \lambda_{11} (\phi_1^\dagger \phi_1)^2+ \lambda_{22} (\phi_2^\dagger \phi_2)^2  + \lambda_{33} (\phi_3^\dagger \phi_3)^2 \nonumber\\
&& + \lambda_{12}  (\phi_1^\dagger \phi_1)(\phi_2^\dagger \phi_2)
 + \lambda_{23}  (\phi_2^\dagger \phi_2)(\phi_3^\dagger \phi_3) + \lambda_{31} (\phi_3^\dagger \phi_3)(\phi_1^\dagger \phi_1) \nonumber\\
&& + \lambda'_{12} (\phi_1^\dagger \phi_2)(\phi_2^\dagger \phi_1) 
 + \lambda'_{23} (\phi_2^\dagger \phi_3)(\phi_3^\dagger \phi_2) + \lambda'_{31} (\phi_3^\dagger \phi_1)(\phi_1^\dagger \phi_3)  \nonumber
\eea
which is symmetric under any phase rotation of doublets.

We find all possible extrema of the potential by requiring
\be 
\biggl(\frac{\partial V}{\partial \phi_i} \biggr)_{\phi_i = \langle \phi_i \rangle }= 0 , \qquad \biggl(\frac{\partial V}{\partial \phi^\dagger_i} \biggr)_{\phi_i = \langle \phi_i \rangle }= 0, \qquad i=1,2,3
\ee
which results in the following solutions
\be 
(0,0,v), \quad (0,v,v), \quad (v,v,v),
\ee
where in each case permutations are allowed and in general doublets could acquire non-equal VEVs ($v_1 \neq v_2 \neq v_3$). In the following sections, however, the conditions are derived for the presented VEV alignment and the results do not apply to permuted VEV alignments. 

To find the conditions on the parameters which are required for the above points to be minima of the potential, we construct the Hessian
\be 
 \left|\begin{array}{ccc} 
\frac{{\partial}^2 V}{\partial\phi_1 \partial \phi^\dagger_1} & 
\frac{{\partial}^2 V}{\partial\phi_1 \partial \phi^\dagger_2} & 
\frac{{\partial}^2 V}{\partial\phi_1 \partial \phi^\dagger_3}\\[3mm]
\frac{{\partial}^2 V}{\partial\phi_2 \partial \phi^\dagger_1} & 
\frac{{\partial}^2 V}{\partial\phi_2 \partial \phi^\dagger_2} & 
\frac{{\partial}^2 V}{\partial\phi_2 \partial \phi^\dagger_3} \\[3mm]
\frac{{\partial}^2 V}{\partial\phi_3 \partial \phi^\dagger_1} & 
\frac{{\partial}^2 V}{\partial\phi_3 \partial \phi^\dagger_2} & 
\frac{{\partial}^2 V}{\partial\phi_3 \partial \phi^\dagger_3} 
\end{array}\right| >0
\ee
and require it to be positive definite (see e.g. \cite{Unwin:2011rn}). 

We find the following results:
\begin{enumerate}

\item
\textbf{Point} $(0,0,\frac{v}{\sqrt{2}})$ breaks the symmetry of the potential to $U_1(1)$ and becomes the minimum at 
\be
v^2= \frac{\mu^2_3}{\lambda_{33}} \nonumber
\ee
provided the following conditions are satisfied:
\bea 
&& \bullet\quad  -\mu^2_1  +\frac{1}{2}(\lambda_{31}+\lambda'_{31} )v^2   >0  \\
&& \bullet\quad  -\mu^2_2  +\frac{1}{2}(\lambda_{23}+\lambda'_{23} )v^2   >0  \nonumber\\
&& \bullet\quad   \mu^2_3    >0  \nonumber
\eea
However, all these conditions are already required for the positivity of mass eigenstates at point $(0,0,\frac{v}{\sqrt{2}})$. Therefore, we conclude that this point is always a minimum of the potential.

\item
\textbf{Point} $(0,\frac{v}{\sqrt{2}},\frac{v}{\sqrt{2}})$ breaks the symmetry of the potential to a $Z_2$ generated by $(1,-1,-1)$. This point becomes the minimum of the potential at 
\be 
v^2= \frac{2\mu^2_3}{2\lambda_{33}+\lambda_{23}+\lambda'_{23}}= \frac{2\mu^2_2}{2\lambda_{22}+\lambda_{23}+\lambda'_{23}} \nonumber
\ee
when the following conditions are satisfied:
\bea 
&& \bullet\quad  \lambda_{22} >0 \\
&& \bullet\quad  \lambda_{33} > 0  \nonumber\\
&& \bullet\quad  -2\mu^2_1 + (\lambda_{12} + \lambda'_{12}+ \lambda_{31}+ \lambda'_{31})v^2 >0 \nonumber\\
&& \bullet\quad  4\lambda_{22}\lambda_{33} > (\lambda_{23}+\lambda'_{23})^2    \nonumber
\eea

\item
\textbf{Point} $(\frac{v}{\sqrt{2}},\frac{v}{\sqrt{2}},\frac{v}{\sqrt{2}})$ breaks the symmetry of the potential completely and becomes the minimum at 
\bea
v^2&=& \frac{2\mu^2_3}{2\lambda_{33}+\lambda_{23}+\lambda'_{23} + \lambda_{31}+\lambda'_{31}} \nonumber\\
&=& \frac{2\mu^2_2}{2\lambda_{22}+\lambda_{12}+\lambda'_{12}+\lambda_{23}+\lambda'_{23}} \nonumber\\
&=&  \frac{2\mu^2_1}{2\lambda_{11}+\lambda_{12}+\lambda'_{12}+\lambda_{31}+\lambda'_{31}} \nonumber
\eea 
with the following conditions:
\bea 
&& \bullet\quad   4\lambda_{11}\lambda_{22} >  (\lambda_{12} + \lambda'_{12})^2   \\
&& \bullet\quad  4\lambda_{11}\lambda_{22}\lambda_{33} + (\lambda_{12} + \lambda'_{12})(\lambda_{31} + \lambda'_{32})(\lambda_{23} + \lambda'_{23}) >  \nonumber\\
&& \qquad \lambda_{11} (\lambda_{23} + \lambda'_{23})^2 + \lambda_{22} (\lambda_{31} + \lambda'_{31})^2 +\lambda_{33} (\lambda_{12} + \lambda'_{12})^2   \nonumber
\eea

\end{enumerate}

\underline{\textbf{Higgs mass spectrum for} $(0,0,\frac{v}{\sqrt{2}})$}

The VEV alignment $(0,0,\frac{v}{\sqrt{2}})$ breaks the $U(1) \times U(1)$ symmetry of the potential to 
$U_1(1)$ where the fields from the first and the second doublets are assigned $U_1(1)$ quantum numbers $-1$ and $1$ respectively, and the fields from the third doublet get $U_1(1)$ quantum number zero. 
We can write the generator as:
\be 
U_1(1) = \alpha(-1,1,0).
\label{U1}
\ee
Writing the generator of the group in this form facilitates the task of assigning $U_1(1)$ charges to the doublets. 
We require all SM fields to have the same $U_1(1)$ charge as the active doublet i.e. zero so that the whole Lagrangian is $U_1(1)$-symmetric. 
Having defined the doublet as in (\ref{explicit-fields}) the lightest neutral field from the first and the second doublet which is stabilised by the remaining $U_1(1)$ symmetry is a viable DM candidate.

Expanding the potential around the vacuum point $(0,0,\frac{v}{\sqrt{2}})$, with 
\be 
v^2=\frac{\mu_3^2}{\lambda_{33}},
\ee
results in the following Higgs mass spectrum:
\begin{footnotesize}
\bea
&& \textbf{A}_3  : \quad m^2=0 \\
&& \textbf{H}^\pm_3  : \quad m^2=0 \nonumber\\
&& \textbf{H}_3 : \quad m^2= 2\mu_3^2 \nonumber\\
&& \textbf{H}^\pm_2 : \quad m^2= -\mu^2_2 +\frac{1}{2}\lambda_{23} v^2  \nonumber\\
&& \textbf{H}^\pm_1 : \quad m^2= -\mu^2_1 +\frac{1}{2}\lambda_{31}v^2  \nonumber\\
&& \textbf{H}_2  : \quad m^2= -\mu^2_2  +\frac{1}{2}(\lambda_{23}+\lambda'_{23} )v^2  \nonumber\\
&& \textbf{A}_2  : \quad m^2= -\mu^2_2  +\frac{1}{2}(\lambda_{23}+\lambda'_{23} )v^2 \nonumber\\
&& \textbf{H}_1  : \quad m^2= -\mu^2_1  +\frac{1}{2}(\lambda_{31}+\lambda'_{31} )v^2 \nonumber\\
&& \textbf{A}_1  : \quad m^2= -\mu^2_1  +\frac{1}{2}(\lambda_{31}+\lambda'_{31} )v^2 \nonumber
\eea
\end{footnotesize}
where the fields appearing in the doublets in (\ref{explicit-fields}) are the same as the mass eigenstates (shown in bold text).

Note that the fields from the third doublet play the role of the SM-Higgs doublet fields, namely a massless neutral Goldstone boson $(\textbf{A}_3)$, two massless charged Goldstone bosons $(\textbf{H}^\pm_3)$ and the SM-Higgs boson $(\textbf{H}_3)$.

Positivity of the mass eigenstates enforces the following extra conditions on the parameters of the potential:
\bea
&&  \bullet\quad  \lambda_{33} > 0 \\
&&  \bullet\quad  \mu^2_3 > 0 \nonumber\\
&&  \bullet\quad  2\mu^2_1 < \lambda_{31}v^2  \nonumber\\
&&  \bullet\quad  2\mu^2_1 < (\lambda_{31}+{\lambda'}_{31})v^2 \nonumber\\
&&  \bullet\quad  2\mu^2_2 < \lambda_{23}v^2 \nonumber\\
&&  \bullet\quad  2\mu^2_2 < (\lambda_{23}+{\lambda'}_{23})v^2 \nonumber
\eea

\subsection{U(1) symmetric 3HDM potential}
A potential symmetric under the $U_1(1)$ group in (\ref{U1}) contains the following terms\footnote{A general $U_2(1)$ symmetric 3HDM potential contains:
\bea
V_{U_2(1)} &=&  -\mu_{23}^2(\phi_2^\dagger\phi_3) + \lambda_{1} (\phi_1^\dagger\phi_1)(\phi_2^\dagger\phi_3) + \lambda_{2} (\phi_2^\dagger\phi_2) (\phi_2^\dagger\phi_3) \nonumber\\
&&+ \lambda_{3} (\phi_3^\dagger\phi_3)(\phi_2^\dagger\phi_3)
+ \lambda_{4}(\phi_2^\dagger\phi_3)^2 + \lambda_{5}(\phi_2^\dagger\phi_1)(\phi_1^\dagger\phi_3) + h.c. \nonumber
\eea}:
\be 
V_{U_1(1)}= \lambda_1 (\phi_1^\dagger\phi_3)(\phi_2^\dagger\phi_3) + h.c. 
\label{U_1(1)-3HDM}
\ee
in addition to $V_0$ in Eq.~(\ref{U(1)XU(1)-3HDM}).

The possible minima of this potential are:
\begin{enumerate}

\item
\textbf{Point} $(0,0,\frac{v}{\sqrt{2}})$ respects the symmetry of the potential and becomes the minimum at 
\be
v^2= \frac{\mu^2_3}{\lambda_{33}} \nonumber
\ee
provided the following conditions are satisfied:
\bea 
&& \bullet\quad  -\mu^2_1  +\frac{1}{2}(\lambda_{31}+\lambda'_{31} )v^2   >0  \\
&& \bullet\quad  -\mu^2_2  +\frac{1}{2}(\lambda_{23}+\lambda'_{23} )v^2   >0  \nonumber\\
&& \bullet\quad   \mu^2_3    >0    \nonumber
\eea
where the third condition is already required for the positivity of mass eigenstates at point $(0,0,\frac{v}{\sqrt{2}})$.

\item
\textbf{Point} $(0,\frac{v}{\sqrt{2}},\frac{v}{\sqrt{2}})$ only becomes the minimum if $\lambda_1 =0$ which reduces the symmetry of the potential to $U(1) \times U(1)$. We conclude that this point is not a minimum of the $U(1)$-symmetric potential. 

\item
\textbf{Point} $(\frac{v}{\sqrt{2}},\frac{v}{\sqrt{2}},\frac{v}{\sqrt{2}})$ breaks the symmetry of the potential completely and becomes the minimum at 
\bea
v^2&=& \frac{2\mu^2_3}{2\lambda_{33}+\lambda_{23}+\lambda'_{23} + \lambda_{31}+\lambda'_{31}+ 2\lambda_1} \nonumber\\
&=& \frac{2\mu^2_2}{2\lambda_{22}+\lambda_{12}+\lambda'_{12}+\lambda_{23}+\lambda'_{23} + \lambda_1} \nonumber\\
&=&  \frac{2\mu^2_1}{2\lambda_{11}+\lambda_{12}+\lambda'_{12}+\lambda_{31}+\lambda'_{31} + \lambda_1} \nonumber
\eea 
with the following conditions:
\bea 
&& \bullet\quad  2\lambda_{11}> \lambda_1 \\
&& \bullet\quad  2\lambda_{22}> \lambda_1 \nonumber\\
&& \bullet\quad  \lambda_{33} > \lambda_1  \nonumber\\
&& \bullet\quad   (\lambda_1 -2\lambda_{11})(\lambda_1 -2\lambda_{22}) > (\lambda_{12} + \lambda'_{12})^2  \nonumber\\
&& \bullet\quad  2 (\lambda_{12}+\lambda'_{12})(\lambda_{31}+\lambda'_{31}+2\lambda_{1})(\lambda_{23}+\lambda'_{23}+2\lambda_{1}) \nonumber\\
&& \qquad + ( 2 \lambda_{11}-\lambda_{1})( 2\lambda_{22}-\lambda_{1} )( 2\lambda_{33}-2\lambda_{1} ) \quad > \quad
 ( 2 \lambda_{33}-2\lambda_1) (\lambda_{12} + \lambda'_{12})^2 \nonumber\\
&& \qquad + ( 2\lambda_{11}-\lambda_1) (\lambda_{23} + \lambda'_{23}+2\lambda_1)^2 + ( 2\lambda_{22}-\lambda_1 ) (\lambda_{31} + \lambda'_{31} + 2\lambda_1)^2  \nonumber
\eea

\end{enumerate}

\underline{\textbf{Higgs mass spectrum for} $(0,0,\frac{v}{\sqrt{2}})$}

Defining the doublets similarly to the previous case and expanding the potential around the vacuum point $(0,0,\frac{v}{\sqrt{2}})$, with 
\be 
v^2=\frac{\mu_3^2}{\lambda_{33}},
\ee
results in a mass spectrum of the following form:
\begin{footnotesize}
\bea
&& \textbf{A}_3  : \quad m^2=0 \\
&& \textbf{H}^\pm_3  : \quad m^2=0 \nonumber\\
&& \textbf{H}_3 : \quad m^2= 2\mu_3^2 \nonumber\\
&& \textbf{H}^\pm_2 : \quad m^2= -\mu^2_2 +\frac{1}{2}\lambda_{23} v^2  \nonumber\\
&& \textbf{H}^\pm_1 : \quad m^2= -\mu^2_1 +\frac{1}{2}\lambda_{31}v^2  \nonumber\\
&& \textbf{H}_2 \equiv \frac{aH^0_{2}+ H^0_{1}}{\sqrt{1+a^2}}: \quad m^2= \frac{1}{2}(X-\sqrt{Y}) \nonumber\\
&& \textbf{H}_1 \equiv \frac{bH^0_{2}+ H^0_{1}}{\sqrt{1+b^2}} : \quad m^2= \frac{1}{2}(X+\sqrt{Y}) \nonumber\\
&& \textbf{A}_2 \equiv \frac{-aA^0_{2}+ A^0_{1}}{\sqrt{1+a^2}} : \quad m^2= \frac{1}{2}(X-\sqrt{Y}) \nonumber\\
&& \textbf{A}_1 \equiv \frac{-bA^0_{2}+ A^0_{1}}{\sqrt{1+b^2}} : \quad m^2= \frac{1}{2}(X+\sqrt{Y})  \nonumber\\
&&  \mbox{where} \quad X=  -\mu_1^2-\mu_2^2 + \frac{1}{2}(\lambda_{23} + \lambda_{31} + \lambda'_{23} + \lambda'_{31})v^2  \nonumber\\
&& \qquad \qquad Y= (\lambda_1 v^2)^2+ \left[ \mu_1^2 -\mu_2^2 + \frac{1}{2}(\lambda_{23} - \lambda_{31} + \lambda'_{23} - \lambda'_{31})v^2  \right]^2 \nonumber\\
&& \qquad \qquad a= \frac{1}{\lambda_1 v^2} \left[\mu_1^2 -\mu_2^2 + \frac{1}{2}(\lambda_{23} - \lambda_{31} + \lambda'_{23} - \lambda'_{31})v^2 -\sqrt{Y}  \right] \nonumber\\
&& \qquad \qquad b= \frac{1}{\lambda_1 v^2} \left[\mu_1^2 -\mu_2^2 + \frac{1}{2}(\lambda_{23} - \lambda_{31} + \lambda'_{23} - \lambda'_{31})v^2 + \sqrt{Y}  \right]  \nonumber
\eea
\end{footnotesize}

The lightest neutral eigenstate among the ones from the first and the second doublet, $\textbf{H}_1$, $\textbf{H}_2$ and $\textbf{A}_1$, $\textbf{A}_2$, which is stabilised by the conserved $U_1(1)$ symmetry is a viable DM candidate.

\subsection{U(1)$\times$Z$_2$ symmetric 3HDM potential}
In addition to $V_0$ in Eq.~(\ref{U(1)XU(1)-3HDM}) the $U_2(1) \times Z_2$-symmetric 3HDM potential contains the following term:
\be 
V_{U(1) \times Z_2}=  \lambda_1 (\phi_2^\dagger\phi_3)^2 + h.c. 
\label{U(1)XZ_2-3HDM}
\ee
This term is symmetric under
\be
U_2(1) =
\mbox{diag}(e^{-2i\beta /3}, e^{i\beta / 3}, e^{i\beta / 3}), \quad \mbox{and} \quad   Z_2= \mbox{diag}(-1,-1,1).
\ee

The possible minima of this potential are:
\begin{enumerate}
\item
\textbf{Point} $(0,0,\frac{v}{\sqrt{2}})$ breaks the symmetry of the potential to $Z_2$ and becomes the minimum at 
\be
v^2= \frac{\mu^2_3}{\lambda_{33}} \nonumber
\ee
provided the following conditions are satisfied:
\bea 
&& \bullet\quad   -\mu^2_2  +\frac{1}{2}(\lambda_{23}+\lambda'_{23} )v^2   >0  \\
&& \bullet\quad   -\mu^2_1  +\frac{1}{2}(\lambda_{31}+\lambda'_{31} )v^2   >0  \nonumber\\
&& \bullet\quad   \mu^2_3    >0   \nonumber
\eea
The last two conditions are already required for the positivity of mass eigenstates at point $(0,0,\frac{v}{\sqrt{2}})$.

\item
\textbf{Point} $(0,\frac{v}{\sqrt{2}},\frac{v}{\sqrt{2}})$ breaks the symmetry of the potential to a $Z_2$ generated by $(1,-1,-1)$ and becomes the minimum at 
\be 
v^2= \frac{2\mu^2_3}{2\lambda_{33}+\lambda_{23}+\lambda'_{23}+\lambda_{31}+\lambda'_{31}+2\lambda_1}
= \frac{2\mu^2_2}{2\lambda_{22}+\lambda_{12}+\lambda'_{12}+\lambda_{23}+\lambda'_{23}+2\lambda_1} \nonumber
\ee
when the following conditions are satisfied:
\bea 
&& \bullet\quad  -2\mu^2_1 + (\lambda_{12} + \lambda'_{12}+ \lambda_{31}+ \lambda'_{31})v^2 >0  \\
&& \bullet\quad  2\lambda_{22}-2\lambda_1-\lambda_{12} - \lambda'_{12} >0  \nonumber\\
&& \bullet\quad  2\lambda_{33}-2\lambda_1-\lambda_{31} - \lambda'_{31} >0  \nonumber\\
&& \bullet\quad  (2\lambda_{22}-2\lambda_1-\lambda_{12} - \lambda'_{12})(2\lambda_{33}-2\lambda_1-\lambda_{31} - \lambda'_{31})>(4\lambda_1+\lambda_{23}+ \lambda'_{23})^2    \nonumber
\eea

\item
\textbf{Point} $(\frac{v}{\sqrt{2}},\frac{v}{\sqrt{2}},\frac{v}{\sqrt{2}})$ breaks the symmetry of the potential completely and becomes the minimum at 
\bea
v^2&=& \frac{2\mu^2_3}{2\lambda_{33}+\lambda_{23}+\lambda'_{23} + \lambda_{31}+\lambda'_{31} +2\lambda_1} \nonumber\\
&=& \frac{2\mu^2_2}{2\lambda_{22}+\lambda_{12}+\lambda'_{12}+\lambda_{23}+\lambda'_{23}+2\lambda_1} \nonumber\\
&=&  \frac{2\mu^2_1}{2\lambda_{11}+\lambda_{12}+\lambda'_{12}+\lambda_{31}+\lambda'_{31}} \nonumber
\eea 
with the following conditions:
\bea 
&& \bullet\quad \lambda_{11}>0 \\
&& \bullet\quad  \lambda_{22}> \lambda_1  \nonumber\\
&& \bullet\quad  \lambda_{33} > \lambda_1   \nonumber\\
&& \bullet\quad  4\lambda_{11}(\lambda_{22} -\lambda_{1}) > (\lambda_{12} + \lambda'_{12})^2  \nonumber\\
&& \bullet\quad  (\lambda_{12}+\lambda'_{12})(\lambda_{31}+\lambda'_{31}+8\lambda_1)(\lambda_{23}+\lambda'_{23}) + 4 \lambda_{11}(\lambda_{22} -\lambda_{1})(\lambda_{33} -\lambda_{1}) \quad >  \nonumber\\
&& \qquad (-\lambda_1 + \lambda_{33}) (\lambda_{12} + \lambda'_{12})^2 + \lambda_{11} (\lambda_{23} + \lambda'_{23}+8\lambda_1)^2+ (-\lambda_1 + \lambda_{22}) (\lambda_{31} + \lambda'_{31} )^2  \nonumber
\eea

\end{enumerate}

\underline{\textbf{Higgs mass spectrum for} $(0,0,\frac{v}{\sqrt{2}})$}

Expanding the potential around the vacuum point $(0,0,\frac{v}{\sqrt{2}})$, with 
\be 
v^2=\frac{\mu_3^2}{\lambda_{33}}
\ee
the mass spectrum appears as follows:
\begin{footnotesize}
\bea
&& \textbf{A}_3  : \quad m^2=0 \\
&& \textbf{H}^\pm_3  : \quad m^2=0 \nonumber\\
&& \textbf{H}_3 : \quad m^2= 2\mu_3^2 \nonumber\\
&& \textbf{H}^\pm_2 : \quad m^2= -\mu^2_2 +\frac{1}{2}\lambda_{23} v^2  \nonumber\\
&& \textbf{H}^\pm_1 : \quad m^2= -\mu^2_1 +\frac{1}{2}\lambda_{31}v^2  \nonumber\\
&& \textbf{H}_2  : \quad m^2= -\mu^2_2  +\frac{1}{2}(\lambda_{23}+\lambda'_{23} +2\lambda_1)v^2  \nonumber\\
&& \textbf{A}_2  : \quad m^2= -\mu^2_2  +\frac{1}{2}(\lambda_{23}+\lambda'_{23} -2\lambda_1)v^2 \nonumber\\
&& \textbf{H}_1  : \quad m^2= -\mu^2_1  +\frac{1}{2}(\lambda_{31}+\lambda'_{31} )v^2 \nonumber\\
&& \textbf{A}_1  : \quad m^2= -\mu^2_1  +\frac{1}{2}(\lambda_{31}+\lambda'_{31} )v^2 \nonumber
\eea
\end{footnotesize}

The lightest neutral field from the first and the second doublet which is stabilised by the remaining $Z_2$ symmetry is a viable DM candidate.

\subsection{Z$_2$ symmetric 3HDM potential}

Constructing the $Z_2$-symmetric part of the potential depends on the generator of the group. The $Z_2$ generator which forbids FCNCs has the following form
\be 
a=  \mathrm{diag}\left(-1, -1, 1 \right). 
\ee
The terms ensuring the $Z_2$ group generated by $a$ are\footnote{Note that the only symmetry group of this potential is the $Z_2$ group generated by $a$. However, this is not the only $Z_2$ symmetric potential that can be written. A potential with the following terms added to $V_0$ is also symmetric only under the $Z_2$ group generated by $a$:
\be 
V'_{Z_2} =  \lambda_{1}(\phi_1^\dagger\phi_2)^2 + \lambda_2(\phi_2^\dagger\phi_3)^2 + \lambda_3(\phi_3^\dagger\phi_1)^2 +  \lambda_{4}(\phi_1^\dagger\phi_2)\left[ (\phi_1^\dagger\phi_1)+ (\phi_2^\dagger\phi_2) \right] + \lambda_5 (\phi_1^\dagger\phi_3)(\phi_2^\dagger\phi_3) + h.c. \nonumber
\ee}
\be 
V_{Z_2} = V_0 -\mu'^2_{12}(\phi_1^\dagger\phi_2) + \lambda_{1}(\phi_1^\dagger\phi_2)^2 + \lambda_2(\phi_2^\dagger\phi_3)^2 + \lambda_3(\phi_3^\dagger\phi_1)^2  + h.c. 
\label{Z_2-3HDM}
\ee
which need to be added to $V_0$ in Eq.~(\ref{U(1)XU(1)-3HDM}) to result in a uniquely $Z_2$-symmetric 3HDM potential.


The possible minima of this potential are:
\begin{enumerate}
\item
\textbf{Point} $(0,0,\frac{v}{\sqrt{2}})$ respects the symmetry of the potential and becomes the minimum at 
\be
v^2= \frac{\mu^2_3}{\lambda_{33}} \nonumber
\ee
provided the following conditions are satisfied:
\bea 
&& \bullet\quad   -\mu^2_2  +\frac{1}{2}(\lambda_{23}+\lambda'_{23} )v^2  >0  \\
&& \bullet\quad  -\mu^2_1  +\frac{1}{2}(\lambda_{31}+\lambda'_{31} )v^2   >0  \nonumber\\
&& \bullet\quad  \biggl(-\mu^2_1  +\frac{1}{2}(\lambda_{31}+\lambda'_{31} )v^2 \biggr) \biggl( -\mu^2_2  +\frac{1}{2}(\lambda_{23}+\lambda'_{23} )v^2 \biggr) > (\mu'^2_{12})^2  \nonumber\\
&& \bullet\quad   \mu^2_3    >0   \nonumber
\eea
The last condition is already required for the positivity of mass eigenstates at point $(0,0,\frac{v}{\sqrt{2}})$.

\item
\textbf{Point} $(0,\frac{v}{\sqrt{2}},\frac{v}{\sqrt{2}})$ can only be a minimum of this potential if $\mu'^2_{12}=0$ which makes the potential $Z_2 \times Z_2$-symmetric. We therefore conclude that this point is not a minimum of the $Z_2$-symmetric potential.

\item
\textbf{Point} $(\frac{v}{\sqrt{2}},\frac{v}{\sqrt{2}},\frac{v}{\sqrt{2}})$ breaks the symmetry of the potential completely and becomes the minimum at 
\bea
v^2 &=& \frac{2\mu^2_3}{2\lambda_{33}+\lambda_{23}+\lambda'_{23} + \lambda_{31}+\lambda'_{31} +2\lambda_2+2\lambda_3} \nonumber\\
&=& \frac{2\mu^2_2 +2\mu'^2_{12}}{2\lambda_{22}+\lambda_{12}+\lambda'_{12}+\lambda_{23}+\lambda'_{23}+2\lambda_1+2\lambda_2} \nonumber\\
&=& \frac{2\mu^2_1 +2\mu'^2_{12}}{2\lambda_{11}+\lambda_{12}+\lambda'_{12}+\lambda_{31}+\lambda'_{31} +2\lambda_1+2\lambda_3} \nonumber
\eea 
with the following conditions:
\bea 
&& \bullet\quad  \mu'^2_{12} > (\lambda_1 + \lambda_3 -\lambda_{11})v^2   \\
&& \bullet\quad  \mu'^2_{12} > (\lambda_1 + \lambda_2 -\lambda_{22})v^2   \nonumber\\
&& \bullet\quad  \mu'^2_{12} > (\lambda_2 + \lambda_3 -\lambda_{33})v^2   \nonumber\\
&& \bullet\quad  4\biggl(\mu'^2_{12}-(\lambda_{1}+\lambda_3-\lambda_{11})v^2\biggr) 
   \biggl(\mu'^2_{12}-(\lambda_{1} +\lambda_{2}-\lambda_{22})v^2 \biggr) >   \nonumber\\
&& \qquad \biggl(-2\mu'^2_{12}+(4\lambda_1+\lambda_{12} + \lambda'_{12})v^2 \biggr)^2  \nonumber\\
&&  \bullet\quad \biggl( \mu'^2_{12} - (\lambda_1 + \lambda_3 -\lambda_{11})v^2 \biggr) 
    \biggl( \mu'^2_{12} - (\lambda_1 + \lambda_2 -\lambda_{22})v^2 \biggr)
    \biggl( \mu'^2_{12} - (\lambda_2 + \lambda_3 -\lambda_{33})v^2 \biggr)  \nonumber\\
&&  \qquad + 2 \biggl(-\mu'^2_{12}+(\frac{4\lambda_1+\lambda_{12} + \lambda'_{12}}{2})v^2 \biggr)
      \biggl(\frac{4\lambda_3+\lambda_{31} + \lambda'_{31}}{2} \biggr)
      \biggl(\frac{4\lambda_2+\lambda_{23} + \lambda'_{23}}{2} \biggr) v^4 >   \nonumber\\
&& \qquad \biggl( \mu'^2_{12} - (\lambda_1 + \lambda_3 -\lambda_{11})v^2 \biggr)
      \biggl(-\mu'^2_{12}+(\frac{4\lambda_1+\lambda_{12} + \lambda'_{12}}{2})v^2 \biggr)^2  \nonumber\\
&& \qquad   + \biggl( \mu'^2_{12} - (\lambda_1 + \lambda_2 -\lambda_{22})v^2 \biggr)
      \biggl(\frac{4\lambda_3+\lambda_{31} + \lambda'_{31}}{2} \biggr)^2 v^4 \nonumber\\
&& \qquad + \biggl( \mu'^2_{12} - (\lambda_2 + \lambda_3 -\lambda_{33})v^2 \biggr) 
  \biggl(\frac{4\lambda_2+\lambda_{23} + \lambda'_{23}}{2} \biggr)^2 v^4  \nonumber
\eea

\end{enumerate}

\underline{\textbf{Higgs mass spectrum for} $(0,0,\frac{v}{\sqrt{2}})$}

Expanding the potential around the vacuum point $(0,0,\frac{v}{\sqrt{2}})$, with 
\be 
v^2=\frac{\mu_3^2}{\lambda_{33}},
\ee
the mass spectrum appears as follows:
\begin{footnotesize}
\bea
&& \textbf{A}_3  : \quad m^2=0 \nonumber\\
&& \textbf{H}^\pm_3  : \quad m^2=0 \nonumber\\
&& \textbf{H}_3 : \quad m^2= 2\mu_3^2 \nonumber\\
&& \textbf{H}_2 \equiv \frac{aH^0_{2}+ H^0_{1}}{\sqrt{1+a^2}} : \quad m^2= \frac{1}{2}(X-\sqrt{Y}) \nonumber\\
&& \textbf{H}_1 \equiv \frac{bH^0_{2}+ H^0_{1}}{\sqrt{1+b^2}} : \quad m^2= \frac{1}{2}(X+\sqrt{Y}) \nonumber\\
&& \qquad \qquad  \mbox{where} \quad X=  -\mu_1^2-\mu_2^2 + \frac{1}{2}(\lambda_{23} + \lambda_{31} + \lambda'_{23} + \lambda'_{31} +2\lambda_2 + 2\lambda_3)v^2  \nonumber\\
&& \qquad \qquad Y= 4\mu^4_{12}+ \left[ \mu_1^2 -\mu_2^2 + \frac{1}{2}(\lambda_{23} - \lambda_{31} + \lambda'_{23} - \lambda'_{31} +2\lambda_2 - 2\lambda_3)v^2  \right]^2 \nonumber\\
&& \qquad \qquad a= \frac{1}{-2\mu^2_{12}} \left[\mu_1^2 -\mu_2^2 + \frac{1}{2}(\lambda_{23} - \lambda_{31} + \lambda'_{23} - \lambda'_{31} + 2\lambda_2 - 2\lambda_3)v^2 -\sqrt{Y}  \right]  \nonumber\\
&& \qquad \qquad b= \frac{1}{-2\mu^2_{12}} \left[\mu_1^2 -\mu_2^2 + \frac{1}{2}(\lambda_{23} - \lambda_{31} + \lambda'_{23} - \lambda'_{31} +2\lambda_2 -2\lambda_3)v^2 + \sqrt{Y}  \right] \nonumber\\
&& \textbf{H}^\pm_2 = \frac{a'H^\pm_{2}+ H^\pm_{1}}{\sqrt{1+a'^2}} : \quad m^2= \frac{1}{2}(X'-\sqrt{Y'}) \nonumber\\
&& \textbf{H}^\pm_1 = \frac{b'H^\pm_{2}+ H^\pm_{1}}{\sqrt{1+b'^2}} : \quad m^2= \frac{1}{2}(X'+\sqrt{Y'}) \nonumber\\
&& \qquad \qquad \mbox{where} \quad X'=  -\mu_2^2-\mu_1^2 + \frac{1}{2}(\lambda_{31} + \lambda_{23})v^2  \nonumber\\
&& \qquad \qquad Y'= 4\mu^4_{12}+ \left[ \mu_2^2 -\mu_1^2 + \frac{1}{2}(\lambda_{31} - \lambda_{23})v^2  \right]^2 \nonumber\\
&& \qquad \qquad a'= \frac{1}{-2\mu^2_{12}} \left[\mu_2^2 -\mu_1^2 + \frac{1}{2}(\lambda_{31} - \lambda_{23})v^2 -\sqrt{Y'}  \right] \nonumber\\
&& \qquad \qquad b'= \frac{1}{-2\mu^2_{12}} \left[\mu_2^2 -\mu_1^2 + \frac{1}{2}(\lambda_{31} - \lambda_{23})v^2 + \sqrt{Y'}  \right]  \nonumber\\
&& \textbf{A}_2 \equiv \frac{a''A^0_{2}+ A^0_{1}}{\sqrt{1+a''^2}} : \quad m^2= \frac{1}{2}(X''-\sqrt{Y''}) \nonumber\\
&& \textbf{A}_1 \equiv \frac{b''A^0_{2}+ A^0_{1}}{\sqrt{1+b''^2}} : \quad m^2= \frac{1}{2}(X''+\sqrt{Y''})  \nonumber\\
&& \qquad \qquad  \mbox{where} \quad X''=  -\mu_1^2-\mu_2^2 + \frac{1}{2}(\lambda_{23} + \lambda_{31} + \lambda'_{23} + \lambda'_{31} -2\lambda_2 - 2\lambda_3)v^2  \nonumber\\
&& \qquad \qquad Y''= 4\mu^4_{12}+ \left[ \mu_1^2 -\mu_2^2 + \frac{1}{2}(\lambda_{23} - \lambda_{31} + \lambda'_{23} - \lambda'_{31} -2\lambda_2 + 2\lambda_3)v^2  \right]^2 \nonumber\\
&& \qquad \qquad a''= \frac{1}{-2\mu^2_{12}} \left[\mu_1^2 -\mu_2^2 + \frac{1}{2}(\lambda_{23} - \lambda_{31} + \lambda'_{23} - \lambda'_{31} - 2\lambda_2 + 2\lambda_3)v^2 -\sqrt{Y''}  \right]  \nonumber\\
&& \qquad \qquad b''= \frac{1}{-2\mu^2_{12}} \left[\mu_1^2 -\mu_2^2 + \frac{1}{2}(\lambda_{23} - \lambda_{31} + \lambda'_{23} - \lambda'_{31} -2\lambda_2 +2\lambda_3)v^2 + \sqrt{Y''}  \right]  \nonumber
\eea
\end{footnotesize}

The lightest neutral field from the first or the second doublet, stabilised by the conserved $Z_2$ symmetry, is a viable DM candidate.

\subsection{Z$_3$ symmetric 3HDM potential}
The $Z_3$-symmetric 3HDM potential contains the following terms:
\be
V_{Z_3} = \lambda_{1}(\phi_1^\dagger\phi_2)(\phi_3^\dagger\phi_2) + 
\lambda_{2}(\phi_2^\dagger\phi_3)(\phi_1^\dagger\phi_3) + 
\lambda_{3}(\phi_3^\dagger\phi_1)(\phi_2^\dagger\phi_1) + h.c. 
\label{Z_3-3HDM}
\ee
in addition to $V_0$ in Eq.~(\ref{U(1)XU(1)-3HDM}).


This group is generated by $a = \mathrm{diag}(\omega,\omega^2,1)$, where $\omega=exp(2i\pi/3)$.

The possible minima of this potential are:
\begin{enumerate}
\item
\textbf{Point} $(0,0,\frac{v}{\sqrt{2}})$ respects the symmetry of the potential and becomes the minimum at 
\be
v^2= \frac{\mu^2_3}{\lambda_{33}} \nonumber
\ee
provided the following conditions are satisfied:
\bea 
&& \bullet\quad   -\mu^2_2  +\frac{1}{2}(\lambda_{23}+\lambda'_{23} )v^2   >0  \\
&& \bullet\quad   -\mu^2_1  +\frac{1}{2}(\lambda_{31}+\lambda'_{31} )v^2   >0  \nonumber\\
&& \bullet\quad    \mu^2_3    >0   \nonumber
\eea
The last condition is already required for the positivity of mass eigenstates at point $(0,0,\frac{v}{\sqrt{2}})$.

\item
\textbf{Point} $(0,\frac{v}{\sqrt{2}},\frac{v}{\sqrt{2}})$ breaks the symmetry of the potential completely. This point becomes the minimum only when $\lambda_1=-\lambda_2$ (in the more general case of $v_2 \neq v_3$ the condition $v_2/v_3 = -\lambda_2/\lambda_1$ is required) which does not lead to an extra symmetry of the potential. The minimisation requires 
\be 
v^2= \frac{2\mu^2_3}{2\lambda_{33}+\lambda_{23}+\lambda'_{23}} = \frac{2\mu^2_2}{2\lambda_{22}+\lambda_{23}+\lambda'_{23}} \nonumber
\ee
and the following conditions be satisfied:
\bea
&&  \bullet\quad  \lambda_{22}>0 \\
&&  \bullet\quad  \lambda_{33}>0  \nonumber\\ 
&&  \bullet\quad -2\mu^2_1 + (\lambda_{12} + \lambda'_{12}+ \lambda_{31}+ \lambda'_{31})v^2 >0  \nonumber\\
&&  \bullet\quad \lambda_{22} \biggl(-4\mu^2_1 + 2(\lambda_{12} + \lambda'_{12}+ \lambda_{31}+ \lambda'_{31})v^2 \biggr) >\lambda^2_1 v^2  \nonumber\\
&& \bullet\quad  \lambda_{22}\lambda_{33} \biggl(-2\mu^2_1 + (\lambda_{12} + \lambda'_{12}+ \lambda_{31}+ \lambda'_{31})v^2 \biggr) + \lambda_1\lambda_2(\lambda_{23}+\lambda'_{23})v^2 >  \nonumber\\
&& \qquad \frac{1}{2} \lambda_{33}\lambda^2_1 v^2 + 2 \lambda_{22}\lambda^2_2 v^2  + \frac{1}{4} (\lambda_{23}+\lambda'_{23})^2  \biggl(-2\mu^2_1 + 2(\lambda_{12} + \lambda'_{12}+ \lambda_{31}+ \lambda'_{31})v^2 \biggr)  \nonumber
\eea

\item
\textbf{Point} $(\frac{v}{\sqrt{2}},\frac{v}{\sqrt{2}},\frac{v}{\sqrt{2}})$ breaks the symmetry of the potential completely and becomes the minimum at 
\bea
v^2&=& \frac{2\mu^2_3}{2\lambda_{33}+\lambda_{23}+\lambda'_{23} + \lambda_{31}+\lambda'_{31} +\lambda_1+ 2\lambda_2 +\lambda_3} \nonumber\\
&=& \frac{2\mu^2_2}{2\lambda_{22}+\lambda_{12}+\lambda'_{12}+\lambda_{23}+\lambda'_{23}+2\lambda_1 + \lambda_2 +\lambda_3} \nonumber\\
&=&  \frac{2\mu^2_1}{2\lambda_{11}+\lambda_{12}+\lambda'_{12}+\lambda_{31}+\lambda'_{31} +\lambda_1 + \lambda_2 +2\lambda_3} \nonumber
\eea 
with the following conditions:
\bea 
&& \bullet \quad 2\lambda_{11}> \lambda_1+\lambda_2+2\lambda_3  \\
&& \bullet \quad 2\lambda_{22} > 2\lambda_1 + \lambda_2 + \lambda_3   \nonumber\\
&& \bullet \quad 2\lambda_{33}> \lambda_1+2 \lambda_2+\lambda_3   \nonumber\\
&& \bullet \quad (\lambda_{1}+\lambda_2+2\lambda_3-2\lambda_{11})(2\lambda_1+\lambda_{2} +\lambda_{3}-2\lambda_{22}) >
  (\lambda_1+2\lambda_3+\lambda_{12} + \lambda'_{12})^2  \nonumber\\
&& \bullet \quad   (\lambda_{1}+\lambda_2+2\lambda_3-2\lambda_{11})
    (2\lambda_1+\lambda_{2} +\lambda_{3}-2\lambda_{22})
    (\lambda_1+2\lambda_{2} +\lambda_{3}-2\lambda_{33})  \nonumber\\
&& \qquad + 2(\lambda_1+2\lambda_3+\lambda_{12} + \lambda'_{12})
    (2\lambda_2+\lambda_3+\lambda_{31} + \lambda'_{31})
    (2\lambda_1+\lambda_2+\lambda_{23} + \lambda'_{23}) >  \nonumber\\
&& \qquad - (\lambda_1+2\lambda_{2} +\lambda_{3}-2\lambda_{33})(\lambda_1+2\lambda_3+\lambda_{12} +   	\lambda'_{12})^2   \nonumber\\
&&  \qquad - (\lambda_{1}+\lambda_2+2\lambda_3-2\lambda_{11})(2\lambda_1+\lambda_2+\lambda_{23} + \lambda'_{23})^2  \nonumber\\
&& \qquad - (2\lambda_1+\lambda_{2} +\lambda_{3}-2\lambda_{22})(2\lambda_2+\lambda_3+\lambda_{31} + \lambda'_{31})  \nonumber
\eea

\end{enumerate}

\underline{\textbf{Higgs mass spectrum for} $(0,0,\frac{v}{\sqrt{2}})$}

Expanding the potential around the vacuum point $(0,0,\frac{v}{\sqrt{2}})$, with 
\be 
v^2=\frac{\mu_3^2}{\lambda_{33}},
\ee
the mass spectrum is the same as the $U(1)$-symmetric case with the slight difference in the definition of $a$, $b$ and $Y$ (replace $\lambda_1$ by $\lambda_2$).

The lightest neutral field from the first or the second doublet, stabilised by the conserved $Z_3$ symmetry, is a viable DM candidate.

\subsection{Z$_4$ symmetric 3HDM potential}
The most general $Z_4$-symmetric potential has two parts, $V_0$ in Eq.~(\ref{U(1)XU(1)-3HDM}) and the following terms:
\bea
V_{Z_4}&=& \lambda_1 (\phi_3^\dagger \phi_1)(\phi_3^\dagger \phi_2)+ \lambda_2 (\phi_1^\dagger \phi_2)^2 + h.c.  
\label{Z_4-3HDM}
\eea
with generator $a = \mathrm{diag}(i,-i,1)$.

The possible minima of this potential are:
\begin{enumerate}
\item
\textbf{Point} $(0,0,\frac{v}{\sqrt{2}})$ respects the symmetry of the potential and becomes the minimum at 
\be
v^2= \frac{\mu^2_3}{\lambda_{33}} \nonumber
\ee
provided the following conditions are satisfied:
\bea 
&& \bullet\quad  -\mu^2_2 +\frac{1}{2}(\lambda_{23}+\lambda'_{23} )v^2   >0  \\
&& \bullet\quad -\mu^2_1  +\frac{1}{2}(\lambda_{31}+\lambda'_{31} )v^2   >0  \nonumber\\
&& \bullet\quad \mu^2_3    >0   \nonumber
\eea
The last condition is already required for the positivity of mass eigenstates at point $(0,0,\frac{v}{\sqrt{2}})$.

\item
\textbf{Point} $(0,\frac{v}{\sqrt{2}},\frac{v}{\sqrt{2}})$ only becomes the minimum when $\lambda_1=0$ which leads to the potential being symmetric under $U_1(1)$ and $Z_2$ generated by $(-1,-1,1)$. We therefore conclude that this VEV alignment is not realised in the $Z_4$-symmetric potential. 

\item
\textbf{Point} $(\frac{v}{\sqrt{2}},\frac{v}{\sqrt{2}},\frac{v}{\sqrt{2}})$ breaks the symmetry of the potential completely and becomes the minimum at 
\bea
v^2&=& \frac{2\mu^2_3}{2\lambda_{33}+\lambda_{23}+\lambda'_{23} + \lambda_{31}+\lambda'_{31} +2\lambda_1} \nonumber\\
&=& \frac{2\mu^2_2}{2\lambda_{22}+\lambda_{12}+\lambda'_{12}+\lambda_{23}+\lambda'_{23}+\lambda_1 +2\lambda_2} \nonumber\\
&=&  \frac{2\mu^2_1}{2\lambda_{11}+\lambda_{12}+\lambda'_{12}+\lambda_{31}+\lambda'_{31} +\lambda_1 +2\lambda_2} \nonumber
\eea 
with the following conditions:
\bea 
&& \bullet\quad 2\lambda_{11}> \lambda_1 +2\lambda_2   \\
&& \bullet\quad 2\lambda_{22} > \lambda_1 + 2\lambda_2    \nonumber\\
&& \bullet\quad \lambda_{33}> \lambda_1   \nonumber\\
&& \bullet\quad (\lambda_{1}+2\lambda_2-2\lambda_{11})(\lambda_1+2\lambda_{2} -2\lambda_{22}) > (4\lambda_2+\lambda_{12} + \lambda'_{12})^2 \nonumber
\eea

\end{enumerate}

\underline{\textbf{Higgs mass spectrum for} $(0,0,\frac{v}{\sqrt{2}})$}

Expanding the potential around the vacuum point $(0,0,\frac{v}{\sqrt{2}})$, with 
\be 
v^2=\frac{\mu_3^2}{\lambda_{33}},
\ee
the mass spectrum is the same as the $U(1)$-symmetric case.

The lightest neutral field from the first or the second doublet, stabilised by the conserved $Z_4$ symmetry, is a viable DM candidate.

\subsection{Z$_2\times$Z$_2$ symmetric 3HDM potential}
The most general $Z_2 \times Z_2$-symmetric potential consists of $V_0$ in Eq.~(\ref{U(1)XU(1)-3HDM}) and
\bea
V_{Z_2 \times Z_2}&=&  \lambda_1 (\phi_1^\dagger \phi_2)^2 + \lambda_2(\phi_2^\dagger \phi_3)^2 + \lambda_3 (\phi_3^\dagger \phi_1)^2 + h.c. \nonumber
\label{Z_2XZ_2-3HDM}
\eea
generated by independent sign flips of the three doublets: $a_1 = \mathrm{diag}(-1,1,1)$ 
and $a_2 = \mathrm{diag}(1,-1,1)$.

The possible minima of this potential are:
\begin{enumerate}
\item
\textbf{Point} $(0,0,\frac{v}{\sqrt{2}})$ respects the symmetry of the potential and becomes the minimum at 
\be
v^2= \frac{\mu^2_3}{\lambda_{33}} \nonumber
\ee
provided the following conditions are satisfied:
\bea 
&& \bullet\quad  -\mu^2_2  +\frac{1}{2}(\lambda_{23}+\lambda'_{23} )v^2   >0  \\
&& \bullet\quad  -\mu^2_1  +\frac{1}{2}(\lambda_{31}+\lambda'_{31} )v^2   >0  \nonumber\\
&& \bullet\quad   \mu^2_3    >0   \nonumber
\eea
The last condition is already required for the positivity of mass eigenstates at point $(0,0,\frac{v}{\sqrt{2}})$.

\item
\textbf{Point} $(0,\frac{v}{\sqrt{2}},\frac{v}{\sqrt{2}})$ breaks the symmetry of the potential to $Z_2$ generated by $(-1,1,1)$ and becomes the minimum at 
\bea 
v^2 &=& \frac{2\mu^2_3}{2\lambda_{33}+\lambda_{23}+\lambda'_{23}+\lambda_{31}+\lambda'_{31}+2\lambda_2+2\lambda_3} \nonumber\\
&=& \frac{2\mu^2_2}{2\lambda_{22}+\lambda_{12}+\lambda'_{12}+\lambda_{23}+\lambda'_{23}+2\lambda_1+2\lambda_2} \nonumber
\eea
when the following conditions are satisfied:
\bea
&& \bullet\quad  -2\mu^2_1  +(\lambda_{12}+\lambda'_{12}+\lambda_{31}+\lambda'_{31} )v^2  >0  \\
&& \bullet\quad   2\lambda_{22} -2\lambda_1 -2\lambda_2 - \lambda_{12}- \lambda'_{12} >0  \nonumber\\
&& \bullet\quad   2\lambda_{33} -2\lambda_2 -2\lambda_3 - \lambda_{31}- \lambda'_{31} >0  \nonumber\\
&& \bullet\quad  (2\lambda_{22} -2\lambda_1 -2\lambda_2 - \lambda_{12}- \lambda'_{12} )
 ( 2\lambda_{33} -2\lambda_2 -2\lambda_3 - \lambda_{31}- \lambda'_{31} ) > \nonumber\\
&& \qquad (\lambda_{23}+\lambda'_{23}+ 4\lambda_2)^2 v^4 \nonumber
\eea
This VEV alignment for the softly broken $Z_2 \times Z_2$-symmetric 3HDM has been studied in detail in \cite{Grzadkowski:2010au}.

\item
\textbf{Point} $(\frac{v}{\sqrt{2}},\frac{v}{\sqrt{2}},\frac{v}{\sqrt{2}})$ breaks the symmetry of the potential completely and becomes the minimum at 
\bea
v^2&=& \frac{2\mu^2_3}{2\lambda_{33}+\lambda_{23}+\lambda'_{23} + \lambda_{31}+\lambda'_{31} +2\lambda_2 +2\lambda_3} \nonumber\\
&=& \frac{2\mu^2_2}{2\lambda_{22}+\lambda_{12}+\lambda'_{12}+\lambda_{23}+\lambda'_{23} +2\lambda_1 +2\lambda_2} \nonumber\\
&=&  \frac{2\mu^2_1}{2\lambda_{11}+\lambda_{12}+\lambda'_{12}+\lambda_{31}+\lambda'_{31} +2\lambda_1 +2\lambda_3} \nonumber
\eea 
with the following conditions:
\bea 
&& \bullet\quad \lambda_{11}> \lambda_1+\lambda_3    \\
&& \bullet \quad \lambda_{22} > \lambda_1 + \lambda_2  \nonumber\\
&& \bullet \quad \lambda_{33}> \lambda_2+\lambda_3  \nonumber\\
&& \bullet \quad 4(\lambda_{1}+\lambda_3-\lambda_{11})(\lambda_{1} +\lambda_{2}-\lambda_{22}) >   (2\lambda_1+\lambda_{12} + \lambda'_{12})^2   \nonumber\\
&& \bullet \quad 4(\lambda_{11} -\lambda_1-\lambda_3)
                (\lambda_{22} -\lambda_1-\lambda_2)
                (\lambda_{33} -\lambda_2-\lambda_3)  \nonumber\\
&& \qquad     + (2\lambda_1+\lambda_{12} + \lambda'_{12})
                (2\lambda_3+\lambda_{31} + \lambda'_{31})
                (2\lambda_2+\lambda_{23} + \lambda'_{23}) >  \nonumber\\
&&  \qquad      (\lambda_{33} -\lambda_2-\lambda_3)(2\lambda_1+\lambda_{12} + \lambda'_{12})^2
               +(\lambda_{11} -\lambda_1-\lambda_3)(2\lambda_2+\lambda_{23} + \lambda'_{23})^2  \nonumber\\
&& \qquad      +(\lambda_{22} -\lambda_1-\lambda_2)(2\lambda_3+\lambda_{31} + \lambda'_{31})^2  \nonumber
\eea

\end{enumerate}

\underline{\textbf{Higgs mass spectrum for} $(0,0,\frac{v}{\sqrt{2}})$}

Expanding the potential around the vacuum point $(0,0,\frac{v}{\sqrt{2}})$, with 
\be 
v^2=\frac{\mu_3^2}{\lambda_{33}},
\ee
the mass spectrum appears as follows:
\begin{footnotesize}
\bea
&& \textbf{A}_3  : \quad m^2=0 \\
&& \textbf{H}^\pm_3  : \quad m^2=0 \nonumber\\
&& \textbf{H}_3 : \quad m^2= 2\mu_3^2 \nonumber\\
&& \textbf{H}^\pm_2 : \quad m^2= -\mu^2_2 +\frac{1}{2}\lambda_{23} v^2  \nonumber\\
&& \textbf{H}^\pm_1 : \quad m^2= -\mu^2_1 +\frac{1}{2}\lambda_{31}v^2  \nonumber\\
&& \textbf{H}_2  : \quad m^2= -\mu^2_2  +\frac{1}{2}(\lambda_{23}+\lambda'_{23} +2\lambda_2)v^2  \nonumber\\
&& \textbf{A}_2  : \quad m^2= -\mu^2_2  +\frac{1}{2}(\lambda_{23}+\lambda'_{23} -2\lambda_2)v^2 \nonumber\\
&& \textbf{H}_1  : \quad m^2= -\mu^2_1  +\frac{1}{2}(\lambda_{31}+\lambda'_{31} +2\lambda_3 )v^2 \nonumber\\
&& \textbf{A}_1  : \quad m^2= -\mu^2_1  +\frac{1}{2}(\lambda_{31}+\lambda'_{31} -2\lambda_3)v^2 \nonumber
\eea
\end{footnotesize}

The lightest neutral field from the first or the second doublet which is stabilised by the conserved $Z_2 \times Z_2$ symmetry is a viable DM candidate.


\section{Analysis of non-Abelian finite 3HDM}\label{finite-unitary}

In this section we study all non-Abelian finite symmetries in 3HDM potentials. Table \ref{non-Abelian-table} lists all these symmetry groups, their generators and their corresponding potentials.

\begin{table} [h]
\begin{footnotesize}
\begin{center}
\begin{tabular}{|m{1.5cm} || m{5.5cm}| m{7cm}|} \hline
\mbox{Symmetry}          & \mbox{Generators}   &  \mbox{Potential}   \\[3mm] \hline  \hline
$D_6$             & $\mbox{diag}(\omega,\omega^2,1), \quad \phi_1 \leftrightarrow - \phi_2$   & $V'_0 + \lambda_{1}\left[(\phi_2^\dagger\phi_1)(\phi_3^\dagger\phi_1) - 
                   (\phi_1^\dagger\phi_2)(\phi_3^\dagger\phi_2) \right] + \lambda_{2}(\phi_1^\dagger\phi_3)(\phi_2^\dagger\phi_3)+h.c. $  \\[3mm] \hline
$D_8$             & $\mbox{diag}(i,-i,1), \quad \phi_1 \leftrightarrow - \phi_2$  & $V'_0 + \lambda_1(\phi^\dagger_3 \phi_1)(\phi^\dagger_3 \phi_2) + \lambda_2(\phi^\dagger_1 \phi_2)^2 +h.c. $  \\[3mm] \hline  
$A_4$             & $\mbox{diag}(1,-1,-1),   \newline  \phi_1 \rightarrow \phi_2 \rightarrow \phi_3 \rightarrow \phi_1$   & $V'_0 + \lambda_{1}\left[(\phi_1^\dagger\phi_2)^2  + 
                   (\phi_2^\dagger\phi_3)^2 + (\phi_3^\dagger\phi_1)^2\right] + h.c.$  \\[3mm] \hline  
$S_4$             & $\phi_1 \leftrightarrow - \phi_2, \quad \phi_1 \rightarrow \phi_2 \rightarrow \phi_3 \rightarrow \phi_1$  &  $V'_0 + \lambda_{1}\left[(\phi_1^\dagger\phi_2)^2 +       				   			(\phi_2^\dagger\phi_3)^2 + (\phi_3^\dagger\phi_1)^2\right] + h.c.$  \\[3mm] \hline  
$\Delta(54)/ Z_3$ & $\mbox{diag}(\omega,\omega^2,1)$, \quad
$\phi_3 \rightarrow - \phi_3$,  
\newline
$\phi_1 \rightarrow \phi_2 \rightarrow \phi_3 \rightarrow \phi_1$,  \quad
$\phi_1\leftrightarrow \phi_2$ & $ V'_0 + 					\lambda_{1} \biggl[(\phi_1^\dagger\phi_2)(\phi_1^\dagger\phi_3) + (\phi_2^\dagger\phi_3)(\phi_2^\dagger\phi_1) +(\phi_3^\dagger\phi_1)(\phi^\dagger_3\phi_2) \biggr] +h.c.$  					\\[3mm] \hline  
$\Sigma(36)$      & $\mbox{diag}(1, \omega,\omega^2), \quad d \ (\mbox{see Eq.~\ref{Sigma-generators}})$     & $V'_0 $ \\[3mm] \hline  
\end{tabular}
\end{center}
\end{footnotesize}
\caption{\footnotesize All non-Abelian finite symmetries realisable in the scalar sector of 3HDMs. Note that in each case there are different constraints on the parameters of the phase invariant part of the potential, $V'_0$, which are presented in the main text. The generators of the $\Sigma (36)$ symmetric potential are defined in detail in Eq.~(\ref{Sigma-generators}).}
\label{non-Abelian-table}
\end{table}

\subsection{D$_6 \simeq$ Z$_3 \rtimes$Z$_2$ symmetric 3HDM potential}
The generic $D_6$-symmetric potential has the following form:
\bea
V_{D_6} &=& - \mu^2_{12} (\phi_1^\dagger \phi_1+\phi_2^\dagger \phi_2) - \mu^2_3(\phi_3^\dagger \phi_3) \\
&& + \lambda_{11} \left[(\phi_1^\dagger \phi_1)^2+ (\phi_2^\dagger \phi_2)^2 \right] + \lambda_{33} \left[ (\phi_3^\dagger \phi_3)^2 \right] \nonumber\\
&&  + \lambda_{12} \left[ (\phi_1^\dagger \phi_1)(\phi_2^\dagger \phi_2) \right] 
 + \lambda_{13} \left[ (\phi_3^\dagger \phi_3)(\phi_1^\dagger \phi_1) + (\phi_3^\dagger \phi_3)(\phi_2^\dagger \phi_2) \right] \nonumber\\
&&  + \lambda'_{12} \left[ (\phi_1^\dagger \phi_2)(\phi_2^\dagger \phi_1) \right] 
 + \lambda'_{13} \left[ (\phi_2^\dagger \phi_3)(\phi_3^\dagger \phi_2) + (\phi_3^\dagger \phi_1)(\phi_1^\dagger \phi_3) \right] \nonumber\\
&& + \lambda_1 \biggl[(\phi_2^\dagger \phi_1)(\phi_3^\dagger \phi_1) - (\phi_1^\dagger \phi_2)(\phi_3^\dagger \phi_2) \biggr] + \lambda_2 \biggl[(\phi_1^\dagger \phi_3)(\phi_2^\dagger \phi_3)  \biggr] + h.c. \nonumber
\label{D_6-3HDM}
\eea
with real $\lambda_1$ and complex $\lambda_2$.This group is generated by
\be
a =\left(\begin{array}{ccc} \omega & 0 & 0 \\ 0 & \omega^2 & 0 \\ 0 & 0 & 1 \end{array}\right),\quad \omega=exp(2i\pi/3), \qquad b = \left(\begin{array}{ccc} 0 & -1 & 0 \\ -1 & 0 & 0 \\ 0 & 0 & 1 \end{array}\right) \nonumber
\ee
with $a^3=1$ and $b^2=1$ \cite{Ishimori:2010au}.


The possible minima of this potential are:
\begin{enumerate}
\item
\textbf{Point} $(0,0,\frac{v}{\sqrt{2}})$ respects the symmetry of the potential and becomes the minimum at 
\be
v^2= \frac{\mu^2_3}{\lambda_{33}} \nonumber
\ee
provided the following conditions are satisfied:
\bea 
&& \bullet\quad   -\mu^2_{12}  +\frac{1}{2}(\lambda_{23}+\lambda'_{23} )v^2   >0  \\
&& \bullet\quad    \mu^2_3    >0   \nonumber
\eea
The last condition is already required for the positivity of mass eigenstates at point $(0,0,\frac{v}{\sqrt{2}})$.

\item
\textbf{Point} $(0,\frac{v}{\sqrt{2}},\frac{v}{\sqrt{2}})$ breaks the symmetry of the potential completely and becomes the minimum only when $\lambda_1=\lambda_2$ (in the more general case of $v_2 \neq v_3$ the condition $v_2/v_3 = \lambda_2/\lambda_1$ is required) which does not lead to an extra symmetry of the potential. The minimisation requires 
\be 
v^2= \frac{2\mu^2_3}{2\lambda_{33}+\lambda_{13}+\lambda'_{13}} = \frac{2\mu^2_{12}}{2\lambda_{11}+\lambda_{13}+\lambda'_{13}} \nonumber
\ee
and the following conditions be satisfied:
\bea
&&  \bullet\quad  \lambda_{11}>0 \\
&&  \bullet\quad  \lambda_{33}>0  \nonumber\\ 
&&  \bullet\quad -2\mu^2_{12} + (\lambda_{12} + \lambda'_{12}+ \lambda_{13}+ \lambda'_{13})v^2 >0  \nonumber\\
&& \bullet\quad  \lambda_{11}\lambda_{33} \biggl(-2\mu^2_{12} + (\lambda_{12} + \lambda'_{12}+ \lambda_{13}+ \lambda'_{13})v^2 \biggr) - \lambda_1\lambda_2(\lambda_{13}+\lambda'_{13})v^2 >  \nonumber\\
&& \qquad \frac{1}{2} \lambda_{33}\lambda^2_1 v^2 + 2 \lambda_{11}\lambda^2_2 v^2  + \frac{1}{4} (\lambda_{13}+\lambda'_{13})^2  \biggl(-2\mu^2_{12} + 2(\lambda_{12} + \lambda'_{12}+ \lambda_{13}+ \lambda'_{13})v^2 \biggr)  \nonumber
\eea

\item
\textbf{Point} $(\frac{v}{\sqrt{2}},\frac{v}{\sqrt{2}},\frac{v}{\sqrt{2}})$ becomes the minimum at 
\bea
v^2&=& \frac{2\mu^2_3}{2\lambda_{33}+2\lambda_{13}+2\lambda'_{13} + 2\lambda_2 } \nonumber\\
&=& \frac{2\mu^2_{12}}{2\lambda_{11}+\lambda_{12}+\lambda'_{12}+\lambda_{13}+\lambda'_{13} \pm \lambda_1 + \lambda_2} \nonumber
\eea 
which means that $\lambda_1=0$, however this condition leads to an extra $Z_2$ (the exchange of $\phi_1$ and $\phi_2$) symmetry of the potential. Therefore, we conclude that the point $(\frac{v}{\sqrt{2}},\frac{v}{\sqrt{2}},\frac{v}{\sqrt{2}})$ cannot be a minimum of this potential.

\end{enumerate}

\underline{\textbf{Higgs mass spectrum for} $(0,0,\frac{v}{\sqrt{2}})$}

Expanding the potential around this vacuum point, with 
\be 
v^2=\frac{\mu_3^2}{\lambda_{33}},
\ee
the mass spectrum appears as follows:
\begin{footnotesize}
\bea
&& \textbf{A}_3  : \quad m^2=0 \\
&& \textbf{H}^\pm_3  : \quad m^2=0 \nonumber\\
&& \textbf{H}_3 : \quad m^2= 2\mu_3^2 \nonumber\\
&& \textbf{H}^\pm_2 : \quad m^2= -\mu^2_{12} +\frac{1}{2}\lambda_{13} v^2  \nonumber\\
&& \textbf{H}^\pm_1 : \quad m^2= -\mu^2_{12} +\frac{1}{2}\lambda_{13}v^2  \nonumber\\
&& \textbf{H}_2 \equiv \frac{H^0_{2}- H^0_{1}}{\sqrt{2}} : \quad m^2= -\mu^2_{12} +\frac{1}{2}(\lambda_{13}+\lambda'_{13}-\lambda_2) v^2 \nonumber\\
&& \textbf{H}_1 \equiv \frac{H^0_{2}+ H^0_{1}}{\sqrt{2}} : \quad m^2= -\mu^2_{12} +\frac{1}{2}(\lambda_{13}+\lambda'_{13}+ \lambda_2) v^2 \nonumber\\
&& \textbf{A}_2 \equiv \frac{A^0_{2}- A^0_{1}}{\sqrt{2}} : \quad m^2= -\mu^2_{12} +\frac{1}{2}(\lambda_{13}+\lambda'_{13}-\lambda_2) v^2 \nonumber\\
&& \textbf{A}_1 \equiv \frac{A^0_{2}+ A^0_{1}}{\sqrt{2}} : \quad m^2= -\mu^2_{12} +\frac{1}{2}(\lambda_{13}+\lambda'_{13} +\lambda_2) v^2 \nonumber
\eea
\end{footnotesize}

The lightest neutral field from the first or the second doublet, stabilised by the conserved $D_6$ symmetry, is a viable DM candidate.

\subsection{D$_8 \simeq$ Z$_4 \rtimes$Z$_2$ symmetric 3HDM potential}
The generic $D_8$-symmetric 3HDM potential has the following form:
\bea
V_{D_8}&=& - \mu^2_{12} (\phi_1^\dagger \phi_1+\phi_2^\dagger \phi_2) - \mu^2_3(\phi_3^\dagger \phi_3) \\
&&+ \lambda_{11} \left[(\phi_1^\dagger \phi_1)^2+ (\phi_2^\dagger \phi_2)^2 \right] + \lambda_{33} \left[ (\phi_3^\dagger \phi_3)^2 \right] \nonumber\\
&& + \lambda_{12} \left[ (\phi_1^\dagger \phi_1)(\phi_2^\dagger \phi_2) \right] 
 + \lambda_{13} \left[ (\phi_3^\dagger \phi_3)(\phi_1^\dagger \phi_1) + (\phi_3^\dagger \phi_3)(\phi_2^\dagger \phi_2) \right] \nonumber\\
&& + \lambda'_{12} \left[ (\phi_1^\dagger \phi_2)(\phi_2^\dagger \phi_1) \right] 
 + \lambda'_{13} \left[ (\phi_2^\dagger \phi_3)(\phi_3^\dagger \phi_2) + (\phi_3^\dagger \phi_1)(\phi_1^\dagger \phi_3) \right] \nonumber\\
&&+ \lambda_1 (\phi_3^\dagger \phi_1)(\phi_3^\dagger \phi_2)+ \lambda_2  (\phi_1^\dagger \phi_2)^2 + h.c.  \nonumber
\label{D_8-3HDM}
\eea
This group is generated by:
\be
a = \left(\begin{array}{ccc} i & 0 & 0 \\ 0 & -i & 0 \\ 0 & 0 & 1 \end{array}\right), \qquad b = \left(\begin{array}{ccc} 0 & -1 & 0 \\ -1 & 0 & 0 \\ 0 & 0 & 1 \end{array}\right) \nonumber
\ee
with $a^4=1$, $b^2=1$.

The possible minima of this potential are:
\begin{enumerate}
\item
\textbf{Point} $(0,0,\frac{v}{\sqrt{2}})$ respects the symmetry of the potential and becomes the minimum at 
\be
v^2= \frac{\mu^2_3}{\lambda_{33}} \nonumber
\ee
provided the following conditions are satisfied:
\bea 
&& \bullet\quad  -\mu^2_{12}  +\frac{1}{2}(\lambda_{13}+\lambda'_{13} )v^2   >0  \\
&& \bullet\quad   \mu^2_3    >0   \nonumber
\eea
The last condition is already required for the positivity of mass eigenstates at point $(0,0,\frac{v}{\sqrt{2}})$.

\item
\textbf{Point} $(0,\frac{v}{\sqrt{2}},\frac{v}{\sqrt{2}})$ only becomes the minimum when $\lambda_1=0$ which leads to an extra $Z_2$ (the exchange of $\phi_1$ and $\phi_2$) symmetry of the potential. We therefore conclude that this VEV alignment is not realised in the $D_8$-symmetric potential. 

\item
\textbf{Point} $(\frac{v}{\sqrt{2}},\frac{v}{\sqrt{2}},\frac{v}{\sqrt{2}})$ breaks the symmetry of the potential completely and becomes the minimum at 
\bea
v^2&=& \frac{2\mu^2_3}{2\lambda_{33}+2\lambda_{13}+2\lambda'_{13} +2\lambda_1} \nonumber\\
&=& \frac{2\mu^2_{12}}{2\lambda_{11}+\lambda_{12}+\lambda'_{12}+\lambda_{13}+\lambda'_{13}+\lambda_1 +2\lambda_2}  \nonumber
\eea 
with the following conditions:
\bea 
&& \bullet\quad  \lambda_{33}> \lambda_1   \\
&& \bullet\quad  (4\lambda_2+\lambda_{12} + \lambda'_{12})^2 < (\lambda_{1}+2\lambda_2-2\lambda_{11})^2 \nonumber\\
&& \bullet\quad  2\lambda_{11}> \lambda_1 +2\lambda_2   \nonumber\\
&& \bullet\quad  (\lambda_{1}-\lambda_{33})\biggl[(\lambda_{1}+2\lambda_{2}-2\lambda_{11})^2 - (\lambda_{12}+\lambda'_{12}+4\lambda_{2})^2 \biggr]  < \nonumber\\
&& \qquad  (2\lambda_1 + \lambda_{13}+\lambda'_{13})^2 (\lambda_{12} + \lambda'_{12} +6\lambda_2 -2\lambda_{11}+\lambda_1)  \nonumber
\eea

\end{enumerate}

\underline{\textbf{Higgs mass spectrum for} $(0,0,\frac{v}{\sqrt{2}})$}

Expanding the potential around the vacuum point $(0,0,\frac{v}{\sqrt{2}})$, with 
\be 
v^2=\frac{\mu_3^2}{\lambda_{33}},
\ee
the mass spectrum is the same as the $D_6$-symmetric case with the slight difference in the definition of $a$, $b$ and $Y$ (replace $\lambda_2$ by $\lambda_1$).

The lightest neutral field from the first or the second doublet, stabilised by the conserved $D_8$ symmetry, is a viable DM candidate.

\subsection{A$_4 \simeq $ T = (Z$_2 \times$Z$_2) \rtimes$Z$_3$ symmetric 3HDM potential}

The $A_4$-symmetric 3HDM can be represented by the following potential
\bea
V_{A_4} &=& -\mu^2 \left[(\phi_1^\dagger\phi_1) + (\phi_2^\dagger\phi_2) + (\phi_3^\dagger\phi_3)\right] +
\lambda_{11} \left[(\phi_1^\dagger\phi_1) + (\phi_2^\dagger\phi_2) + (\phi_3^\dagger\phi_3)\right]^2  \\
&& + \lambda_{12} \left[(\phi_1^\dagger\phi_1)(\phi_2^\dagger\phi_2) + (\phi_2^\dagger\phi_2)(\phi_3^\dagger\phi_3) + 
(\phi_3^\dagger\phi_3)(\phi_1^\dagger\phi_1)\right]\nonumber\\
&&+ \lambda'_{12} \left(|\phi_1^\dagger\phi_2|^2 + |\phi_2^\dagger\phi_3|^2 + |\phi_3^\dagger\phi_1|^2\right)  +\lambda_1  \left[(\phi_1^\dagger\phi_2)^2 + (\phi_2^\dagger\phi_3)^2 + (\phi_3^\dagger\phi_1)^2\right] + h.c. \nonumber
\label{A_4-3HDM}
\eea
with complex $\lambda_1$.

This potential is symmetric under:
\bea 
&& a=
\left(\begin{array}{ccc}
1 & 0 & 0 \\
0 & -1 & 0   \\
0 & 0 & -1   \\
\end{array}
\right) , \quad 
b= \left(\begin{array}{ccc}
0 & 1 & 0 \\
0 & 0 & 1    \\
1 & 0 & 0   \\
\end{array}
\right)    \nonumber
\label{usual-A4}
\eea 
with $a^2=1$, $b^3=1$.

The possible minima of this potential are:
\begin{enumerate}
\item
\textbf{Point} $(0,0,\frac{v}{\sqrt{2}})$ breaks the symmetry of the potential to $Z_2$ generated by $g=(-1,-1,1)$ and becomes the minimum at 
\be
v^2= \frac{\mu^2_3}{\lambda_{33}} \nonumber
\ee
provided the following conditions are satisfied:
\bea 
\label{A4-condition}
&& \bullet\quad   -2\mu^2  + (\lambda_{12}+\lambda'_{12}+2\lambda_{11} )v^2   >0  \\
&& \bullet\quad   \mu^2   >0   \nonumber
\eea
The last condition is already required for the positivity of mass eigenstates at point $(0,0,\frac{v}{\sqrt{2}})$.

\item
\textbf{Point} $(0,\frac{v}{\sqrt{2}},\frac{v}{\sqrt{2}})$ becomes the minimum only when $\lambda_1 $ is real which leads to an $S_4$-symmetric potential . Therefore, this point cannot be a minimum of the $A_4$-symmetric potential. Note that in the more general case, the point $(0,v_2,v_3)$ can become a minimum if the following condition is satisfied:
\be 
\left(\frac{v_2}{v_3} \right)^2 = \frac{\lambda_{12}+\lambda'_{12}+2\lambda^*_1}{\lambda_{12}+\lambda'_{12}+2\lambda_1}
\ee

\item
\textbf{Point} $(\frac{v}{\sqrt{2}},\frac{v}{\sqrt{2}},\frac{v}{\sqrt{2}})$ breaks the symmetry of the potential to $S_3$ (permutation of the three doublets) and becomes the minimum at 
\bea
v^2&=& \frac{\mu^2}{3\lambda_{11}+\lambda_{12}+\lambda'_{12} + \lambda_1+ \lambda^*_1} \nonumber
\eea 
with the following conditions:
\begin{small}
\bea 
&& \bullet\quad  -\lambda_{11}> 2\Re\lambda_1    \\
&& \bullet\quad  4(2\Re\lambda_1 +\lambda_{11})^2 > (\lambda_{12} + \lambda'_{12}  +2\lambda_{11}+4\Re\lambda_1)^2  + (4\Im\lambda_1)^2   \nonumber\\
&& \bullet\quad  -4(\lambda_{11}+2\Re\lambda_1 )^3 +3 (\lambda_{11}+2\Re\lambda_1 ) \biggl((\lambda_{12} + \lambda'_{12} +2\lambda_{11}+4\Re\lambda_1)^2  + (4\Im\lambda_1)^2  \biggr) >  \nonumber\\ 
&& \qquad -\biggl((\lambda_{12} + \lambda'_{12} +2\lambda_{11}+4\Re\lambda_1)^2  + (4\Im\lambda_1)^2  \biggr) (\lambda_{12}+ \lambda'_{12} +2\lambda_{11} +4\Re\lambda_1 )  \nonumber
\eea
\end{small}

\end{enumerate}

\underline{\textbf{Higgs mass spectrum for} $(0,0,\frac{v}{\sqrt{2}})$}

Expanding the potential around $(0,0,\frac{v}{\sqrt{2}})$, with 
\be 
v^2 = {\mu^2 \over  \lambda_{11}},
\ee 
we get

\begin{footnotesize}
\bea
&& \textbf{A}_3  : \quad m^2=0 \\
&& \textbf{H}^\pm_3  : \quad m^2=0 \nonumber\\
&& \textbf{H}_3 : \quad m^2= 2\mu^2 \nonumber\\
&& \textbf{H}^\pm_2 : \quad m^2= -\mu^2 +\frac{1}{2}(\lambda_{12}+2\lambda_{11} )v^2  \nonumber\\
&& \textbf{H}^\pm_1 : \quad m^2= -\mu^2 +\frac{1}{2}(\lambda_{12}+2\lambda_{11} ) v^2  \nonumber\\
&& \textbf{H}_2 \equiv \frac{aH^0_{2}+ A^0_{2}}{\sqrt{1+a^2}}  : 
\quad m^2= -\mu^2  +\frac{1}{2} \left(\lambda_{12}+\lambda'_{12} +2\lambda_{11} - \sqrt{\Re^2\lambda_1 + \Im^2\lambda_1}\right)v^2  \nonumber\\
&& \textbf{A}_2 \equiv \frac{bH^0_{2}+ A^0_{2}}{\sqrt{1+b^2}}  : 
\quad m^2= -\mu^2 +\frac{1}{2}\left(\lambda_{12}+\lambda'_{12} +2\lambda_{11} + \sqrt{\Re^2\lambda_1 + \Im^2\lambda_1}\right)v^2   \nonumber\\
&& \textbf{H}_1 \equiv \frac{aH^0_{1}+ A^0_{1}}{\sqrt{1+a^2}}  : 
\quad m^2= -\mu^2 +\frac{1}{2}\left(\lambda_{12}+\lambda'_{12} +2\lambda_{11} - \sqrt{\Re^2\lambda_1 + \Im^2\lambda_1}\right)v^2     \nonumber\\
&& \textbf{A}_1 \equiv \frac{bH^0_{1}+ A^0_{1}}{\sqrt{1+b^2}}  : 
\quad m^2= -\mu^2  +\frac{1}{2}\left(\lambda_{12}+\lambda'_{12} +2\lambda_{11} + \sqrt{\Re^2\lambda_1 + \Im^2\lambda_1}\right)v^2 \nonumber\\
&&  \mbox{where} \quad  a= \frac{1}{\Im\lambda_1} \left[\Re\lambda_1 - \sqrt{\Re^2\lambda_1 + \Im^2\lambda_1}  \right] \nonumber\\
&& \qquad \qquad b= \frac{1}{\Im\lambda_1} \left[\Re\lambda_1 + \sqrt{\Re^2\lambda_1 + \Im^2\lambda_1}  \right] \nonumber
\eea
\end{footnotesize}

The lightest neutral field from the first or the second doublet, stabilised by the remaining $Z_2$ symmetry, is a viable DM candidate.

\subsection{S$_4 \simeq$ O = (Z$_2 \times $Z$_2) \rtimes $S$_3$ symmetric 3HDM potential}

The $S_4$-symmetric 3HDM can be represented by the following potential
\bea
V_{S_4} &=& -\mu^2 \left[(\phi_1^\dagger\phi_1) + (\phi_2^\dagger\phi_2) + (\phi_3^\dagger\phi_3)\right] +
\lambda_{11} \left[(\phi_1^\dagger\phi_1) + (\phi_2^\dagger\phi_2) + (\phi_3^\dagger\phi_3)\right]^2 \\
&& + \lambda_{12} \left[(\phi_1^\dagger\phi_1)(\phi_2^\dagger\phi_2) + (\phi_2^\dagger\phi_2)(\phi_3^\dagger\phi_3) + 
(\phi_3^\dagger\phi_3)(\phi_1^\dagger\phi_1)\right]\nonumber\\
&& + \lambda'_{12} \left(|\phi_1^\dagger\phi_2|^2 + |\phi_2^\dagger\phi_3|^2 + |\phi_3^\dagger\phi_1|^2\right) +\lambda_1  \left[(\phi_1^\dagger\phi_2)^2 + (\phi_2^\dagger\phi_3)^2 + (\phi_3^\dagger\phi_1)^2\right] + h.c. \nonumber
\label{S_4-3HDM}
\eea
with real $\lambda_1$.

This potential is symmetric under:
\bea 
&& b=
\left(\begin{array}{ccc}
0 & 1 & 0 \\
0 & 0 & 1    \\
1 & 0 & 0   \\
\end{array}
\right) , \quad 
c= \left(\begin{array}{ccc}
0  & -1& 0 \\
-1 & 0 & 0   \\
0  & 0 & 1   \\
\end{array}
\right)    
\label{usual-S4}
\eea 
with $b^3=1$ and $c^2=1$.

The possible minima of this potential are:
\begin{enumerate}
\item
\textbf{Point} $(0,0,\frac{v}{\sqrt{2}})$ breaks the symmetry of the potential to $Z_2$ generated by
\begin{footnotesize}
$ c=\left(\begin{array}{ccc}
0  & -1& 0 \\
-1 & 0 & 0   \\
0  & 0 & 1   \\
\end{array}
\right) $\end{footnotesize}
and becomes the minimum at 
\be
v^2= \frac{\mu^2_3}{\lambda_{33}} \nonumber
\ee
provided the following conditions are satisfied:
\bea 
\label{S4-condition}
&& \bullet\quad   -2\mu^2  + (\lambda_{12}+\lambda'_{12}+2\lambda_{11} )v^2   >0  \\
&& \bullet\quad   \mu^2   >0   \nonumber
\eea
The last condition is already required for the positivity of mass eigenstates at point $(0,0,\frac{v}{\sqrt{2}})$.

\item
\textbf{Point} $(0,\frac{v}{\sqrt{2}},\frac{v}{\sqrt{2}})$ breaks the symmetry of the potential completely and becomes the minimum at
\be 
v^2= \frac{2\mu^2}{3\lambda_{11}+\lambda_{12}+\lambda'_{12}+2\lambda_1} \nonumber
\ee
when the following conditions are satisfied:
\bea
&& \bullet\quad   2\lambda_1+\lambda_{11} < 0  \\
&& \bullet\quad   \lambda_{12} + \lambda'_{12} + 4\lambda_1 < 0  \nonumber\\
&& \bullet\quad   4\lambda_{11}(\lambda_{11} +\lambda_{12} + \lambda'_{12}+ 4\lambda_1) < 0  \nonumber
\eea

\item
\textbf{Point} $(\frac{v}{\sqrt{2}},\frac{v}{\sqrt{2}},\frac{v}{\sqrt{2}})$ breaks the symmetry of the potential to $S_3$ (permutation of the three doublets) and becomes the minimum at 
\bea
v^2&=& \frac{\mu^2}{\lambda_{12}+\lambda'_{12}+3\lambda_{11}  +2\lambda_1} \nonumber
\eea 
with the following conditions:
\bea 
&& \bullet\quad  \lambda_{11}> 2\lambda_1   \\
&& \bullet\quad  (2\lambda_1+\lambda_{12} + \lambda'_{12} +2\lambda_{11})^2 < 4(2\lambda_{1}-\lambda_{11})^2  \nonumber\\
&& \bullet\quad  -4(\lambda_{11}+2\lambda_{1})^3 + (2\lambda_1 + \lambda_{12}+2\lambda_{11}+\lambda'_{12})^3 >  \nonumber\\
&& \qquad 3(-\lambda_{11}-2\lambda_1)(2\lambda_1+\lambda_{12}+2\lambda_{11}+\lambda'_{12})^2 \nonumber
\eea

\end{enumerate}

\underline{\textbf{Higgs mass spectrum for} $(0,0,\frac{v}{\sqrt{2}})$}
 
Expanding the potential around $(0,0,\frac{v}{\sqrt{2}})$, with 
\be 
v^2 = {\mu^2 \over  \lambda_{11}},
\ee 
the mass spectrum gets the following form:
\begin{footnotesize}
\bea
&& \textbf{A}_3  : \quad m^2=0 \\
&& \textbf{H}^\pm_3  : \quad m^2=0 \nonumber\\
&& \textbf{H}_3 : \quad m^2= 2\mu^2 \nonumber\\
&& \textbf{H}^\pm_2 : \quad m^2= -\mu^2 +\frac{1}{2}(\lambda_{12}+2\lambda_{11} ) v^2  \nonumber\\
&& \textbf{H}^\pm_1 : \quad m^2= -\mu^2 +\frac{1}{2}(\lambda_{12}+2\lambda_{11} ) v^2  \nonumber\\
&& \textbf{H}_2  : \quad m^2= -\mu^2  +\frac{1}{2}(\lambda_{12}+\lambda'_{12} +2\lambda_{11} +2\lambda_1)v^2  \nonumber\\
&& \textbf{A}_2  : \quad m^2= -\mu^2  +\frac{1}{2}(\lambda_{12}+\lambda'_{12}+2\lambda_{11}  -2\lambda_1)v^2  \nonumber\\
&& \textbf{H}_1  : \quad m^2= -\mu^2  +\frac{1}{2}(\lambda_{12}+\lambda'_{12}+2\lambda_{11}  +2\lambda_1)v^2  \nonumber\\
&& \textbf{A}_1  : \quad m^2= -\mu^2  +\frac{1}{2}(\lambda_{12}+\lambda'_{12}+2\lambda_{11}  -2\lambda_1)v^2  \nonumber
\eea
\end{footnotesize}

The lightest neutral field from the first or the second doublet, stabilised by the remaining $Z_2$ symmetry, is a viable DM candidate.

\subsection{$\Delta(54)/ $Z$_3 \simeq ($Z$_3 \times $Z$_3) \rtimes $Z$_2$ symmetric 3HDM potential}

A 3HDM potential symmetric under this group has the following form\footnote{Note that the group $\Delta(27)$, which is the full preimage of the group $Z_3 \times Z_3$ in $SU(N)$, is not a realizable symmetry of 3HDM potential since the potential symmetric under $\Delta(27)$ is automatically symmetric under the larger group $(Z_3 \times Z_3) \rtimes Z_2 = \Delta(54)/Z_3$.}:
\begin{small}
\bea
V_{\Delta(54)/ Z_3} & = &  - \mu^2 \left[\phi_1^\dagger \phi_1 + \phi_2^\dagger \phi_2 +\phi_3^\dagger \phi_3\right]
+ \lambda_{11} \left[\phi_1^\dagger \phi_1+ \phi_2^\dagger \phi_2+\phi_3^\dagger \phi_3\right]^2 \\
&& + \lambda_{12} \left[(\phi_1^\dagger \phi_1)^2
+ (\phi_2^\dagger \phi_2)^2+(\phi_3^\dagger \phi_3)^2
- (\phi_1^\dagger \phi_1)(\phi_2^\dagger \phi_2) 
- (\phi_2^\dagger \phi_2)(\phi_3^\dagger \phi_3)
- (\phi_3^\dagger \phi_3)(\phi_1^\dagger \phi_1)\right]\nonumber\\
&& + \lambda'_{12} \left[|\phi_1^\dagger \phi_2|^2 
+ |\phi_2^\dagger \phi_3|^2 
+ |\phi_3^\dagger \phi_1|^2\right] \nonumber\\
&&+ \lambda_1 \left[(\phi_1^\dagger \phi_2)(\phi_1^\dagger \phi_3) 
+ (\phi_2^\dagger \phi_3)(\phi_2^\dagger \phi_1) 
+ (\phi_3^\dagger \phi_1)(\phi_3^\dagger \phi_2)\right]+ h.c. \nonumber
\label{Delta-3HDM}
\eea
\end{small}
with real $\mu^2$, $\lambda_{11}$, $\lambda_{12}$, $\lambda'_{12}$ and complex $\lambda_1$.
This group is generated by
\be
a = \left(\begin{array}{ccc} \omega & 0 & 0 \\ 0 & \omega^2 & 0 \\ 0 & 0 & 1 \end{array}\right),\quad
b = \left(\begin{array}{ccc} 0 & 1 & 0 \\ 0 & 0 & 1 \\ 1 & 0 & 0 \end{array}\right), \quad 
c = \left(\begin{array}{ccc} 0 & 1 & 0 \\ 1 & 0 & 0 \\ 0 & 0 & -1 \end{array}\right), \quad 
\omega=exp(2i\pi/3) \nonumber
\ee
with $a^3=1$, $b^3=1$ and $c^2=1$.

The possible minima of this potential are:
\begin{enumerate}
\item
\textbf{Point} $(0,0,\frac{v}{\sqrt{2}})$ breaks the symmetry of the potential to $Z_3$ generated by $a=(\omega, \omega^2, 1)$ and becomes the minimum at 
\be
v^2= \frac{\mu^2}{\lambda_{11}} \nonumber
\ee
provided the following conditions are satisfied:
\bea 
\label{Delta-condition}
&& \bullet\quad   -\lambda_{12}+\lambda'_{12}    >0 \\
&& \bullet\quad    \lambda_{11}+2\lambda_{12}    >0 \nonumber
\eea

\item
\textbf{Point} $(0,\frac{v}{\sqrt{2}},\frac{v}{\sqrt{2}})$ breaks the symmetry of the potential completely and becomes the minimum only when $\lambda_1$ is real, which leads to a larger symmetry of the potential, or when $v_2=-v_3$. So, we study the conditions for the point $(0,\frac{v}{\sqrt{2}},-\frac{v}{\sqrt{2}})$ as the minimum of the potential:
\be 
v^2= \frac{2\mu^2}{4\lambda_{11}+\lambda_{12}+\lambda'_{12}}  \nonumber
\ee
where the following conditions need to be satisfied:
\bea
&& \bullet\quad  -3\lambda_{12}+\lambda'_{12}   >0  \\
&& \bullet\quad   2\lambda_{11} + \lambda_{12}  >0  \nonumber\\
&& \bullet\quad (-3\lambda_{12}+\lambda'_{12})(2\lambda_{11} + \lambda_{12}) > 4|\lambda_1|^2  \nonumber\\
&& \bullet\quad (-3\lambda_{12}+\lambda'_{12})(2\lambda_{11} + \lambda_{12})^2 + 4(\lambda^2_1+ {\lambda^*_1}^2 )(2\lambda_{11}-\lambda_{12}+\lambda'_{12}) > \nonumber\\
&& \qquad (-3\lambda_{12}+\lambda'_{12}) (2\lambda_{11}-\lambda_{12}+\lambda'_{12})^2 +8 |\lambda_1|^2( 2\lambda_{11}+\lambda_{12}) \nonumber
\eea

\item
\textbf{Point} $(\frac{v}{\sqrt{2}},\frac{v}{\sqrt{2}},\frac{v}{\sqrt{2}})$ breaks the symmetry of the potential to $S_3$ (permutation of the three doublets) and becomes the minimum at 
\bea
v^2&=& \frac{\mu^2}{3\lambda_{11} +\lambda'_{12} +  \lambda_1 + \lambda^*_1} \nonumber
\eea 
with the following conditions:
\bea 
&& \bullet\quad  \lambda_{11}+\lambda_{12}> 2\Re\lambda_1    \\
&& \bullet\quad  4(\lambda_{11}+\lambda_{12}-2\Re\lambda_1 )^2 >
(4\Re\lambda_1  +2\lambda_{11}-\lambda_{12} + \lambda'_{12})^2   \nonumber\\
&& \bullet\quad  4(\lambda_{11}+\lambda_{12}-2\Re\lambda_1 )^3 + (4\Re\lambda_1  - \lambda_{12}+\lambda'_{12}+2\lambda_{11})^3 >  \nonumber\\
&& \qquad 3(\lambda_{11}+\lambda_{12}-2\Re\lambda_1 )(4\Re\lambda_1  -\lambda_{12}+\lambda'_{12}+2\lambda_{11})^2 \nonumber
\eea

\end{enumerate}

\underline{\textbf{Higgs mass spectrum for} $(0,0,\frac{v}{\sqrt{2}})$}

The mass spectrum, with 
\be 
v^2= \frac{\mu^2}{\lambda_{11}+\lambda_{12}},
\ee
has the following form:
\begin{footnotesize}
\bea
&& \textbf{A}_3  : \quad m^2=0 \\
&& \textbf{H}^\pm_3   : \quad m^2=0 \nonumber\\
&& \textbf{H}_3 : \quad m^2= 2\mu^2 \nonumber\\
&& \textbf{H}^\pm_2  : \quad m^2= -\mu^2 +\frac{2\lambda_{11}-\lambda_{12}}{2} v^2  \nonumber\\
&& \textbf{H}^\pm_1  : \quad m^2= -\mu^2 +\frac{2\lambda_{11}-\lambda_{12}}{2} v^2  \nonumber\\
&& \textbf{H}'_2 \equiv \frac{-H^0_{2} + H^0_{1} +A^0_{2}+ A^0_{1}}{2} : \quad m^2= -\mu^2 + \frac{1}{2}(2\lambda_{11} -\lambda_{12} + \lambda'_{12} - \Re\lambda_1 - \Im\lambda_1) v^2  \nonumber\\
&& \textbf{H}'_1 \equiv \frac{H^0_{2} - H^0_{1} +A^0_{2}+ A^0_{1}}{2} : \quad m^2= -\mu^2 + \frac{1}{2}(2\lambda_{11} -\lambda_{12} + \lambda'_{12} - \Re\lambda_1 + \Im\lambda_1) v^2  \nonumber\\
&& \textbf{A}'_2 \equiv \frac{H^0_{2} + H^0_{1} -A^0_{2}+ A^0_{1}}{2} : \quad m^2= -\mu^2 + \frac{1}{2}(2\lambda_{11} -\lambda_{12} + \lambda'_{12} + \Re\lambda_1 + \Im\lambda_1) v^2  \nonumber\\
&& \textbf{A}'_1 \equiv \frac{H^0_{2} + H^0_{1} +A^0_{2}- A^0_{1}}{2} : \quad m^2= -\mu^2 + \frac{1}{2}(2\lambda_{11} -\lambda_{12} + \lambda'_{12} + \Re\lambda_1 - \Im\lambda_1) v^2  \nonumber
\eea
\end{footnotesize}

The lightest neutral field from the first or the second doublet, stabilised by the remaining $Z_3$ symmetry, is a viable DM candidate.

\subsection{$\Sigma(36) \simeq ($Z$_3 \times $Z$_3) \rtimes $Z$_4$ symmetric 3HDM potential}

A $\Sigma(36)$-symmetric 3HDM potential has the following form:
\bea
V_{\Sigma(36)} & = & - \mu^2 \left(\phi_1^\dagger \phi_1+ \phi_2^\dagger \phi_2+\phi_3^\dagger \phi_3 \right)
 + \lambda_{11} \left(\phi_1^\dagger \phi_1+ \phi_2^\dagger \phi_2+\phi_3^\dagger \phi_3 \right)^2 \\
&&+ \lambda'_{12} \left(|\phi_1^\dagger \phi_2 - \phi_2^\dagger \phi_3|^2 
+ |\phi_2^\dagger \phi_3 - \phi_3^\dagger \phi_1|^2 
+ |\phi_3^\dagger \phi_1 - \phi_1^\dagger \phi_2|^2\right). \nonumber
\eea
This group is generated by arbitrary permutations of the three doublets, and by:
\be
a = \left(\begin{array}{ccc} 1 & 0 & 0 \\ 0 & \omega & 0 \\ 0 & 0 & \omega^2 \end{array}\right),\quad
d =\frac{1}{\omega^2-\omega} \left(\begin{array}{ccc} 1 & 1 & 1 \\ 1 & \omega^2 & \omega \\ 1 & \omega & \omega^2 \end{array}\right), \quad \omega=exp(2i\pi/3)  
\label{Sigma-generators}
\ee
with $a^3=1$ and $d^4=1$.

The possible minima of this potential are:
\begin{enumerate}
\item
\textbf{Point} $(0,0,\frac{v}{\sqrt{2}})$ breaks the symmetry of the potential to $Z_3$ generated by $a=(\omega, \omega^2, 1)$ and becomes the minimum at 
\be
v^2= \frac{\mu^2}{\lambda_{11}} \nonumber
\ee
provided the following conditions are satisfied:
\bea 
\label{Sigma-condition}
&& \bullet\quad   \lambda'_{12}  >0  \\
&& \bullet\quad    \mu^2    >0   \nonumber
\eea
The last condition is already required for the positivity of mass eigenstates at point $(0,0,\frac{v}{\sqrt{2}})$.

\item
\textbf{Point} $(0,\frac{v}{\sqrt{2}},\frac{v}{\sqrt{2}})$ breaks the symmetry of the potential completely and becomes the minimum at 
\be 
v^2= \frac{\mu^2}{2\lambda_{11}+\lambda'_{12}} \nonumber
\ee
when the following conditions are satisfied:
\bea
&& \bullet\quad    12\lambda_{11} > \lambda'_{12} > 0   \\
&& \bullet\quad   8(\lambda_{11})^2 - \frac{27}{4}\lambda_{11}\lambda'_{12} -2(\lambda'_{12})^2 > 0  \nonumber
\eea

\item
\textbf{Point} $(\frac{v}{\sqrt{2}},\frac{v}{\sqrt{2}},\frac{v}{\sqrt{2}})$ breaks the symmetry of the potential completely and becomes the minimum at 
\bea
v^2&=& \frac{\mu^2}{3\lambda_{11}} \nonumber
\eea 
with the following conditions:
\bea 
&& \bullet\quad   \lambda_{11}+2\lambda'_{12}> 0   \\
&& \bullet\quad   4(\lambda_{11}+2\lambda'_{12})^2 >  (2\lambda_{11}-\lambda'_{12})^2 \nonumber\\
&& \bullet\quad   4(\lambda_{11}+2\lambda'_{12})^3 + (2\lambda_{11}-\lambda'_{12})^3 > 3(\lambda_{11}+2\lambda'_{12})(2\lambda_{11}-\lambda'_{12})^2 \nonumber
\eea

\end{enumerate}

\underline{\textbf{Higgs mass spectrum for} $(0,0,\frac{v}{\sqrt{2}})$}

Expanding the potential around the vacuum point $(0,0,\frac{v}{\sqrt{2}})$, with 
\be 
v^2=\frac{\mu^2}{\lambda_{11}},
\label{minimization-condition}
\ee
the mass spectrum appears as follows:

\begin{footnotesize}
\bea
&& \textbf{A}_3  : \quad m^2=0 \\
&& \textbf{H}^\pm_3  : \quad m^2=0 \nonumber\\
&& \textbf{H}_3 : \quad m^2= 2\mu^2 \nonumber\\
&& \textbf{H}^\pm_2 : \quad m^2= -\mu^2 + \lambda_{11}v^2 \nonumber\\
&& \textbf{H}^\pm_1 : \quad m^2= -\mu^2 + \lambda_{11}v^2  \nonumber\\
&& \textbf{H}_2 \equiv \frac{H^0_{2}- H^0_{1}}{\sqrt{2}} : \quad m^2= -\mu^2 + \frac{1}{2}(2\lambda_{11} +3\lambda'_{12}) v^2  \nonumber\\
&& \textbf{H}_1 \equiv \frac{H^0_{2}+ H^0_{1}}{\sqrt{2}} : \quad m^2= -\mu^2 + \frac{1}{2}(2\lambda_{11} +\lambda'_{12}) v^2  \nonumber\\
&& \textbf{A}_2 \equiv \frac{A^0_{2}- A^0_{1}}{\sqrt{2}} : \quad m^2= -\mu^2 + \frac{1}{2}(2\lambda_{11} +\lambda'_{12}) v^2  \nonumber\\
&& \textbf{A}_1 \equiv \frac{A^0_{2}+ A^0_{1}}{\sqrt{2}} : \quad m^2= -\mu^2 + \frac{1}{2}(2\lambda_{11} +3\lambda'_{12} ) v^2  \nonumber
\eea
\end{footnotesize}

It is interesting to note that the minimisation condition (\ref{minimization-condition}) results in $m^2_{\textbf{H}^\pm_1}$ and $m^2_{\textbf{H}^\pm_2}$ vanishing at tree-level. However, this is accidental (there is no symmetry reason for their vanishing) and so we expect them to acquire small masses at higher order.

The viable DM candidate in this case is the lightest neutral field from the first or the second doublet, stabilised by the remaining $Z_3$ symmetry.


\section{Analysis of 6HDMs}
\label{6HDM}
We define each doublet $\phi_i$ as a doublet of doublets
\be
\phi_i = \doublet{H^i_u}{H^i_d} \nonumber
\ee
where $H^i_u$ and $H^i_d$ are defined in the following way:
\be 
H^i_u = \doublet{$\begin{scriptsize}$ H^+_{iu} $\end{scriptsize}$}{\frac{{H^0_{iu}+iA^0_{iu}}}{\sqrt{2}}},\quad 
H^i_d = \doublet{$\begin{scriptsize}$ H^+_{id} $\end{scriptsize}$}{\frac{{H^0_{id}+iA^0_{id}}}{\sqrt{2}}} \nonumber
\ee
We add the following MSSM-inspired term to avoid extra massless particles:
\be 
\mu''^2 \left(H^\dagger_{1u}H_{1d} + H^\dagger_{2u}H_{2d} + H^\dagger_{3u}H_{3d}   \right) + h.c.
\ee
Note that these terms do not break the symmetry of the potential, since they only mix the intra-doublet fields $H_{iu}$ and $H_{id}$ and not the doublets $\phi_i$.

Extending the $(0,0,\frac{v}{\sqrt{2}})$ minimum to 6 Higgs doublets results in a VEV alignment of the form $(0,0,0,0,\frac{v}{\sqrt{2}},\frac{v}{\sqrt{2}})$, with two active doublets:
\be 
\langle H_{3u} \rangle = \langle H_{3d} \rangle = v \nonumber
\ee
and four inert doublets;
\be  
\langle H_{1u} \rangle = \langle H_{1d} \rangle = \langle H_{2u} \rangle = \langle H_{2d} \rangle= 0. \nonumber
\ee 
In the following sections we present the mass spectrum in each case.

\subsection{U(1)$\times$U(1) symmetric 6HDM potential}

The mass spectrum of the $U(1) \times U(1)$ symmetric 6HDM potential around the minimum point $(0,0,0,0,\frac{v}{\sqrt{2}},\frac{v}{\sqrt{2}})$ with
\be 
v^2=\frac{\mu_3^2-\mu'^2}{2\lambda_{33}}
\ee
has the following form:
\begin{footnotesize}
\bea
&& \textbf{G}_3 = \frac{A^0_{3u}+ A^0_{3d}}{\sqrt{2}} : \quad m^2=0 \\
&& \textbf{G}^\pm_3 = \frac{H^\pm_{3u}+ H^\pm_{3d}}{\sqrt{2}}  : \quad m^2=0 \nonumber\\
&& \textbf{h}_3 = \frac{H^0_{3u}+ H^0_{3d}}{\sqrt{2}} : \quad m^2= 2\mu_3^2 -2\mu'^2 \nonumber\\
&& \textbf{H}_3 = \frac{H^0_{3u}- H^0_{3d}}{\sqrt{2}} : \quad m^2= -2\mu'^2 \nonumber\\
&& \textbf{A}_3 = \frac{A^0_{3u}- A^0_{3d}}{\sqrt{2}} : \quad m^2= -2\mu'^2 \nonumber\\
&& \textbf{H}^\pm_3 = \frac{H^\pm_{3u}- H^\pm_{3d}}{\sqrt{2}} : \quad m^2= -2\mu'^2  \nonumber\\
&& \textbf{h}_2 = \frac{H^0_{2u}+ H^0_{2d}}{\sqrt{2}} : \quad m^2= -\mu^2_2+\mu'^2 +(\lambda_{23}+\lambda'_{23}) v^2  \nonumber\\
&& \textbf{G}_2 = \frac{A^0_{2u}+ A^0_{2d}}{\sqrt{2}}  : \quad m^2= -\mu^2_2+\mu'^2 +(\lambda_{23}+\lambda'_{23}) v^2  \nonumber\\
&& \textbf{G}^\pm_2 = \frac{H^\pm_{2u}+ H^\pm_{2d}}{\sqrt{2}} : \quad m^2=-\mu^2_2+\mu'^2 +\lambda_{23} v^2  \nonumber\\
&& \textbf{H}_2 = \frac{H^0_{2u}- H^0_{2d}}{\sqrt{2}} : \quad m^2= -\mu^2_2 - \mu'^2 +\lambda_{23} v^2  \nonumber\\
&& \textbf{A}_2 = \frac{A^0_{2u}- A^0_{2d}}{\sqrt{2}} : \quad m^2= -\mu^2_2 - \mu'^2 +\lambda_{23} v^2  \nonumber\\
&& \textbf{H}^\pm_2 = \frac{H^\pm_{2u}- H^\pm_{2d}}{\sqrt{2}} : \quad m^2= -\mu^2_2 - \mu'^2 +\lambda_{23} v^2  \nonumber\\
&& \textbf{h}_1 = \frac{H^0_{1u}+ H^0_{1d}}{\sqrt{2}}  : \quad m^2= -\mu^2_1+\mu'^2 +(\lambda_{31}+\lambda'_{31}) v^2  \nonumber\\
&& \textbf{G}_1 = \frac{A^0_{1u}+ A^0_{1d}}{\sqrt{2}} : \quad m^2= -\mu^2_1+\mu'^2 +(\lambda_{31}+\lambda'_{31}) v^2  \nonumber\\
&& \textbf{G}^\pm_1 = \frac{H^\pm_{1u}+ H^\pm_{1d}}{\sqrt{2}} : \quad m^2= -\mu^2_1+\mu'^2 +\lambda_{31}v^2  \nonumber\\
&& \textbf{H}_1 = \frac{H^0_{1u}- H^0_{1d}}{\sqrt{2}} : \quad m^2= -\mu^2_1 - \mu'^2 +\lambda_{31}v^2  \nonumber\\
&& \textbf{A}_1 = \frac{A^0_{1u}- A^0_{1d}}{\sqrt{2}} : \quad m^2= -\mu^2_1 - \mu'^2 +\lambda_{31}v^2  \nonumber\\
&& \textbf{H}^\pm_1 = \frac{H^\pm_{1u}- H^\pm_{1d}}{\sqrt{2}}  : \quad m^2= -\mu^2_1 - \mu'^2 +\lambda_{31}v^2  \nonumber
\eea
\end{footnotesize}

\subsection{U(1) symmetric 6HDM potential}

The mass spectrum of the $U(1)$ symmetric 6HDM potential around the minimum point $(0,0,0,0,\frac{v}{\sqrt{2}},\frac{v}{\sqrt{2}})$ with
\be 
v^2=\frac{\mu_3^2-\mu'^2}{2\lambda_{33}}
\ee
has the following form:
\begin{footnotesize}
\bea
&& \textbf{G}_3 = \frac{A^0_{3u}+ A^0_{3d}}{\sqrt{2}}  : \quad m^2=0 \\
&& \textbf{G}^\pm_3 = \frac{H^\pm_{3u}+ H^\pm_{3d}}{\sqrt{2}}  : \quad m^2=0 \nonumber\\
&& \textbf{h}_3 = \frac{H^0_{3u}+ H^0_{3d}}{\sqrt{2}} : \quad m^2= 2\mu_3^2-2\mu'^2 \nonumber\\
&& \textbf{H}_3 = \frac{H^0_{3u}- H^0_{3d}}{\sqrt{2}}  : \quad m^2= -2\mu'^2 \nonumber\\
&& \textbf{A}_3 = \frac{A^0_{3u}- A^0_{3d}}{\sqrt{2}}  : \quad m^2= -2\mu'^2 \nonumber\\
&& \textbf{H}^\pm_3 = \frac{H^\pm_{3u}- H^\pm_{3d}}{\sqrt{2}} : \quad m^2= -2\mu'^2  \nonumber\\
&& \textbf{H}^\pm_2 = \frac{H^\pm_{2u}- H^\pm_{2d}}{\sqrt{2}}  : \quad m^2= -\mu^2_2 - \mu'^2 +\lambda_{23} v^2  \nonumber\\
&& \textbf{G}^\pm_2 = \frac{H^\pm_{2u}+ H^\pm_{2d}}{\sqrt{2}}  : \quad m^2=-\mu^2_2+\mu'^2 +\lambda_{23} v^2  \nonumber\\
&& \textbf{H}^\pm_1 = \frac{H^\pm_{1u}- H^\pm_{1d}}{\sqrt{2}} : \quad m^2= -\mu^2_1 - \mu'^2 +\lambda_{31}v^2  \nonumber\\
&& \textbf{G}^\pm_1 = \frac{H^\pm_{1u}+ H^\pm_{1d}}{\sqrt{2}} : \quad m^2= -\mu^2_1+\mu'^2 +\lambda_{31}v^2  \nonumber\\
&& \textbf{A}_2 = \frac{A^0_{2u}- A^0_{2d}}{\sqrt{2}} : \quad m^2= -\mu^2_2 - \mu'^2 +\lambda_{23} v^2  \nonumber\\
&& \textbf{A}_1 = \frac{A^0_{1u}- A^0_{1d}}{\sqrt{2}} : \quad m^2= -\mu^2_1 - \mu'^2 +\lambda_{31}v^2  \nonumber\\
&& \textbf{G}'_2 = \frac{-aA^0_{2u}- aA^0_{2d}+A^0_{1u}+ A^0_{1d}}{\sqrt{2+2a^2}} : \quad m^2= X - \sqrt{Y}  \nonumber\\
&& \textbf{G}'_1 = \frac{-bA^0_{2u}- bA^0_{2d}+A^0_{1u}+ A^0_{1d}}{\sqrt{2+2b^2}} : \quad m^2= X + \sqrt{Y}  \nonumber\\
&& \textbf{H}_2 = \frac{H^0_{2u}- H^0_{2d}}{\sqrt{2}} : \quad m^2= -\mu^2_2 - \mu'^2 +\lambda_{23} v^2  \nonumber\\
&& \textbf{H}_1 = \frac{H^0_{1u}- H^0_{1d}}{\sqrt{2}} : \quad m^2= -\mu^2_1 - \mu'^2 +\lambda_{31}v^2  \nonumber\\
&& \textbf{h}'_2 = \frac{aH^0_{2u}+ aH^0_{2d}+H^0_{1u}+ H^0_{1d}}{\sqrt{2+2a^2}} : \quad m^2= X - \sqrt{Y}  \nonumber\\
&& \textbf{h}'_1 = \frac{bH^0_{2u}+ bH^0_{2d}+H^0_{1u}+ H^0_{1d}}{\sqrt{2+2b^2}} : \quad m^2= X + \sqrt{Y}  \nonumber\\
&&  \mbox{where} \quad X= \frac{1}{2}\left[2\mu'^2 -\mu_1^2-\mu_2^2 + (\lambda_{23} + \lambda_{31} + \lambda'_{23} + \lambda'_{31})v^2 \right] \nonumber\\
&& \qquad \qquad Y= (\lambda_1 v^2)^2+ \frac{1}{4}\left[\mu_1^2 -\mu_2^2 + (\lambda_{23} - \lambda_{31} + \lambda'_{23} - \lambda'_{31})v^2  \right]^2 \nonumber\\
&& \qquad \qquad a= \frac{1}{\lambda_1 v^2} \left[\mu_1^2 -\mu_2^2 + (\lambda_{23} - \lambda_{31} + \lambda'_{23} - \lambda'_{31})v^2 -\sqrt{Y}  \right] \nonumber\\
&& \qquad \qquad b= \frac{1}{\lambda_1 v^2} \left[\mu_1^2 -\mu_2^2 + (\lambda_{23} - \lambda_{31} + \lambda'_{23} - \lambda'_{31})v^2 + \sqrt{Y}  \right]  \nonumber
\eea
\end{footnotesize}

\subsection{U(1)$\times$Z$_2$ symmetric 6HDM potential}
The mass spectrum of the $U(1) \times Z_2$ symmetric 6HDM potential around the minimum point $(0,0,0,0,\frac{v}{\sqrt{2}},\frac{v}{\sqrt{2}})$ with
\be 
v^2=\frac{\mu_3^2-\mu'^2}{2\lambda_{33}}
\ee
has the following form:
\begin{footnotesize}
\bea
&& \textbf{G}_3 = \frac{A^0_{3u}+ A^0_{3d}}{\sqrt{2}}  : \quad m^2=0 \\
&& \textbf{G}^\pm_3 = \frac{H^\pm_{3u}+ H^\pm_{3d}}{\sqrt{2}}   : \quad m^2=0 \nonumber\\
&& \textbf{h}_3 = \frac{H^0_{3u}+ H^0_{3d}}{\sqrt{2}} : \quad m^2= 2\mu_3^2 -2\mu'^2 \nonumber\\
&& \textbf{H}_3 = \frac{H^0_{3u}- H^0_{3d}}{\sqrt{2}} : \quad m^2= -2\mu'^2 \nonumber\\
&& \textbf{A}_3 = \frac{A^0_{3u}- A^0_{3d}}{\sqrt{2}} : \quad m^2= -2\mu'^2 \nonumber\\
&& \textbf{H}^\pm_3 = \frac{H^\pm_{3u}- H^\pm_{3d}}{\sqrt{2}}  : \quad m^2= -2\mu'^2  \nonumber\\
&& \textbf{h}_2 = \frac{H^0_{2u}+ H^0_{2d}}{\sqrt{2}} : \quad m^2= -\mu^2_2+\mu'^2 +(\lambda_{23}+\lambda'_{23} + \lambda_1) v^2  \nonumber\\
&& \textbf{G}_2 = \frac{A^0_{2u}+ A^0_{2d}}{\sqrt{2}} : \quad m^2= -\mu^2_2+\mu'^2 +(\lambda_{23}+\lambda'_{23} -\lambda_1) v^2  \nonumber\\
&& \textbf{G}^\pm_2 = \frac{H^\pm_{2u}+ H^\pm_{2d}}{\sqrt{2}}  : \quad m^2=-\mu^2_2+\mu'^2 +\lambda_{23} v^2  \nonumber\\
&& \textbf{H}_2 = \frac{H^0_{2u}- H^0_{2d}}{\sqrt{2}} : \quad m^2= -\mu^2_2 - \mu'^2 +\lambda_{23} v^2  \nonumber\\
&& \textbf{A}_2 = \frac{A^0_{2u}- A^0_{2d}}{\sqrt{2}} : \quad m^2= -\mu^2_2 - \mu'^2 +\lambda_{23} v^2  \nonumber\\
&& \textbf{H}^\pm_2 = \frac{H^\pm_{2u}- H^\pm_{2d}}{\sqrt{2}} : \quad m^2= -\mu^2_2 - \mu'^2 +\lambda_{23} v^2  \nonumber\\
&& \textbf{h}_1 = \frac{H^0_{1u}+ H^0_{1d}}{\sqrt{2}} : \quad m^2= -\mu^2_1+\mu'^2 +(\lambda_{31}+\lambda'_{31}) v^2  \nonumber\\
&& \textbf{G}_1 = \frac{A^0_{1u}+ A^0_{1d}}{\sqrt{2}} : \quad m^2= -\mu^2_1+\mu'^2 +(\lambda_{31}+\lambda'_{31}) v^2  \nonumber\\
&& \textbf{G}^\pm_1 = \frac{H^\pm_{1u}+ H^\pm_{1d}}{\sqrt{2}} : \quad m^2= -\mu^2_1+\mu'^2 +\lambda_{31}v^2  \nonumber\\
&& \textbf{H}_1 = \frac{H^0_{1u}- H^0_{1d}}{\sqrt{2}} : \quad m^2= -\mu^2_1 - \mu'^2 +\lambda_{31}v^2  \nonumber\\
&& \textbf{A}_1 = \frac{A^0_{1u}- A^0_{1d}}{\sqrt{2}} : \quad m^2= -\mu^2_1 - \mu'^2 +\lambda_{31}v^2  \nonumber\\
&& \textbf{H}^\pm_1 = \frac{H^\pm_{1u}- H^\pm_{1d}}{\sqrt{2}}  : \quad m^2= -\mu^2_1 - \mu'^2 +\lambda_{31}v^2  \nonumber
\eea
\end{footnotesize}

\subsection{Z$_2$ symmetric 6HDM potential}
The mass spectrum of the $Z_2$ symmetric 6HDM potential around the minimum point $(0,0,0,0,\frac{v}{\sqrt{2}},\frac{v}{\sqrt{2}})$ with
\be 
v^2=\frac{\mu_3^2-\mu'^2}{2\lambda_{33}}
\ee
has the following form:
\begin{footnotesize}
\bea
&& \textbf{G}_3 = \frac{A^0_{3u}+ A^0_{3d}}{\sqrt{2}} : \quad m^2=0 \\
&& \textbf{G}^\pm_3 = \frac{H^\pm_{3u}+ H^\pm_{3d}}{\sqrt{2}} : \quad m^2=0 \nonumber\\
&& \textbf{h}_3 = \frac{H^0_{3u}+ H^0_{3d}}{\sqrt{2}} : \quad m^2= 2\mu_3^2 -2\mu'^2 \nonumber\\
&& \textbf{H}_3 = \frac{H^0_{3u}- H^0_{3d}}{\sqrt{2}} : \quad m^2= -2\mu'^2 \nonumber\\
&& \textbf{A}_3 = \frac{A^0_{3u}- A^0_{3d}}{\sqrt{2}} : \quad m^2= -2\mu'^2 \nonumber\\
&& \textbf{H}^\pm_3 = \frac{H^\pm_{3u}- H^\pm_{3d}}{\sqrt{2}} : \quad m^2= -2\mu'^2  \nonumber\\
&& \textbf{h}'_2 = \frac{XH^0_{2u} -XH^0_{2d} -H^0_{1u} +H^0_{1d}}{\sqrt{2+2X^2}} : \quad m^2= \alpha_1 +\alpha_2 -\beta_1 -\beta_2 -\sqrt{\Delta} \nonumber\\
&& \textbf{H}'_2 = \frac{-YH^0_{2u} +YH^0_{2d} -H^0_{1u} +H^0_{1d}}{\sqrt{2+2Y^2}} : \quad m^2= \alpha_1 +\alpha_2 -\beta_1 -\beta_2 +\sqrt{\Delta} \nonumber\\
&& \textbf{h}'_1 = \frac{-WH^0_{2u} -WH^0_{2d} +H^0_{1u} +H^0_{1d}}{\sqrt{2+2W^2}} : \quad m^2= \alpha_1 +\alpha_2 +\beta_1 +\beta_2 -\sqrt{\Delta} \nonumber\\
&& \textbf{H}'_1 = \frac{ZH^0_{2u} +ZH^0_{2d} +H^0_{1u} +H^0_{1d}}{\sqrt{2+2Z^2}}  : \quad m^2= \alpha_1 +\alpha_2 +\beta_1 +\beta_2 +\sqrt{\Delta} \nonumber\\
&&  \mbox{where} \quad \alpha_1 = -\frac{\mu^2_2}{2}+ \frac{1}{4}(2\lambda_{23} + \lambda'_{23} +2\lambda_2)v^2 \nonumber\\[2mm]
&& \qquad \qquad \alpha_2=  -\frac{\mu^2_1}{2}+ \frac{1}{4}(2\lambda_{31} + \lambda'_{31} +2\lambda_3)v^2 \nonumber\\[2mm]
&& \qquad \qquad \beta_1=   \frac{\mu'^2}{2}+ \frac{1}{4}(\lambda'_{23} +2\lambda_2)v^2 \nonumber\\[2mm]
&& \qquad \qquad \beta_2=   \frac{\mu'^2}{2}+ \frac{1}{4}(\lambda'_{31} +2\lambda_3)v^2 \nonumber\\[2mm]
&& \qquad \qquad \gamma=   -\frac{\mu^2_{12}}{2} \nonumber\\[2mm]
&& \textbf{G}'_2 = \frac{X'A^0_{2u} -X'A^0_{2d} -A^0_{1u} +A^0_{1d}}{\sqrt{2+2X'^2}} : \quad m^2= \alpha'_1 +\alpha'_2 -\beta'_1 -\beta'_2 -\sqrt{\Delta'} \nonumber\\
&& \textbf{A}'_2 = \frac{-Y'A^0_{2u} -Y'A^0_{2d} -A^0_{1u} +A^0_{1d}}{\sqrt{2+2Y'^2}} : \quad m^2= \alpha'_1 +\alpha'_2 -\beta'_1 -\beta'_2 +\sqrt{\Delta'} \nonumber\\
&& \textbf{G}'_1 = \frac{-W'A^0_{2u} -W'A^0_{2d} +A^0_{1u} +A^0_{1d}}{\sqrt{2+2W'^2}} : \quad m^2= \alpha'_1 +\alpha'_2 +\beta'_1 +\beta'_2 -\sqrt{\Delta'} \nonumber\\
&& \textbf{A}'_1 = \frac{Z'A^0_{2u} +Z'A^0_{2d} +A^0_{1u} +A^0_{1d}}{\sqrt{2+2Z'^2}} : \quad m^2= \alpha'_1 +\alpha'_2 +\beta'_1 +\beta'_2 +\sqrt{\Delta'} \nonumber\\
&&  \mbox{where} \quad \alpha'_1 = -\frac{\mu^2_2}{2}+ \frac{1}{4}(2\lambda_{23} + \lambda'_{23} -2\lambda_2)v^2 \nonumber\\[2mm]
&& \qquad \qquad \alpha'_2=  -\frac{\mu^2_1}{2}+ \frac{1}{4}(2\lambda_{31} + \lambda'_{31} -2\lambda_3)v^2 \nonumber\\[2mm]
&& \qquad \qquad \beta'_1=   \frac{\mu'^2}{2}+ \frac{1}{4}(\lambda'_{23} -2\lambda_2)v^2 \nonumber\\[2mm]
&& \qquad \qquad \beta'_2=   \frac{\mu'^2}{2}+ \frac{1}{4}(\lambda'_{31} -2\lambda_3)v^2 \nonumber\\[2mm]
&& \qquad \qquad \gamma'=   -\frac{\mu^2_{12}}{2} \nonumber\\[2mm]
&& \textbf{H}'^\pm_2 = \frac{X''H^\pm_{2u} -X''H^\pm_{2d} -H^\pm_{1u} +H^\pm_{1d}}{\sqrt{2+2X''^2}} : \quad m^2= \alpha''_1 +\alpha''_2 -\beta''_1 -\beta''_2 -\sqrt{\Delta''} \nonumber\\
&& \textbf{G}'^\pm_2 = \frac{-Y''H^\pm_{2u} +Y''H^\pm_{2d} -H^\pm_{1u} +H^\pm_{1d}}{\sqrt{2+2Y''^2}} : \quad m^2= \alpha''_1 +\alpha''_2 -\beta''_1 -\beta''_2 +\sqrt{\Delta''} \nonumber\\
&& \textbf{H}'^\pm_1 = \frac{-W''H^\pm_{2u} -W''H^\pm_{2d} +H^\pm_{1u} +H^\pm_{1d}}{\sqrt{2+2W''^2}} : \quad m^2= \alpha''_1 +\alpha''_2 +\beta''_1 +\beta''_2 -\sqrt{\Delta''} \nonumber\\
&& \textbf{G}'^\pm_1 = \frac{Z''H^\pm_{2u} +Z''H^\pm_{2d} +H^\pm_{1u} +H^\pm_{1d}}{\sqrt{2+2Z''^2}} : \quad m^2= \alpha''_1 +\alpha''_2 +\beta''_1 +\beta''_2 +\sqrt{\Delta''} \nonumber\\
&&  \mbox{where} \quad \alpha''_1 = -\mu^2_2+ \lambda_{23} v^2 \nonumber\\[2mm]
&& \qquad \qquad \alpha''_2=  -\mu^2_1 + \lambda_{31} v^2 \nonumber\\[2mm]
&& \qquad \qquad \beta''_1= \beta''_2=  \mu'^2 , \qquad \gamma''=   -\mu^2_{12} \nonumber\\[2mm]
&&  \mbox{and} \quad X = \frac{1}{2\gamma} \left[-\alpha_1 + \alpha_2 + \beta_1 -\beta_2 + \sqrt{\Delta}  \right]  \nonumber\\[2mm]
&& \qquad \qquad Y=   \frac{1}{2\gamma} \left[\alpha_1 - \alpha_2 - \beta_1 +\beta_2 + \sqrt{\Delta}  \right]  \nonumber\\[2mm]
&& \qquad \qquad W=   \frac{1}{2\gamma} \left[-\alpha_1 + \alpha_2 - \beta_1 +\beta_2 + \sqrt{\Delta}  \right]  \nonumber\\[2mm]
&& \qquad \qquad Z=   \frac{1}{2\gamma} \left[\alpha_1 - \alpha_2 + \beta_1 -\beta_2 + \sqrt{\Delta}  \right]  \nonumber\\[2mm]
&& \qquad \qquad \Delta= (\alpha_1 - \alpha_2 - \beta_1 +\beta_2 )^2 + 4\gamma^2  \nonumber\\[2mm] \nonumber
\eea
\end{footnotesize}
and $X',Y',W',Z'$ and $X'',Y'',W'',Z''$ are defined in the same way as $X,Y,W,Z$ with the corresponding primed and double-primed $\alpha$, $\beta$ and $\gamma$s.

\subsection{Z$_3$ symmetric 6HDM potential}
The mass spectrum of the $Z_3$ symmetric 6HDM potential around the minimum point $(0,0,0,0,\frac{v}{\sqrt{2}},\frac{v}{\sqrt{2}})$ with
\be 
v^2=\frac{\mu_3^2-\mu'^2}{2\lambda_{33}}
\ee
is the same as the $U(1)$-symmetric case with a slight difference (replace $\lambda_1$ by  $\lambda_2$).

\subsection{Z$_4$ symmetric 6HDM potential}
The mass spectrum of the $Z_4$ symmetric 6HDM potential around the minimum point $(0,0,0,0,\frac{v}{\sqrt{2}},\frac{v}{\sqrt{2}})$ with
\be 
v^2=\frac{\mu_3^2-\mu'^2}{2\lambda_{33}}
\ee
is identical to that of the $U(1)$-symmetric case.

\subsection{Z$_2\times$Z$_2$ symmetric 6HDM potential}

The mass spectrum of the $Z_2 \times Z_2$ symmetric 6HDM potential around the minimum point $(0,0,0,0,\frac{v}{\sqrt{2}},\frac{v}{\sqrt{2}})$ with
\be 
v^2=\frac{\mu_3^2-\mu'^2}{2\lambda_{33}}
\ee
has the following form:
\begin{footnotesize}
\bea
&& \textbf{G}_3 = \frac{A^0_{3u}+ A^0_{3d}}{\sqrt{2}} : \quad m^2=0 \\
&& \textbf{G}^\pm_3 = \frac{H^\pm_{3u}+ H^\pm_{3d}}{\sqrt{2}} : \quad m^2=0 \nonumber\\
&& \textbf{h}_3 = \frac{H^0_{3u}+ H^0_{3d}}{\sqrt{2}} : \quad m^2= 2\mu_3^2 -2\mu'^2 \nonumber\\
&& \textbf{H}_3 = \frac{H^0_{3u}- H^0_{3d}}{\sqrt{2}} : \quad m^2= -2\mu'^2 \nonumber\\
&& \textbf{A}_3 = \frac{A^0_{3u}- A^0_{3d}}{\sqrt{2}} : \quad m^2= -2\mu'^2 \nonumber\\
&& \textbf{H}^\pm_3 = \frac{H^\pm_{3u}- H^\pm_{3d}}{\sqrt{2}} : \quad m^2= -2\mu'^2  \nonumber\\
&& \textbf{h}_2 = \frac{H^0_{2u}+ H^0_{2d}}{\sqrt{2}} : \quad m^2= -\mu^2_2+\mu'^2 +(\lambda_{23}+\lambda'_{23} + \lambda_2) v^2  \nonumber\\
&& \textbf{G}_2 = \frac{A^0_{2u}+ A^0_{2d}}{\sqrt{2}} : \quad m^2= -\mu^2_2+\mu'^2 +(\lambda_{23}+\lambda'_{23} -\lambda_2) v^2  \nonumber\\
&& \textbf{G}^\pm_2 = \frac{H^\pm_{2u}+ H^\pm_{2d}}{\sqrt{2}} : \quad m^2=-\mu^2_2+\mu'^2 +\lambda_{23} v^2  \nonumber\\
&& \textbf{H}_2 = \frac{H^0_{2u}- H^0_{2d}}{\sqrt{2}} : \quad m^2= -\mu^2_2 - \mu'^2 +\lambda_{23} v^2  \nonumber\\
&& \textbf{A}_2 = \frac{A^0_{2u}- A^0_{2d}}{\sqrt{2}} : \quad m^2= -\mu^2_2 - \mu'^2 +\lambda_{23} v^2  \nonumber\\
&& \textbf{H}^\pm_2 = \frac{H^\pm_{2u}- H^\pm_{2d}}{\sqrt{2}} : \quad m^2= -\mu^2_2 - \mu'^2 +\lambda_{23} v^2  \nonumber\\
&& \textbf{h}_1 = \frac{H^0_{1u}+ H^0_{1d}}{\sqrt{2}} : \quad m^2= -\mu^2_1+\mu'^2 +(\lambda_{31}+\lambda'_{31} + \lambda_3) v^2  \nonumber\\
&& \textbf{G}_1 = \frac{A^0_{1u}+ A^0_{1d}}{\sqrt{2}} : \quad m^2= -\mu^2_1+\mu'^2 +(\lambda_{31}+\lambda'_{31} - \lambda_3) v^2  \nonumber\\
&& \textbf{G}^\pm_1 = \frac{H^\pm_{1u}+ H^\pm_{1d}}{\sqrt{2}} : \quad m^2= -\mu^2_1+\mu'^2 +\lambda_{31}v^2  \nonumber\\
&& \textbf{H}_1 = \frac{H^0_{1u}- H^0_{1d}}{\sqrt{2}}  : \quad m^2= -\mu^2_1 - \mu'^2 +\lambda_{31}v^2  \nonumber\\
&& \textbf{A}_1 = \frac{A^0_{1u}- A^0_{1d}}{\sqrt{2}}  : \quad m^2= -\mu^2_1 - \mu'^2 +\lambda_{31}v^2  \nonumber\\
&& \textbf{H}^\pm_1 = \frac{H^\pm_{1u}- H^\pm_{1d}}{\sqrt{2}} : \quad m^2= -\mu^2_1 - \mu'^2 +\lambda_{31}v^2  \nonumber
\eea
\end{footnotesize}

\subsection{D$_6$ symmetric 6HDM potential}

The mass spectrum of the $D_6$ symmetric 6HDM potential around the minimum point $(0,0,0,0,\frac{v}{\sqrt{2}},\frac{v}{\sqrt{2}})$ with
\be 
v^2=\frac{\mu_3^2-\mu'^2}{2\lambda_{33}}
\ee
has the following form:
\begin{footnotesize}
\bea
&& \textbf{G}_3 = \frac{A^0_{3u}+ A^0_{3d}}{\sqrt{2}} : \quad m^2=0 \\
&& \textbf{G}^\pm_3 = \frac{H^\pm_{3u}+ H^\pm_{3d}}{\sqrt{2}}  : \quad m^2=0 \nonumber\\
&& \textbf{h}_3 = \frac{H^0_{3u}+ H^0_{3d}}{\sqrt{2}} : \quad m^2= 2\mu_3^2 -2\mu'^2 \nonumber\\
&& \textbf{H}_3 = \frac{H^0_{3u}- H^0_{3d}}{\sqrt{2}} : \quad m^2= -2\mu'^2 \nonumber\\
&& \textbf{A}_3 = \frac{A^0_{3u}- A^0_{3d}}{\sqrt{2}} : \quad m^2= -2\mu'^2 \nonumber\\
&& \textbf{H}^\pm_3 = \frac{H^\pm_{3u}- H^\pm_{3d}}{\sqrt{2}} : \quad m^2= -2\mu'^2  \nonumber\\
&& \textbf{G}^\pm_2 = \frac{H^\pm_{2u}+ H^\pm_{2d}}{\sqrt{2}} : \quad m^2= -\mu^2_{12}+\mu'^2 +\lambda_{13} v^2  \nonumber\\
&& \textbf{G}^\pm_1 = \frac{H^\pm_{1u}+ H^\pm_{1d}}{\sqrt{2}} : \quad m^2= -\mu^2_{12}+\mu'^2 +\lambda_{13} v^2  \nonumber\\
&& \textbf{H}^\pm_2 = \frac{H^\pm_{2u}- H^\pm_{2d}}{\sqrt{2}} : \quad m^2= -\mu^2_{12}-\mu'^2 +\lambda_{13} v^2  \nonumber\\
&& \textbf{H}^\pm_1 = \frac{H^\pm_{1u}- H^\pm_{1d}}{\sqrt{2}} : \quad m^2= -\mu^2_{12}-\mu'^2 +\lambda_{13} v^2  \nonumber\\
&& \textbf{A}_2 = \frac{A^0_{2u}- A^0_{2d}}{\sqrt{2}} : \quad m^2= -\mu^2_{12}-\mu'^2 +\lambda_{13} v^2  \nonumber\\
&& \textbf{A}_1 = \frac{A^0_{1u}- A^0_{1d}}{\sqrt{2}} : \quad m^2= -\mu^2_{12}-\mu'^2 +\lambda_{13} v^2  \nonumber\\
&& \textbf{H}_2 = \frac{H^0_{2u}- H^0_{2d}}{\sqrt{2}} : \quad m^2= -\mu^2_{12}-\mu'^2 +\lambda_{13} v^2  \nonumber\\
&& \textbf{H}_1 = \frac{H^0_{1u}- H^0_{1d}}{\sqrt{2}} : \quad m^2= -\mu^2_{12}-\mu'^2 +\lambda_{13} v^2  \nonumber\\
&& \textbf{G}'_2 = \frac{A^0_{2u}+ A^0_{2d}+A^0_{1u}+ A^0_{1d}}{2}: \quad m^2= -\mu^2_{12}+\mu'^2 +(\lambda_{13}+\lambda'_{13} - \lambda_2) v^2  \nonumber\\
&& \textbf{G}'_1 = \frac{-A^0_{2u}- A^0_{2d}+A^0_{1u}+ A^0_{1d}}{\sqrt{2}}: \quad m^2= -\mu^2_{12}+\mu'^2 +(\lambda_{13}+\lambda'_{13} +\lambda_2) v^2   \nonumber\\
&& \textbf{h}'_2 = \frac{H^0_{2u}+H^0_{2d}+H^0_{1u}+ H^0_{1d}}{2}: \quad m^2= -\mu^2_{12}+\mu'^2 +(\lambda_{13}+\lambda'_{13} - \lambda_2) v^2  \nonumber\\
&& \textbf{h}'_1 = \frac{-H^0_{2u}-H^0_{2d}+H^0_{1u}+ H^0_{1d}}{2}: \quad m^2= -\mu^2_{12}+\mu'^2 +(\lambda_{13}+\lambda'_{13} + \lambda_2) v^2  \nonumber
\eea
\end{footnotesize}

\subsection{D$_8$ symmetric 6HDM potential}

The mass spectrum of the $D_8$ symmetric 6HDM potential around the minimum point $(0,0,0,0,\frac{v}{\sqrt{2}},\frac{v}{\sqrt{2}})$ with
\be 
v^2=\frac{\mu_3^2-\mu'^2}{2\lambda_{33}}
\ee
is the same as the $D_6$-symmetric case with the slight difference in the definition of $a$, $b$ and $Y$ (replace $\lambda_2$ by $\lambda_1$).

\subsection{A$_4$ symmetric 6HDM potential}

The mass spectrum of the $A_4$ symmetric 6HDM potential around the minimum point $(0,0,0,0,\frac{v}{\sqrt{2}},\frac{v}{\sqrt{2}})$ with
\be 
v^2=\frac{\mu_3^2-\mu'^2}{2\lambda_{11}}
\ee
has the following form:
\begin{footnotesize}
\bea
&& \textbf{G}_3 = \frac{A^0_{3u}+ A^0_{3d}}{\sqrt{2}}  : \quad m^2=0 \\
&& \textbf{G}^\pm_3 = \frac{H^\pm_{3u}+ H^\pm_{3d}}{\sqrt{2}}  : \quad m^2=0 \nonumber\\
&& \textbf{h}_3 = \frac{H^0_{3u}+ H^0_{3d}}{\sqrt{2}}  : \quad m^2= 2\mu^2 -2\mu'^2 \nonumber\\
&& \textbf{A}_3 = \frac{A^0_{3u}- A^0_{3d}}{\sqrt{2}}  : \quad m^2= -2\mu'^2 \nonumber\\
&& \textbf{H}^\pm_3 = \frac{H^\pm_{3u}- H^\pm_{3d}}{\sqrt{2}} : \quad m^2= -2\mu'^2  \nonumber\\
&& \textbf{H}_3 = \frac{H^0_{3u}- H^0_{3d}}{\sqrt{2}}  : \quad m^2=  -2\mu'^2 \nonumber\\
&& \textbf{G}^\pm_2 = \frac{H^\pm_{2u}+ H^\pm_{2d}}{\sqrt{2}}  : \quad m^2= -\mu^2 +\mu'^2+ (\lambda_{12} +2\lambda_{11}) v^2 \nonumber\\
&& \textbf{G}^\pm_1 = \frac{H^\pm_{1u}+ H^\pm_{1d}}{\sqrt{2}}  : \quad m^2= -\mu^2 +\mu'^2+ (\lambda_{12} +2\lambda_{11}) v^2 \nonumber\\
&& \textbf{H}^\pm_2 = \frac{H^\pm_{2u}- H^\pm_{2d}}{\sqrt{2}}  : \quad m^2= -\mu^2 -\mu'^2+ (\lambda_{12} +2\lambda_{11}) v^2 \nonumber\\
&& \textbf{H}^\pm_1 = \frac{H^\pm_{1u}- H^\pm_{1d}}{\sqrt{2}}  : \quad m^2= -\mu^2 -\mu'^2+ (\lambda_{12} +2\lambda_{11}) v^2 \nonumber\\
&& \textbf{H}_2 = \frac{H^0_{2u}- H^0_{2d}}{\sqrt{2}} : \quad m^2= \mu^2 -\mu'^2+ \frac{1}{2}(\lambda_{12} +2\lambda_{11}) v^2 \nonumber\\
&& \textbf{H}_1 = \frac{H^0_{1u}- H^0_{1d}}{\sqrt{2}} : \quad m^2= \mu^2 -\mu'^2+ \frac{1}{2}(\lambda_{12} +2\lambda_{11}) v^2 \nonumber\\
&& \textbf{A}_2 = \frac{A^0_{2u}- A^0_{2d}}{\sqrt{2}} : \quad m^2= \mu^2 -\mu'^2+ \frac{1}{2}(\lambda_{12} +2\lambda_{11}) v^2 \nonumber\\
&& \textbf{A}_1 = \frac{A^0_{1u}- A^0_{1d}}{\sqrt{2}} : \quad m^2= \mu^2 -\mu'^2+ \frac{1}{2}(\lambda_{12} +2\lambda_{11}) v^2 \nonumber\\
&& \textbf{h}'_2 = \frac{aH^0_{2u}+aH^0_{2d}+A^0_{2u}+A^0_{2d}}{\sqrt{2+2a^2}} : \quad m^2= -\mu^2 +\mu'^2+ (\lambda_{12} +2\lambda_{11} + \lambda'_{12} + \Re\lambda_1 - 2\sqrt{2}\Im\lambda_1) v^2 \nonumber\\
&& \textbf{h}'_1 = \frac{aH^0_{1u}+aH^0_{1d}+A^0_{1u}+A^0_{1d}}{\sqrt{2+2a^2}} : \quad m^2= -\mu^2 +\mu'^2+ (\lambda_{12} +2\lambda_{11} + \lambda'_{12} + \Re\lambda_1 - 2\sqrt{2}\Im\lambda_1) v^2 \nonumber\\
&& \textbf{G}'_2 = \frac{bH^0_{2u}+bH^0_{2d}+A^0_{2u}+A^0_{2d}}{\sqrt{2+2b^2}} : \quad m^2= -\mu^2 +\mu'^2+ (\lambda_{12} +2\lambda_{11} + \lambda'_{12} + \Re\lambda_1 + 2\sqrt{2}\Im\lambda_1) v^2 \nonumber\\
&& \textbf{G}'_1 = \frac{bH^0_{1u}+bH^0_{1d}+A^0_{1u}+A^0_{1d}}{\sqrt{2+2b^2}} : \quad m^2= -\mu^2 +\mu'^2+ (\lambda_{12} +2\lambda_{11} + \lambda'_{12} + \Re\lambda_1 + 2\sqrt{2}\Im\lambda_1) v^2 \nonumber\\
&&  \mbox{where} \quad a= (1-\sqrt{2}) \quad \mbox{and} \quad b= (1+\sqrt{2}) \nonumber
\eea
\end{footnotesize}

\subsection{S$_4$ symmetric 6HDM potential}

The mass spectrum of the $S_4$ symmetric 6HDM potential around the minimum point $(0,0,0,0,\frac{v}{\sqrt{2}},\frac{v}{\sqrt{2}})$ with
\be 
v^2=\frac{\mu_3^2-\mu'^2}{2\lambda_{11}}
\ee
has the following form:
\begin{footnotesize}
\bea
&& \textbf{G}_3 = \frac{A^0_{3u}+ A^0_{3d}}{\sqrt{2}}  : \quad m^2=0 \\
&& \textbf{G}^\pm_3 = \frac{H^\pm_{3u}+ H^\pm_{3d}}{\sqrt{2}}  : \quad m^2=0 \nonumber\\
&& \textbf{h}_3 = \frac{H^0_{3u}+ H^0_{3d}}{\sqrt{2}}  : \quad m^2= 2\mu^2 -2\mu'^2 \nonumber\\
&& \textbf{A}_3 = \frac{A^0_{3u}- A^0_{3d}}{\sqrt{2}}  : \quad m^2= -2\mu'^2 \nonumber\\
&& \textbf{H}^\pm_3 = \frac{H^\pm_{3u}- H^\pm_{3d}}{\sqrt{2}}  : \quad m^2= -2\mu'^2  \nonumber\\
&& \textbf{H}_3 = \frac{H^0_{3u}- H^0_{3d}}{\sqrt{2}}  : \quad m^2= -2\mu'^2 \nonumber\\
&& \textbf{G}^\pm_2 = \frac{H^\pm_{2u}+ H^\pm_{2d}}{\sqrt{2}} : \quad m^2= -\mu^2 +\mu'^2 +(\lambda_{12}+2\lambda_{11}) v^2 \nonumber\\
&& \textbf{G}^\pm_1 = \frac{H^\pm_{1u}+ H^\pm_{1d}}{\sqrt{2}} : \quad m^2= -\mu^2 +\mu'^2 +(\lambda_{12}+2\lambda_{11}) v^2 \nonumber\\
&& \textbf{H}^\pm_2 = \frac{H^\pm_{2u}- H^\pm_{2d}}{\sqrt{2}} : \quad m^2= -\mu^2 -\mu'^2 +(\lambda_{12}+2\lambda_{11}) v^2 \nonumber\\
&& \textbf{H}^\pm_1 = \frac{H^\pm_{1u}- H^\pm_{1d}}{\sqrt{2}} : \quad  m^2= -\mu^2 -\mu'^2 +(\lambda_{12}+2\lambda_{11}) v^2 \nonumber\\
&& \textbf{h}_2 = \frac{H^0_{2u}+ H^0_{2d}}{\sqrt{2}} : \quad  m^2= -\mu^2 + \mu'^2 +(\lambda_{12}+2\lambda_{11} +\lambda'_{12} + \lambda_1) v^2 \nonumber\\
&& \textbf{h}_1 = \frac{H^0_{1u}+ H^0_{1d}}{\sqrt{2}} : \quad m^2= -\mu^2 + \mu'^2 +(\lambda_{12}+2\lambda_{11} +\lambda'_{12} + \lambda_1) v^2 \nonumber\\
&& \textbf{A}_2 = \frac{A^0_{2u}- A^0_{2d}}{\sqrt{2}} : \quad m^2= -\mu^2 - \mu'^2 +(\lambda_{12}+2\lambda_{11}) v^2 \nonumber\\
&& \textbf{A}_1 = \frac{A^0_{1u}- A^0_{1d}}{\sqrt{2}} : \quad m^2= -\mu^2 - \mu'^2 +(\lambda_{12}+2\lambda_{11}) v^2 \nonumber\\
&& \textbf{H}_2 = \frac{H^0_{2u}- H^0_{2d}}{\sqrt{2}} : \quad m^2= -\mu^2 - \mu'^2 +(\lambda_{12}+2\lambda_{11}) v^2 \nonumber\\
&& \textbf{H}_1 = \frac{H^0_{1u}- H^0_{1d}}{\sqrt{2}} : \quad m^2= -\mu^2 - \mu'^2 +(\lambda_{12}+2\lambda_{11}) v^2 \nonumber\\
&& \textbf{G}_2 = \frac{A^0_{2u}+ A^0_{2d}}{\sqrt{2}} : \quad m^2= -\mu^2 + \mu'^2 +(\lambda_{12}+2\lambda_{11} +\lambda'_{12} - \lambda_1) v^2 \nonumber\\
&& \textbf{G}_1 = \frac{A^0_{1u}+ A^0_{1d}}{\sqrt{2}} : \quad m^2= -\mu^2 + \mu'^2 +(\lambda_{12}+2\lambda_{11} +\lambda'_{12} - \lambda_1) v^2 \nonumber
\eea
\end{footnotesize}

\subsection{$\Delta(54)/ $Z$_3$ symmetric 6HDM potential}

The mass spectrum of the $\Delta(54)/ Z_3$ symmetric 6HDM potential around the minimum point $(0,0,0,0,\frac{v}{\sqrt{2}},\frac{v}{\sqrt{2}})$ with
\be 
v^2=\frac{\mu^2-\mu'^2}{2(\lambda_{11}+\lambda_{12})} 
\ee
has the following form:

\begin{footnotesize}
\bea
&& \textbf{G}_3 = \frac{A^0_{3u}+ A^0_{3d}}{\sqrt{2}}  : \quad m^2=0 \\
&& \textbf{G}^\pm_3 = \frac{H^\pm_{3u}+ H^\pm_{3d}}{\sqrt{2}}  : \quad m^2=0 \nonumber\\
&& \textbf{h}_3 = \frac{H^0_{3u}+ H^0_{3d}}{\sqrt{2}}  : \quad m^2= 2\mu^2 -2\mu'^2 \nonumber\\
&& \textbf{H}_3 = \frac{H^0_{3u}- H^0_{3d}}{\sqrt{2}} : \quad m^2= -2\mu'^2 \nonumber\\
&& \textbf{A}_3 = \frac{A^0_{3u}- A^0_{3d}}{\sqrt{2}} : \quad m^2= -2\mu'^2 \nonumber\\
&& \textbf{H}^\pm_3 = \frac{H^\pm_{3u}- H^\pm_{3d}}{\sqrt{2}}  : \quad m^2= -2\mu'^2  \nonumber\\
&& \textbf{H}^\pm_2 = \frac{H^\pm_{2u}- H^\pm_{2d}}{\sqrt{2}}  : \quad m^2= -\mu^2-\mu'^2 +(2\lambda_{11}-\lambda_{12}) v^2  \nonumber\\
&& \textbf{H}^\pm_1 = \frac{H^\pm_{1u}- H^\pm_{1d}}{\sqrt{2}}  : \quad m^2= -\mu^2-\mu'^2 +(2\lambda_{11}-\lambda_{12}) v^2  \nonumber\\
&& \textbf{G}^\pm_2 = \frac{H^\pm_{2u}+ H^\pm_{2d}}{\sqrt{2}}   : \quad m^2= -\mu^2+\mu'^2 +(2\lambda_{11}-\lambda_{12}) v^2  \nonumber\\
&& \textbf{G}^\pm_1 = \frac{H^\pm_{1u}+ H^\pm_{1d}}{\sqrt{2}}  : \quad m^2=-\mu^2+\mu'^2 +(2\lambda_{11}-\lambda_{12}) v^2  \nonumber\\
&& \textbf{H}_2 = \frac{H^0_{2u}- H^0_{2d}}{\sqrt{2}}  : \quad m^2= -\mu^2-\mu'^2 +(2\lambda_{11}-\lambda_{12}) v^2 \nonumber\\ 
&& \textbf{A}_2 = \frac{A^0_{2u}- A^0_{2d}}{\sqrt{2}}  : \quad m^2= -\mu^2-\mu'^2 +(2\lambda_{11}-\lambda_{12}) v^2 \nonumber\\ 
&& \textbf{H}_1 = \frac{H^0_{1u}- H^0_{1d}}{\sqrt{2}}  : \quad m^2= -\mu^2-\mu'^2 +(2\lambda_{11}-\lambda_{12}) v^2 \nonumber\\ 
&& \textbf{A}_1 = \frac{A^0_{1u}- A^0_{1d}}{\sqrt{2}}  : \quad m^2= -\mu^2-\mu'^2 +(2\lambda_{11}-\lambda_{12}) v^2 \nonumber\\
&& \textbf{h}'_2 = \frac{W H^0_{2u}+ W H^0_{2d} -X H^0_{1u} - X H^0_{1d} + A^0_{2u} + A^0_{2d}}{\sqrt{2+ X^2+2 W^2}} : \nonumber\\
&& \qquad \qquad m^2= -\mu^2+\mu'^2 +(2\lambda_{11}-\lambda_{12}+\lambda'_{12} - \frac{1}{2}\sqrt{\Re^2\lambda_1 + \Im^2\lambda_1})v^2 \nonumber\\[3mm]
&& \textbf{h}'_1 = \frac{WH^0_{2u}+ WH^0_{2d}+XH^0_{1u} +XH^0_{1d}+A^0_{2u}+A^0_{2d}}{\sqrt{2+X^2+2W^2}} : \nonumber\\
&&  \qquad \qquad m^2= -\mu^2+\mu'^2 +(2\lambda_{11}-\lambda_{12}+\lambda'_{12} + \frac{1}{2}\sqrt{\Re^2\lambda_1 + \Im^2\lambda_1})v^2   \nonumber\\[3mm]
&& \textbf{G}'_2 = \frac{-W'H^0_{1u} -W'H^0_{1d}-(Y+Z)A^0_{2u}+ (Y-Z) A^0_{2d}+A^0_{1u}+A^0_{1d}}{\sqrt{2+(Y+Z)^2+(Y-Z)^2+2W'^2}} : \nonumber\\
&&  \qquad \qquad m^2= -\mu^2+\mu'^2 +(2\lambda_{11}-\lambda_{12}+\lambda'_{12} - \frac{1}{4}\sqrt{2\Re^2\lambda_1 + 4\Im^2\lambda_1})v^2   \nonumber\\[3mm]
&& \textbf{G}'_1 = \frac{W'H^0_{1u} +W'H^0_{1d}-(Y-Z)A^0_{2u}+ (Y+Z) A^0_{2d}+A^0_{1u}+A^0_{1d}}{\sqrt{2+(Y+Z)^2+(Y-Z)^2+2W'^2}} : \nonumber\\
&&  \qquad \qquad m^2= -\mu^2+\mu'^2 +(2\lambda_{11}-\lambda_{12}+\lambda'_{12} + \frac{1}{4}\sqrt{2\Re^2\lambda_1 + 4\Im^2\lambda_1})v^2  \nonumber\\
&&  \mbox{where} \quad W = \frac{\Re\lambda_1}{\Im\lambda_1} \qquad \mbox{and} \qquad W' = \frac{\Im\lambda_1}{\Re\lambda_1} \nonumber\\[2mm]
&& \qquad \qquad X= \frac{\sqrt{\Re^2\lambda_1 + \Im^2\lambda_1}}{\Im\lambda_1}  \nonumber\\[2mm]
&& \qquad \qquad Y= \frac{\Re\lambda_1 \left(2{\mu'^2 / v^2} +\lambda'_{12} \right)}{-2\left(2{\mu'^2 / v^2} +\lambda'_{12} \right)^2  +\Re^2\lambda_1 + 2\Im^2\lambda_1} \nonumber\\[2mm]
&& \qquad \qquad Z= \frac{\left(2 {\mu'^2 / v^2} +\lambda'_{12} \right)^2\sqrt{2\Re^2\lambda_1 + 4\Im^2\lambda_1}}{\left[-2\left(2 {\mu'^2 / v^2} +\lambda'_{12} \right)^2  +\Re^2\lambda_1 + 2\Im^2\lambda_1 \right] \Re\lambda_1} \nonumber
\eea
\end{footnotesize}

\subsection{$\Sigma(36)$ symmetric 6HDM potential}

The mass spectrum of the $\Sigma(36)$ symmetric 6HDM potential around the minimum point $(0,0,0,0,\frac{v}{\sqrt{2}},\frac{v}{\sqrt{2}})$ with
\be 
v^2=\frac{\mu^2-\mu'^2}{2\lambda_{11}}
\ee
has the following form:
\begin{footnotesize}
\bea
&& \textbf{G}_3 = \frac{A^0_{3u}+ A^0_{3d}}{\sqrt{2}} : \quad m^2=0 \\
&& \textbf{G}^\pm_3 = \frac{H^\pm_{3u}+ H^\pm_{3d}}{\sqrt{2}} : \quad m^2=0 \nonumber\\
&& \textbf{G}^\pm_2 = \frac{H^\pm_{2u}+ H^\pm_{2d}}{\sqrt{2}} : \quad m^2= -\mu^2+\mu'^2+2\lambda_{11} v^2 \nonumber\\
&& \textbf{G}^\pm_1 = \frac{H^\pm_{1u}+ H^\pm_{1d}}{\sqrt{2}} : \quad m^2= -\mu^2+\mu'^2+2\lambda_{11} v^2 \nonumber\\
&& \textbf{h}_3 = \frac{H^0_{3u}+ H^0_{3d}}{\sqrt{2}} : \quad m^2= 2\mu^2 -2\mu'^2 \nonumber\\
&& \textbf{H}_3 = \frac{H^0_{3u}- H^0_{3d}}{\sqrt{2}} : \quad m^2= -2\mu'^2 \nonumber\\
&& \textbf{A}_3 = \frac{A^0_{3u}- A^0_{3d}}{\sqrt{2}} : \quad m^2= -2\mu'^2 \nonumber\\
&& \textbf{H}_2 = \frac{H^0_{2u}- H^0_{2d}}{\sqrt{2}} : \quad m^2= -2\mu'^2  \nonumber\\ 
&& \textbf{A}_2 = \frac{A^0_{2u}- A^0_{2d}}{\sqrt{2}} : \quad m^2= -2\mu'^2  \nonumber\\
&& \textbf{H}_1 = \frac{H^0_{1u}- H^0_{1d}}{\sqrt{2}} : \quad m^2= -2\mu'^2  \nonumber\\
&& \textbf{A}_1 = \frac{A^0_{1u}- A^0_{1d}}{\sqrt{2}} : \quad m^2= -2\mu'^2 \nonumber\\
&& \textbf{H}^\pm_3 = \frac{H^\pm_{3u}- H^\pm_{3d}}{\sqrt{2}} : \quad m^2= -2\mu'^2  \nonumber\\
&& \textbf{H}^\pm_2 = \frac{H^\pm_{2u}- H^\pm_{2d}}{\sqrt{2}} : \quad m^2= -2\mu'^2  \nonumber\\
&& \textbf{H}^\pm_1 = \frac{H^\pm_{1u}- H^\pm_{1d}}{\sqrt{2}} : \quad m^2= -2\mu'^2 \nonumber\\
&& \textbf{h}_2 = \frac{H^0_{2u}+ H^0_{2d}}{\sqrt{2}}  : \quad m^2= 2\lambda'_{12} v^2  \nonumber\\
&& \textbf{G}_2 = \frac{A^0_{2u}+ A^0_{2d}}{\sqrt{2}}  : \quad m^2= 2\lambda'_{12} v^2 \nonumber\\
&& \textbf{h}_1 = \frac{H^0_{1u}+ H^0_{1d}}{\sqrt{2}}  : \quad m^2= 2\lambda'_{12} v^2 \nonumber\\
&& \textbf{G}_1 = \frac{A^0_{1u}+ A^0_{1d}}{\sqrt{2}}  : \quad m^2= 2\lambda'_{12} v^2 \nonumber
\eea
\end{footnotesize}

\section{Conclusion}
\label{conclusion}

In this paper we have considered 3HDMs which are the next simplest class of models, following the
well studied 2HDMs. We have argued that 3HDMs are ready for serious investigation since their possible
symmetries have been largely identified. Furthermore, they may shed light on the flavour problem, in the sense
that their symmetries may be identified as family symmetries which also describe the three families of quarks and leptons.

We have catalogued and studied 3HDMs in terms of all possible allowed symmetries
(continuous and discrete Abelian and discrete non-Abelian).
We have analysed the potential in each case, and derived the conditions under which 
the vacuum alignments $(0,0,v)$, $(0,v,v)$ and $(v,v,v)$ are minima of the potential.
For the alignment $(0,0,v)$, relevant for DM models, we have calculated the corresponding 
physical Higgs boson mass spectrum. 

Motivated by SUSY, we have extended the analysis
to the case of three up-type Higgs doublets and three down-type Higgs doublets (six doublets in total),
for the case of $\tan \beta = 1$.
Many of the results are also applicable to flavon
models where the three Higgs doublets are replaced by three electroweak singlets.

In conclusion, following the discovery of a Higgs boson by the LHC, it is clear that Nature admits at least
one Higgs doublet for the purpose of breaking electroweak symmetry. However it is not yet clear if there are 
two or more Higgs doublets which are relevant in Nature. We have systematically studied the case of 3HDMs,
whose symmetries may shed light on the flavour problem. If SUSY is relevant, then we have shown
how the analysis may be straightforwardly extended to 6HDMs.

\section{Acknowledgement}
VK acknowledges numerous useful discussions with Andrew G. Akeroyd and Alexander Merle. VK is supported by The Leverhulme Trust (London, UK) via a Visiting Fellowship. SM is financed in part through the NExT Institute and from the STFC Consolidated ST/J000396/1. SFK also acknowledges partial support from the STFC Consolidated ST/J000396/1 and EU ITN grants UNILHC 237920 and INVISIBLES 289442.

\appendix

\section{Finding all realisable symmetries}\label{methodology}

\subsection{Abelian symmetries}
As mentioned in section \ref{syms-in-3hdm}, the group of physically distinct unitary reparametrisation transformations respected by the kinetic terms $G_0$ in 3HDMs is
\be 
G_0= PSU(3) \simeq SU(3)/\Z_3. \nonumber
\ee
To find all realisable Abelian subgroups of $G_0$, one first needs to construct the maximal Abelian subgroups of $PSU(3)$ and then explore the realisable subgroups of the maximal Abelian subgroup.

A maximal Abelian subgroup of $G_0$ is an Abelian group that is not contained in any larger Abelian subgroup of $G_0$. In principal, $G_0$ can have several maximal Abelian subgroups and any subgroup of $G_0$ is either a subgroup of a maximal Abelian subgroup or is itself a maximal Abelian subgroup.

Within $SU(N)$, it is known that all maximal Abelian subgroups are maximal tori \cite{Ivanov:2011ae}:
\be
[U(1)]^{N-1} = U(1)\times U(1) \times \cdots \times U(1).
\label{maximaltorus1}
\ee
All such maximal tori are conjugate to each other\footnote{If $T_1$ and $T_2$ are two maximal tori, there exists $g \in SU(N)$ such that $g^{-1}T_1g = T_2$.}. Therefore, without loss of generality, we could pick one specific maximal torus and study its subgroup. It is convenient to pick the maximal torus represented by phase rotations of individual doublets:
\be
\mbox{diag}\biggl(e^{i\alpha_1},\, e^{i\alpha_2},\, \dots ,\, e^{i\alpha_{N-1}},\, e^{-i\sum\alpha_i}\biggr).
\label{maximaltorus2}
\ee
This transformation can be written as the vector of phases,
\be
\left(\alpha_1,\,\alpha_2,\,\dots,\,\alpha_{N-1}, -\sum\alpha_i\right).
\ee
Therefore, the maximal torus inside $SU(N)$ has the following form:
\be
T_0 =  U(1)_1 \times U(1)_2 \times \cdots \times U(1)_{N-1} \nonumber 
\ee
where
\bea
U(1)_1 & = & \alpha_1(-1,\, 1,\, 0,\, 0,\, \dots,\, 0) \nonumber\\
U(1)_2 & = & \alpha_2(-2,\, 1,\, 1,\, 0,\, \dots,\, 0) \nonumber\\
U(1)_3 & = & \alpha_3(-3,\, 1,\, 1,\, 1,\, \dots,\, 0) \nonumber\\
\vdots &  & \vdots \nonumber\\
U(1)_{N-1} & = & \alpha_{N-1}(-N+1, \, 1,\, 1,\, 1,\, \dots,\, 1) 
\label{groupsUi}
\eea
with all $\alpha_i \in [0,2\pi)$.
However, the center of $SU(N)$, which is generated by $\alpha_{N-1}=2\pi/N$ and results in trivial transformations, is contained in $U(1)_{N-1}$. Therefore we introduce
\be 
\overline{U(1)}_{N-1} = U(1)_{N-1}/Z(SU(N)) = \alpha_{N-1}\left(-{N-1 \over N}, \, {1\over N},\, \dots,\, {1 \over N}\right)
\label{def-of-Ubar}
\ee
where $\alpha_{N-1} \in [0,2\pi)$.

Thus, the maximal torus in $PSU(N)$ appears as follows:
\be
T = U(1)_1\times U(1)_2 \times \cdots \times \overline{U(1)}_{N-1} \label{maximal-torus-PSUN}
\ee

Having constructed the maximal torus within $PSU(N)$, one needs to exhaust the list of realisable Abelian subgroups of $T$, which is done by adding bilinear terms to the most general $T$-symmetric potential:
\be
V_0 = \sum^N_i \left[- |\mu^2_i| (\phi_i^\dagger \phi_i) + \lambda_{ii} (\phi_i^\dagger \phi_i)^2\right] 
+ \sum^N_{ij}\left[\lambda_{ij}(\phi_i^\dagger \phi_i) (\phi_j^\dagger \phi_j) + 
\lambda'_{ij}(\phi_i^\dagger \phi_j) (\phi_j^\dagger \phi_i)\right].
\ee
and checking the symmetries of the resulting potential.

Each bilinear $\phi_a^\dagger \phi_b$ $(a\not = b)$, gets a phase change under $T$ (\ref{maximal-torus-PSUN}): 
\be
(\phi_a^\dagger \phi_b) \to \exp[i(m_1\alpha_1 + m_2\alpha_2 + \dots + m_{N-1}\alpha_{N-1})] ( \phi_a^\dagger \phi_b)
\label{pq-generic-NHDM}
\ee
with integer coefficients $m_1, m_2, \dots, m_{N-1}$. This linear dependence on the angles $\alpha_j$ could be written as 
\be 
\sum^{N-1}_{j=1} m_j \alpha_j .
\ee
The phase transformation properties of a given monomial are fully described by its vector $m_j$, an the phase transformation properties of the potential $V$, which is a collection of $k$ monomials, is characterized
by $k$ vectors $m_{1,j},\, m_{2,j},\, \dots,\, m_{k,j}$.

For a monomial to be invariant under a given transformation defined by phases $\{\alpha_j\}$,
one requires that $$\sum_{j=1}^N m_j \alpha_j = 2 \pi n$$ with some integer $n$. 
For an entire potential of $k$ terms to be invariant under a given phase transformation, one requires:
\be
\sum_{j=1}^N m_{i,j}\alpha_j = 2 \pi n_i\,,\quad \mbox{for all $1 \le i \le k$} \label{systemUN}
\ee
with integer $n_i$s.
In order to solve this set of equations, one could construct the matrix of coefficients $m_{i,j}$ with all integer entries, and diagonalize it, reducing the set of equations to:
\be
m'_{i,i}\alpha'_i = 2\pi  n'_i\,, \quad  \alpha'_i \in [0,2\pi)\,,\quad n'_i \in \Z 
\ee
with non-negative integers $m'_{i,i}$.

\begin{itemize}
\item
If $m_{i,i}=0$, this equation has a solution for any $\alpha_i$; the $i$-th equation contributes a factor $U(1)$ to the 
symmetry group of the potential.
\item
If $m_{i,i} =1$, this equation has no non-trivial solution; the $i$-th equation does not contribute to the symmetry group of the potential.
\item
If $m_{i,i} =d_i > 1$, this equation has $d_i$ solutions $\alpha_i = 2\pi/d_i$; the $i$-th equation contributes the factor $\Z_{d_i}$ to the symmetry group of the potential.
\end{itemize}
The full symmetry group of the potential is then constructed from the direct product of the above factors.

Following this strategy for 3HDMs, we 
\begin{itemize}
\item
construct the maximal torus $T \subset PSU(3)$:
\be
T = U(1)_1\times U(1)_2\,,\quad U(1)_1 = \alpha(-1,\,1,\,0)\,,\quad U(1)_2 = \beta\left(-{2 \over 3},\, {1 \over 3},\, {1 \over 3}\right)
\label{3HDM-maximaltorus}
\ee
where $\alpha,\beta \in [0,2\pi)$,

\item 
write down the most general $T$-symmetrci potential:
\bea
V_0 &=& - \mu^2_{1} (\phi_1^\dagger \phi_1) -\mu^2_2 (\phi_2^\dagger \phi_2) - \mu^2_3(\phi_3^\dagger \phi_3) \\
&&+ \lambda_{11} (\phi_1^\dagger \phi_1)^2+ \lambda_{22} (\phi_2^\dagger \phi_2)^2  + \lambda_{33} (\phi_3^\dagger \phi_3)^2 \nonumber\\
&& + \lambda_{12}  (\phi_1^\dagger \phi_1)(\phi_2^\dagger \phi_2)
 + \lambda_{23}  (\phi_2^\dagger \phi_2)(\phi_3^\dagger \phi_3) + \lambda_{31} (\phi_3^\dagger \phi_3)(\phi_1^\dagger \phi_1) \nonumber\\
&& + \lambda'_{12} (\phi_1^\dagger \phi_2)(\phi_2^\dagger \phi_1) 
 + \lambda'_{23} (\phi_2^\dagger \phi_3)(\phi_3^\dagger \phi_2) + \lambda'_{31} (\phi_3^\dagger \phi_1)(\phi_1^\dagger \phi_3),  \nonumber
\eea

\item 
add to the $T$-symmetric potential any combination (quadratic and quartic) of the following doublets transforming non-trivially under $T$,
\be 
(\phi_1^\dagger \phi_2), \quad (\phi_2^\dagger \phi_3), \quad (\phi_3^\dagger \phi_1)
\ee
and their conjugates,

\item
construct the matrix of coefficients $m_{ij}$ for all the potentials and diagonalize it,

\end{itemize}

and arrive at the full list of subgroups of the maximal torus realisable as the symmetry groups of the potential:
\be
\Z_2,\quad \Z_3,\quad  \Z_4,\quad \Z_2\times \Z_2,\quad U(1),\quad U(1)\times \Z_2,\quad U(1)\times U(1)  \label{list3HDM}
\ee

The only finite Abelian group that is not contained in any maximal torus in $PSU(3)$ is $\Z_3 \times \Z_3$.
It turns out that the $\Z_3 \times \Z_3$-symmetric potential is symmetric under the $(\Z_3 \times \Z_3) \rtimes \Z_2$ group which is non-Abelian. Therefore, according to our definition the symmetry group $\Z_3 \times \Z_3$ is not realisable. However, it needs to be considered when constructing the realisable non-Abelian groups out of the Abelian ones, which is done in the next subsection.

\subsection{Non-Abelian symmetries}
To exhaust the list of finite non-Abelian subgroups $G \subset PSU(3)$, one needs to find all finite Abelian subgroups $A \subset PSU(3)$:
\be 
A: \quad \Z_2,\quad \Z_3,\quad  \Z_4,\quad \Z_2\times \Z_2,\quad \Z_3 \times \Z_3
\label{list-Abelian}
\ee 
which was described in the previous subsection.

Note that the order of all Abelian subgroups $|A|$ has two prime divisors, 2 and 3. The order of all non-Abelian groups $|G|$ can therefore only have the same two prime divisors (Cauchy's theorem). This means that the group $G$ is solvable (Burnside's theorem) and contains a normal self-centralizing Abelian subgroup $A$ from the above list (\ref{list-Abelian}). Therefore, all non-Abelian subgroups $G\subset PSU(3)$ can be constructed by extensions of $A$ by a subgroup of $Aut(A)$ \cite{Ivanov:2012fp}:
\be 
G/A \to Aut(A), \quad \mbox{where} \quad A \lhd G.
\ee 

Let us check the automorphisms of each Abelian subgroup $A\subset PSU(3)$ and the resulting $G$;
\begin{itemize}
\item
$Aut(\Z_2)={1}$, therefore $G=\Z_2$ which is an Abelian group already considered in (\ref{list-Abelian}).

\item
$Aut(\Z_3)=\Z_2$, therefore $G$ is either $\Z_6$ or $D_6 \simeq \Z_3 \rtimes \Z_2$. However, $\Z_6$ is an Abelian group which does not appear in (\ref{list-Abelian}) and as a result is not realisable.

\item
$Aut(\Z_4)=\Z_2$, therefore $G$ is either $D_8\simeq \Z_4 \rtimes \Z_2$ or $Q_8$. However, a $Q_8$-symmetric potential is automatically symmetric under a continuous group of phase rotations and hence non-realisable.

\item
$Aut(\Z_2 \times \Z_2)=S_3$, therefore $G$ is $D_8\simeq (\Z_2 \times \Z_2) \times \Z_2$ or $A_4 \simeq (\Z_2 \times \Z_2) \rtimes \Z_3$ or $S_4 \simeq (\Z_2 \times \Z_2) \rtimes S_3$, where all three groups are realisable.

\item
$Aut(\Z_3 \times \Z_3)=GL_2(3)$, therefore $G$ is either $\Delta(54)/\Z_3 \simeq(\Z_3 \times \Z_3) \rtimes \Z_2$ or $\Sigma(36)\simeq (\Z_3 \times \Z_3) \rtimes \Z_4$, and both groups are realisable.

\end{itemize}

Therefore, we arrive at the full list of non-Abelian subgroups of $PSU(3)$ realisable as symmetry groups of a 3HDM potential:
\be 
D_6, \quad D_8, \quad A_4, \quad S_4, \quad \Delta(54)/\Z_3, \quad \Sigma(36).
\ee

\section{Orbit space}
\label{orbit}

The general renormalisable scalar potential of NHDMs is a combination of gauge-invariant
bilinears $(\phi_a^\dagger \phi_b)$, written compactly as \cite{Lavoura:1994fv,Botella:1994cs,Branco:1999fs}
\be 
V= Y_{ab}(\phi_a^\dagger \phi_b) +Z_{abcd}(\phi_a^\dagger \phi_b)(\phi_c^\dagger \phi_d)
\label{general-potential}
\ee

The space of electroweak-gauge orbits of Higgs fields (the orbit space) was first represented via bilinears for 2HDMs, \cite{Nishi:2006tg,Ferreira:2010yh,ivanov2HDM,2HDMbilinears}, and then extended to $N$ doublets in \cite{Ivanov:2010ww}.
 
The orbit space can be represented as a certain algebraic manifold in the Euclidean space $\mathbb{R}^{N^2}$ of the bilinears. It is convenient to group these bilinears in the following way:
\be
r_0 = \sqrt{{N-1\over 2N}}\sum_a \phi_a^\dagger \phi_a\,,\quad r_i = \sum_{a,b} \phi_a^\dagger \lambda^i_{ab}\phi_b\,,
\quad i = 1, \dots, N^2-1 
\label{rmu}
\ee
where $\lambda^i$ are the generators of $SU(N)$.
In terms of the bilinears, the Higgs potential can then be written as:
\be
V = - M_0 r_0 - M_i r_i + {1 \over 2}\Lambda_{00} r_0^2 + \Lambda_{0i} r_0 r_i + {1 \over 2}\Lambda_{ij} r_i r_j\,.
\label{potential}
\ee

The orbit space in NHDMs was characterised algebraically and geometrically in \cite{Ivanov:2010ww} and obeys the following conditions
\be
r_0 \ge 0\,,\quad {N-2 \over 2(N-1)}r_0^2 \le \vec r^2  \le r_0^2 ,
\ee
lying between two forward cones. It is interesting to note that in the case of 2HDMs ($N=2$) the inner cone disappears and the orbit space fills the entire forward cone $(0 \le \vec r^2  \le r_0^2)$.

It is known that the neutral vacua always lie on the surface of the outer cone $\vec r^2 = r_0^2$,
and the charge-breaking vacua occupy a certain region strictly inside the cone, $\vec r^2 < r_0^2$. 

In the particular case of 3HDMs, the bilinears are as follows:
\begin{small}
\bea
&& r_0 = {(\phi_1^\dagger\phi_1) + (\phi_2^\dagger\phi_2) + (\phi_3^\dagger\phi_3)\over\sqrt{3}}\,,\ 
r_3 = {(\phi_1^\dagger\phi_1) - (\phi_2^\dagger\phi_2) \over 2}\,,\ 
r_8 = {(\phi_1^\dagger\phi_1) + (\phi_2^\dagger\phi_2) - 2(\phi_3^\dagger\phi_3) \over 2\sqrt{3}} \quad
\nonumber\\
&&r_1 = \Re(\phi_1^\dagger\phi_2)\,,\quad 
r_4 = \Re(\phi_3^\dagger\phi_1)\,,\quad 
r_6 = \Re(\phi_2^\dagger\phi_3)\,,\nonumber\\[2mm] 
&&r_2 = \Im(\phi_1^\dagger\phi_2)\,,\quad
r_5 = \Im(\phi_3^\dagger\phi_1)\,,\quad
r_7 = \Im(\phi_2^\dagger\phi_3)\,. \label{ri3HDM}
\eea
\end{small}
The orbit space in 3HDM is defined by 
\be
r_0 \ge 0\,,\quad {1 \over 4} r_0^2\le \vec r^2 \le r_0^2,\quad \sqrt{3}d_{ijk} r_i r_j r_k = {3 \vec r^2 - r_0^2\over 2}r_0\,,
\label{3HDMconditions}
\ee
where $d_{ijk}$ is the fully symmetric $SU(3)$ tensor \cite{Ivanov:2010ww}.

\section{The geometric minimisation method}\label{Appendix-geometric}

The geometric minimisation method was developed for highly symmetric potentials in \cite{Degee:2012sk}, namely for $A_4$ and $S_4$ symmetric 3HDMs. 
Here we briefly introduce the method and show the results in $A_4$, $S_4$, $\Delta(54)/Z_3$ and $\Sigma(36)$ symmetric 3HDM potentials.

As discussed in Appendix~\ref{orbit}, a 3HDM scalar potential can be written as:
\be 
V = - M_0 r_0 - M_i r_i + {1 \over 2}\Lambda_{00} r_0^2 + \Lambda_{0i} r_0 r_i + {1 \over 2}\Lambda_{ij} r_i r_j\,.
\ee
The approach in \cite{Degee:2012sk} is applicable to potentials with sufficiently high symmetry so that $M_i=0$, which is a characteristic of the so called frustrated symmetries \cite{Ivanov:2010zx}.

With $M_i=0$, the potential can be generically written as
\be
V = - M_0 r_0 + r_0^2\sum_{i=0}^k \Lambda_i x_i \quad (\mbox{where} \quad x_i=\frac{r_i r_j}{r_0^2}, \quad x_0=1)
\label{potential2}
\ee
with $k$ different quartic terms where $k$ is usually small for highly symmetric potentials. 

We now calculate all $x_i$, for all possible values of $r$'s inside the orbit space, which will fill a certain region in the space $\RR^k$. Therefore, this region, denoted by $\Gamma$, is the orbit space fitted into the $x_i$ space.

Minimisation of the potential is done by constructing the geometric shape of $\Gamma$, in several steps;
\begin{itemize}
\item 
The potential (\ref{potential2}) is a linear function of $x_i$, therefore a ''steepest descent'' direction $\vec n = - (\Lambda_1\,, \dots\,, \Lambda_k)$ can be introduced in which the potential reduces its value the fastest.
\item
The potential can then be rewritten as
\be
V = - M_0 r_0 + r_0^2\left(\Lambda_0 - \vec n \vec x\right)\,.\label{potential3}
\ee
\item
The points sitting the farthest in the direction of $\vec n$ represent the minima of the potential.
\item 
Having identified these points $x_i$, we find their realisations in terms of fields. 
\end{itemize}
We present the results of applying this method to the frustrated symmetries in the list (\ref{frustrated}) in the following sections.

\subsection{A$_4$-symmetric potential}
Recall that the $A_4$ symmetric potential in terms of the fields has the following form
\bea
V_{A_4} &=& -\mu^2 \left[(\phi_1^\dagger\phi_1) 
+ (\phi_2^\dagger\phi_2) + (\phi_3^\dagger\phi_3)\right] 
+\lambda_{11} \left[(\phi_1^\dagger\phi_1) + (\phi_2^\dagger\phi_2) + (\phi_3^\dagger\phi_3)\right]^2  \\
&& + \lambda_{12} \left[(\phi_1^\dagger\phi_1)(\phi_2^\dagger\phi_2) + (\phi_2^\dagger\phi_2)(\phi_3^\dagger\phi_3) + (\phi_3^\dagger\phi_3)(\phi_1^\dagger\phi_1)\right]\nonumber\\
&& +\lambda'_{12}\left(|\phi_1^\dagger\phi_2|^2 +|\phi_2^\dagger\phi_3|^2+ |\phi_3^\dagger\phi_1|^2\right)+\lambda_1  \left[(\phi_1^\dagger\phi_2)^2 + (\phi_2^\dagger\phi_3)^2 + (\phi_3^\dagger\phi_1)^2\right] + h.c. \nonumber
\label{A4-geometric}
\eea
The potential is rewritten using the definition of $r_i$s in 3HDMs in Eq.~(\ref{ri3HDM}): 
\bea
V&=& -(\sqrt{3}\mu^2) r_{0}
+(\frac{9\lambda_{11}+3\lambda_{12}}{2})r_{0}^2
+(\lambda'_{12}+2\Re\lambda_1)(r_{1}^2+r_{4}^2+r_{6}^2) \nonumber\\
&&+(\lambda'_{12}-2\Re\lambda_1)(r_{2}^2+r_{5}^2+r_{7}^2)
+(\frac{-3\lambda_{11}-\lambda_{12}}{2})(r_{3}^2+r_{8}^2)
\nonumber\\
&&+(4i\Im\lambda_1)(r_{1}r_{2}+r_{4}r_{5}+r_{6}r_{7}) \nonumber
\eea
which could be written in the simplified form:
\be 
V = -M_{0}r_{0}+r_{0}^2(\Lambda_{0} + \Lambda_{1} x + \Lambda_2 y + \Lambda_3 z + \Lambda_4 t) \nonumber
\ee
where
\bea 
&& M_0 = \sqrt{3}\mu^2, \quad \Lambda_{0}=\frac{9\lambda_{11}+3\lambda_{12}}{2}, \quad \Lambda_{1}= \lambda'_{12}+2\Re\lambda_1 \nonumber\\
&& \Lambda_2=\lambda'_{12}-2\Re\lambda_1, \quad \Lambda_3=\frac{-3\lambda_{11}-\lambda_{12}}{2} , \quad \Lambda_4=4i\Im\lambda_1 \nonumber\\
&& x= \frac{r_{1}^2+r_{4}^2+r_{6}^2}{r_{0}^2}, \quad y=\frac{r_{2}^2+r_{5}^2+r_{7}^2}{r_{0}^2}, \quad z= \frac{r_{3}^2+r_{8}^2}{r_{0}^2}   ,\quad t = \frac{r_1r_2 + r_4r_5 + r_6r_7}{r_0^2}. \nonumber
\eea
The neutral part of the orbit space has a complicated shape \cite{Degee:2012sk} with its four vertices corresponding to the four neutral global minimum alignment for this potential, which are the following:
\be
 (0,0,1)\,,\quad  (1,1,1)\,,\quad  ( 1, e^{{i\pi/3}}, e^{-{i\pi/3}})\,,\quad  (0, 1, e^{{i\alpha}}) \nonumber
\ee
where only the relative magnitude of VEVs is given and in each case arbitrary permutation and sign change of doublets are allowed.

The point $(0,0,1)$ becomes the global minimum of the potential (\ref{A4-geometric}) when
\be 
\lambda_{11}+\lambda_{12}>0, \qquad  (2\lambda'_{12} + 3\lambda_{11} + 3\lambda_{12} )^2> 16(\Re^2{\lambda_1}-\Im^2{\lambda_{1}}) 
\ee
in addition to the conditions in (\ref{A4-condition}).

\subsection{S$_4$-symmetric potential}
Recall the $S_4$-symmetric potential:
\bea
V_{S_4} &=& -\mu^2 \left[(\phi_1^\dagger\phi_1) + (\phi_2^\dagger\phi_2) + (\phi_3^\dagger\phi_3)\right] +
\lambda_{11} \left[(\phi_1^\dagger\phi_1) + (\phi_2^\dagger\phi_2) + (\phi_3^\dagger\phi_3)\right]^2 \\
&& + \lambda_{12} \left[(\phi_1^\dagger\phi_1)(\phi_2^\dagger\phi_2) + (\phi_2^\dagger\phi_2)(\phi_3^\dagger\phi_3) + 
(\phi_3^\dagger\phi_3)(\phi_1^\dagger\phi_1)\right]\nonumber\\
&& + \lambda'_{12} \left(|\phi_1^\dagger\phi_2|^2 + |\phi_2^\dagger\phi_3|^2 + |\phi_3^\dagger\phi_1|^2\right) +\lambda_1  \left[(\phi_1^\dagger\phi_2)^2 + (\phi_2^\dagger\phi_3)^2 + (\phi_3^\dagger\phi_1)^2\right] + h.c. \nonumber
\label{S4-geometric}
\eea
which in terms of the $r_i$s has the following form:
\bea
V&=&-(\sqrt{3}\mu^2)r_{0}
+(\frac{9\lambda_{11}+3\lambda_{12}}{2})r_{0}^2
+(\lambda'_{12}+2\lambda_1)(r_{1}^2+r_{4}^2+r_{6}^2) \nonumber\\
&&+(\lambda'_{12}-2\lambda_1)(r_{2}^2+r_{5}^2+r_{7}^2)
+(\frac{-3\lambda_{11}-\lambda_{12}}{2})(r_{3}^2+r_{8}^2)\nonumber
\eea
which could be written in the simplified form:
\be 
V= -M_{0}r_{0}+r_{0}^2(\Lambda_{0} + \Lambda_{1} x + \Lambda_2 y + \Lambda_3 z)
\ee
where
\bea 
&& \Lambda_{1}= \lambda'_{12}+2\lambda_1 , \quad \Lambda_2=\lambda'_{12}-2\lambda_1 
\eea
and the same expression as in the $A_4$ case for $M_{0},\Lambda_{0},\Lambda_3$ and $x,y,z$.

The neutral part of the orbit space has the shape of a trapezoid \cite{Degee:2012sk} with its four vertices corresponding to the four neutral global minimum alignment for this potential, which are the following:
\be
(0,0, 1)\,,\quad ( 1, 1, 1)\,,\quad ( 1, e^{{i\pi/3}}, e^{-{i\pi/3}})\,,
\quad (0, e^{{i\pi/4}}, e^{-{i\pi/4}}) 
\ee
where in each case, arbitrary permutation and sign change of doublets are allowed.

The alignment $(0,0,1)$ becomes the global minimum of the potential (\ref{S4-geometric}) when
\be 
\lambda_{11}>0, \quad \lambda_{11}+\lambda_{12} >0, \quad  3(\lambda_{11}+\lambda_{12})> -\lambda'_{12} \pm 2\lambda_1  
\ee
in addition to the conditions in (\ref{S4-condition}).

\subsection{$\Delta(54)/Z_3$-symmetric potential}
The $\Delta(54)/Z_3$ symmetric potential has the following form, in terms of the fields:
\begin{small}
\bea
V_{\Delta(54)/ Z_3} & = &  - \mu^2 \left[\phi_1^\dagger \phi_1+ \phi_2^\dagger \phi_2+\phi_3^\dagger \phi_3\right]
+ \lambda_{11} \left[\phi_1^\dagger \phi_1+ \phi_2^\dagger \phi_2+\phi_3^\dagger \phi_3\right]^2 \nonumber\\
&&+ \lambda_{12} \left[(\phi_1^\dagger \phi_1)^2+ (\phi_2^\dagger \phi_2)^2+(\phi_3^\dagger \phi_3)^2
- (\phi_1^\dagger \phi_1)(\phi_2^\dagger \phi_2) - (\phi_2^\dagger \phi_2)(\phi_3^\dagger \phi_3)
- (\phi_3^\dagger \phi_3)(\phi_1^\dagger \phi_1)\right]\nonumber\\
&&+ \lambda'_{12} \left[|\phi_1^\dagger \phi_2|^2 + |\phi_2^\dagger \phi_3|^2 + |\phi_3^\dagger \phi_1|^2\right] \nonumber\\
&&+ \lambda_1 \left[(\phi_1^\dagger \phi_2)(\phi_1^\dagger \phi_3) + (\phi_2^\dagger \phi_3)(\phi_2^\dagger \phi_1) + (\phi_3^\dagger \phi_1)(\phi_3^\dagger \phi_2)\right]+ h.c. \nonumber
\label{Delta-geometric}
\eea
\end{small}
and in terms of $r_i$s:
\bea
V&=&-(\sqrt{3}\mu^2)r_{0}+(3\lambda_{11})r_{0}^2+ (3\lambda_{12})(r_{3}^2+r_{8}^2)+ (\lambda'_{12})\biggl[(r_{1}^2+r_{4}^2+r_{6}^2) +(r_{2}^2+r_{5}^2+r_{7}^2) \biggr]\nonumber\\
&& +(2\Re\lambda_{1})\biggl[(r_{1}r_{4}+ r_{4}r_{6}+ r_{6}r_{1}) + (r_{2}r_{5}+ r_{5}r_{7}+ r_{7}r_{2}) \biggr] \nonumber\\
&&   +(2i\Im\lambda_{1}) \biggl[r_{1}r_{5}- r_{2}r_{4}+ r_{5}r_{6} - r_{4}r_{7}+ r_{7}r_{1}- r_{2}r_{6} \biggr] \nonumber\\
&=& -(\sqrt{3}\mu^2)r_{0}+r_{0}^2 \biggl[3\lambda_{11} + 3\lambda_{12} z + \lambda'_{12} (x+y) + 2\Re\lambda_{1} (x'+y') +2i\Im\lambda_{1} t' \biggr]
\eea
where
\bea
&& x'=\frac{r_{1}r_{4}+ r_{4}r_{6}+ r_{6}r_{1}}{r^2_0}, \quad y'=\frac{r_{2}r_{5}+ r_{5}r_{7}+ r_{7}r_{2}}{r^0_2} \nonumber\\
&& t'=\frac{r_{1}r_{5}- r_{2}r_{4}+ r_{5}r_{6} - r_{4}r_{7}+ r_{7}r_{1}- r_{2}r_{6}}{r^2_0}
\eea
and the same expression as in the $A_4$ case for $x,y,z$.

The points satisfying the neutral global minima conditions are the following:
\be
 (0,0,1) ,\quad   (0,0,e^{{i\pi/2}}),\quad  (0,1,e^{{i\pi/2}}) \nonumber
\ee
provided
\be 
|\Re\lambda_1| < |\lambda'_{12} -3\lambda_{12}|.
\ee
In each case of the minimum points, arbitrary permutation and sign change of doublets are allowed.
The alignment $(0,0,1)$ becomes the global minimum of the potential (\ref{Delta-geometric}) when the following conditions, in addition to (\ref{Delta-condition}), are satisfied
\be 
\lambda_{11}>0, \quad \lambda_{12}>0, \quad \lambda'_{12}>0.
\ee

\subsection{$\Sigma(36)$-symmetric potential}

The $\Sigma(36)$ symmetric potential has the following form, in terms of the fields:
\bea
V_{\Sigma(36)} & = & - \mu^2 \left(\phi_1^\dagger \phi_1+ \phi_2^\dagger \phi_2+\phi_3^\dagger \phi_3 \right)
 + \lambda_{11} \left(\phi_1^\dagger \phi_1+ \phi_2^\dagger \phi_2+\phi_3^\dagger \phi_3 \right)^2 \nonumber\\
&&+ \lambda'_{12} \left(|\phi_1^\dagger \phi_2 - \phi_2^\dagger \phi_3|^2 + |\phi_2^\dagger \phi_3 - \phi_3^\dagger \phi_1|^2 + |\phi_3^\dagger \phi_1 - \phi_1^\dagger \phi_2|^2\right) 
\label{Sigma-geometric}
\eea
and in terms of the $r_i$s:
\bea
V&=&-(\sqrt{3}\mu^2)r_{0}+(3\lambda_{11})r_{0}^2 \nonumber\\
&&+ (2\lambda'_{12})\biggl[(r_{1}^2+r_{4}^2+r_{6}^2) -(r_{1}r_{4}+ r_{4}r_{6}+ r_{6}r_{1})
 \nonumber\\
&& \qquad \qquad +(r_{2}^2+r_{5}^2+r_{7}^2)- (r_{2}r_{5}+ r_{5}r_{7}+ r_{7}r_{2}) \biggr] \nonumber\\
&=& -(\sqrt{3}\mu^2) r_{0} + r_{0}^2 \biggl[ 3\lambda_{11} + 2\lambda'_{12} (x-x'+y-y') \biggr]. \nonumber
\eea
The orbit space of this potential has a simple parabolic shape in the 3-dimensional space of $(r_1,r_4,r_6)$ (or equivalently in the space of $(r_2,r_5,r_7)$).

The points satisfying the neutral global minima conditions are the following:
\be
 (0,0,1) ,\quad  (0,0,e^{{i\pi/2}}) ,\quad  (0,1,e^{{i\pi/2}}).
\ee
In each case of the minimum points, arbitrary permutation and sign change of doublets are allowed.
The alignment $(0,0,1)$ becomes the global minimum of the potential (\ref{Sigma-geometric}) when the following condition, in addition to (\ref{Sigma-condition}), is satisfied:
\be 
\lambda_{11}>0 .
\ee

\end{document}